\documentclass{jpp}
\usepackage{graphicx}
\usepackage{epstopdf, epsfig}
\usepackage{hyperref}
\usepackage{amssymb}
\usepackage{amsmath}
\usepackage{xcolor}
\allowdisplaybreaks

\renewcommand\vec\mathbf

\shorttitle{Temperature dependence of shallow-angle magnetic presheaths}
\shortauthor{A. Geraldini, F. I. Parra and F. Militello}

\title{Dependence on ion temperature of shallow-angle magnetic presheaths with adiabatic electrons}

\author{Alessandro Geraldini\aff{1,2,3}
  \corresp{\email{ale.gerald@gmail.com}},
  F. I. Parra\aff{2}
 \and F. Militello\aff{3}}

\affiliation{\aff{1}Institute for Research in Electronics and Applied Physics, University of Maryland, College Park, MD 20740, USA
\aff{2}Rudolf Peierls Centre for Theoretical Physics, University of Oxford, Oxford, OX1 3NP, UK
\aff{3}Culham Centre for Fusion Energy, Culham Science Centre, Abingdon, OX14 3DB, UK}

\begin{document}

\maketitle

\begin{abstract}
The magnetic presheath is a boundary layer occurring when magnetized plasma is in contact with a wall and the angle $\alpha$ between the wall and the magnetic field $\vec{B}$ is oblique.
Here, we consider the fusion-relevant case of a shallow-angle, $\alpha \ll 1$, electron-repelling sheath, with the electron density given by a Boltzmann distribution, valid for $\alpha / \sqrt{\tau+1}  \gg  \sqrt{m_{\text{e}}/m_{\text{i}}}$, where $m_{\text{e}}$ is the electron mass, $m_{\text{i}}$ is the ion mass, $\tau = T_{\text{i}}/ZT_{\text{e}}$, $T_{\text{e}}$ is the electron temperature, $T_{\text{i}}$ is the ion temperature, and $Z$ is the ionic charge state. 
The thickness of the magnetic presheath is of the order of a few ion sound Larmor radii $\rho_{\text{s}} = \sqrt{m_{\text{i}} \left(ZT_{\text{e}} + T_{\text{i}} \right) } / ZeB$, where $e$ is the proton charge and $B = |\vec{B}|$ is the magnitude of the magnetic field.
We study the dependence on $\tau $ of the electrostatic potential and ion distribution function in the magnetic presheath by using a set of prescribed ion distribution functions at the magnetic presheath entrance, parameterized by $\tau$. 
The kinetic model is shown to be asymptotically equivalent to Chodura's fluid model at small ion temperature, $\tau \ll 1$, for $|\ln \alpha| > 3|\ln \tau | \gg 1$.
In this limit, despite the fact that fluid equations give a reasonable approximation to the potential, ion gyro-orbits acquire a spatial extent that occupies a large portion of the magnetic presheath.
At large ion temperature, $\tau \gg 1$, relevant because $T_{\text{i}}$ is measured to be a few times larger than $T_{\text{e}}$ near divertor targets of fusion devices, ions reach the Debye sheath entrance (and subsequently the wall) at a shallow angle whose size is given by $\sqrt{\alpha}$ or $1/\sqrt{\tau}$, depending on which is largest.
\end{abstract}

\section{Introduction}

Plasma-wall interaction is important in systems such as plasma discharges \citep{Lieberman-book}, fusion devices \citep{Stangeby-book}, magnetic filters \citep{Anders-1995-filters}, plasma probes \citep{Hutchinson-book} and thrusters \citep{Martinez-1998}. 
In the context of nuclear fusion research, the plasma-wall interaction at the divertor or limiter targets of fusion devices is directly related to the boundary conditions to be imposed \citep{Loizu-2012} on models of plasma in the open-field line region (the Scrape-Off-Layer).
The heat flux reaching the wall of the device must be minimized and one way to do so is to make the magnetic field lines reach the divertor or limiter target at a shallow angle $\alpha \ll 1$ ($\alpha$ is measured in radians unless otherwise indicated) \citep{Loarte-2007}. 
In typical devices, $\alpha \sim 0.05-0.2 \text{ radians} \left( \sim 3-12^{\circ} \right) $, and in ITER it is expected that $\alpha \sim 0.04 \text{ radians} \sim 2.5^{\circ}$ \citep{Pitts-2009}.
Hence, it is crucial to understand plasma-wall interaction at such small angles in order to address the problem of exhaust in fusion plasmas. 

The magnetic presheath \citep{Chodura-1982} is a boundary layer with a width of a few ion sound Larmor radii, $\rho_{\text{s}} = \sqrt{m_{\text{i}} \left(ZT_{\text{e}} + T_{\text{i}} \right) } / ZeB$, next to the wall, where $T_{\text{i}}$ and $T_{\text{e}}$ are the ion and the electron temperatures respectively, $m_{\text{i}}$ is the ion mass, $Z$ is the ionic charge state, $e$ is the proton charge and $B$ is the magnetic field strength.
This region is characterized by a balance between electric and magnetic forces on the ions.
Closer to the wall, in steady state, there is a non-neutral layer called Debye sheath which typically repels electrons. 
The Debye sheath has a thickness of a few Debye lengths, $\lambda_{\text{D}} = \sqrt{ \epsilon_0 T_{\text{e}} / e^2 n_{\text{e}}  }$, where $n_e$ is the electron density and $\epsilon_0$ is the permittivity of free space, and is characterized by the electric forces dominating the ion dynamics.
The Debye length is generally much smaller than the ion sound gyroradius, $\lambda_{\text{D}} \ll \rho_{\text{s}}$, and therefore the magnetic presheath can be solved as a separate quasineutral system.
Moreover, we assume that ions collide for the last time when they are a distance $d_{\text{coll}} \gg \rho_{\text{s}}$ away from the wall, and therefore the magnetic presheath is collisionless.
The latter assumption is expected to hold in attached divertor regimes of operation, whereas in detached divertors the temperature is so low that the collisional scale may be small enough to make $d_{\text{coll}} \sim \rho_{\text{s}}$ \citep{Tskhakaya-2017}.

Due to their small mass relative to the ions, most electrons are usually repelled by the sheath electric field, and we thus assume electrons to be in thermal equilibrium.
This assumption becomes less valid when the angle between the magnetic field and the wall is very small and when the ion temperature is sufficiently large compared to the electron temperature (as we will see in section \ref{sec-orderings}).
For the ions, many magnetic presheath models use fluid equations, which rely on $T_{\text{i}} = 0$ \citep{Chodura-1982, Riemann-1994, Ahedo-1997, Ahedo-2009}. 
However, in the vicinity of the divertor target of a typical tokamak plasma, the ion temperature is at least as large as the electron temperature, $T_{\text{i}}  \sim T_{\text{e}} $ \citep{Mosetto-2015}, making a kinetic treatment of the ions necessary \citep{Siddiqui-Hershkowitz-2016}.
In this paper, we study the dependence of the magnetic presheath on the parameter
\begin{align} \label{tau}
\tau = \frac{ T_{\text{i}} }{ZT_{\text{e}}} \text{,}
\end{align}
which is of fundamental importance in kinetic models of turbulence.
For $Z=1$, $\tau$ is simply the ratio of ion to electron temperature.
Early attempts to solve the magnetic presheath by retaining the ion distribution function made use of analytical solutions of the ion trajectories \citep{Holland-Fried-Morales-1993, Parks-Lippmann-1994, Cohen-Ryutov-1998, Daube-Riemann-1999} with a variety of assumptions, giving valuable insight into the characteristics of the ion motion in the magnetic presheath.
Later, there were several particle-in-cell (PIC) studies of the Chodura and Debye sheaths \citep{Tskhakaya-2003, Tskhakaya-2004, Khaziev-Curreli-2015}, as well as some kinetic simulations using a Eulerian-Vlasov approach \citep{Coulette-2014, Coulette-Manfredi-2016}. 
Here, we use analytical solutions of the ion trajectories in a magnetic field whose angle with the wall is small \citep{Holland-Fried-Morales-1993, Cohen-Ryutov-1998}. 
An asymptotic theory of magnetic presheaths with $\alpha \ll 1$, and an associated numerical scheme to obtain self-consistent solutions of the electrostatic potential, was presented in detail in \cite{Geraldini-2017} and \cite{Geraldini-2018}.
The method also determines the ion distribution function at the Debye sheath entrance. 
Though only valid for grazing angles, this method has yielded several analytical results, is valid within the current paradigm of plasma exhaust in a fusion device, and is computationally fast.

This paper is structured as follows. 
The orderings and geometry of the magnetic presheath are discussed in Section \ref{sec-orderings}.
We use the shallow-angle ($\alpha \ll 1$) kinetic model described in \cite{Geraldini-2017, Geraldini-2018} which we briefly review in Section~\ref{sec-model}. 
In Section~\ref{sec-limits} we discuss the two limits of small $\tau$ ($\tau \ll 1$) and large $\tau$ ($\tau \gg 1/\alpha $) analytically using our kinetic model.
In particular, we show that our kinetic model is consistent with: the fluid model of \cite{Chodura-1982} for $\tau \ll 1$; a kinetic model that assumes a half-Maxwellian ion distribution function, briefly discussed in Section~3B of \cite{Cohen-Ryutov-1998}, for $\tau \gg 1/\alpha$.
Using a set of boundary conditions that recovers those used in the small and large temperature limits, numerical results of the shallow-angle kinetic model are obtained for finite values of $\tau$.
The boundary conditions and numerical results are presented in Section~\ref{sec-finite}. 
We conclude by summarizing and discussing our results in Section~\ref{sec-disc}.

In order to help the reader keep track of the several symbols used in this paper (many of which were introduced in references \cite{Geraldini-2017} and \cite{Geraldini-2018}), we include a glossary in Appendix \ref{app-symbols}.
For each symbol, the glossary includes a brief verbal definition (or an equation) and a reference to the equation where it first appears in the main text.

\section{Orderings} \label{sec-orderings}

\begin{figure}
\centering
\includegraphics[width=0.47\textwidth]{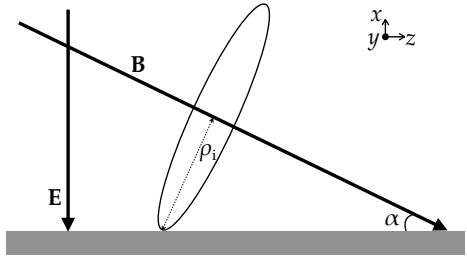}
\caption{
An ion gyro-orbit is shown schematically at a distance of approximately an ion gyroradius $\rho_{\text{i}}$ from the wall (grey horizontal surface).
The magnetic field is constant and the angle between the magnetic field and the wall is small, $\alpha \ll 1$ (in radians). The electric field is directed towards the wall and is a function of the co-ordinate $x$.
}
\label{fig-geometry-elrep}
\end{figure}

Consider a magnetized plasma in steady state, in the region $x\geqslant 0$, in contact with a wall, defined as the plane $x=0$.
We use a set of orthogonal axes, depicted in the top-right corner of figure \ref{fig-geometry-elrep}, with the $x$-axis aligned normal to the wall, and the $y$- and $z$-axes aligned in the two directions parallel to the wall.
The magnetic field is uniform and given by
\begin{align} \label{B-def}
\vec{B} = B \cos \alpha \hat{\vec{z}} - B \sin \alpha \hat{\vec{x}} \text{.}
\end{align}
In equation (\ref{B-def}), $\hat{\vec{x}}$ and $\hat{\vec{z}}$ denote unit vectors parallel to the $x$ and $z$-axes and $\alpha \ll 1$ is the small angle between the magnetic field and the wall.
The components of the ion velocity in the three directions are $v_x$, $v_y$ and $v_z$.
The system is uniform in the plane parallel to the wall, and thus every quantity is independent of the value of $y$ and $z$.
The ion motion can therefore be described using four co-ordinates: $x$, $v_x$, $v_y$, and $v_z$.

We consider a plasma with a single ion species and an electron species. 
An electric field normal to the wall is present to repel the most mobile of the plasma species --- the electrons --- away from the wall,
\begin{align} \label{E-field}
\vec{E} = - \phi'(x) \hat{\vec{x}} \text{,}
\end{align}
where $\phi$ is the electrostatic potential and a prime denotes differentiation with respect to $x$.
The electrostatic potential is assumed to monotonically converge to some value at $x \rightarrow \infty$, and this value is set to be $\phi = 0$.
Moreover, it has been shown that $\phi(x) - \phi(0) \propto \sqrt{x}$ at $x\rightarrow 0$ (see equations (141) and (142) in \cite{Geraldini-2018}), so that the magnetic presheath electric field diverges at the Debye sheath entrance\footnote{This is not a real divergence of the electric field, but is rather a large electric field satisfying $T_{\text{e}}/e\rho_{\text{s}} \ll  \phi'(0) \ll T_{\text{e}}/e\lambda_{\text{D}}$. See, for example, \cite{Riemann-review} for detailed explanations on the use of asymptotic methods for Debye sheaths and for certain types of presheath.}.  
The co-ordinate system and the geometry are depicted in figure \ref{fig-geometry-elrep}.

Since the electric field is present to repel electrons from the wall, the characteristic size of the electrostatic potential $\phi$ is given by
\begin{align}
e \phi \sim T_{\text{e}} \text{.}
\end{align}
Ions gain energies of the order of $Z e \phi \sim Z T_{\text{e}}$; at such energies, they have a velocity of the order of the Bohm speed,
\begin{align} \label{vB}
v_{\text{B}} = \sqrt{\frac{ZT_{\text{e}}}{m_{\text{i}}}} \text{.}
\end{align}
If the energy gained by the ions during this acceleration is smaller than their thermal energy, $ZT_{\text{e}} \lesssim T_{\text{i}}$, the typical ion velocity is the ion thermal speed,
\begin{align} \label{vti}
v_{\text{t,i}} = \sqrt{\frac{2T_{\text{i}} }{m_{\text{i}}}} \text{.}
\end{align}
From equations (\ref{vB}) and (\ref{vti}) it follows that, in general, the ion's speed has a characteristic size equal to the ion sound speed\footnote{Our definition of the ion sound speed is not the most general one, as in fluid treatments this quantity is often defined with an adiabatic constant multiplying the ion temperature.
Since the adiabatic constant is normally of order unity in size, the discrepancy in these definitions does not matter.} $c_{\text{s}}$,
\begin{align} \label{cs}
|\vec{v}| \sim c_s = \sqrt{ v_{\text{B}}^2 + \frac{1}{2}v_{\text{t,i}}^2 } = \sqrt{\frac{ZT_{\text{e}} + T_{\text{i}}}{m_{\text{i}}}} \text{.}
\end{align}
Note that $c_{\text{s}} = v_{\text{B}}$ when $\tau = 0$ and $c_{\text{s}} = \sqrt{T_{\text{i}}/m_{\text{i}}} = v_{\text{t,i}} / \sqrt{2}$ when $\tau = \infty$.

We proceed to argue that the typical size of the magnetic presheath, denoted $d_{\text{mps}}$, is the ion sound gyroradius \citep{Chodura-1982},
\begin{align} \label{rho-s}
\rho_{\text{s}} = \frac{c_s}{\Omega} \text{,}
\end{align} 
where $\Omega = ZeB / m_{\text{i}}$ is the typical ion gyrofrequency.
We consider the two limits $\tau \ll 1$ and $\tau \gg 1$ separately. 
When the ion temperature is much smaller than the electron temperature, $\tau \ll 1$, the only way by which ions can acquire the Bohm velocity $v_{\text{B}}$ in the direction normal to the wall --- necessary to satisfy the Bohm condition at the Debye sheath entrance \citep{Riemann-review} --- is if the electric field becomes large enough that it demagnetizes the ion orbits. 
From the ordering $| \vec{v}| \sim v_{\text{B}}$ for the ion speed and by balancing the magnetic and electric forces, we obtain $\phi' (x) \sim T_{\text{e}}/ e d_{\text{mps}}  \sim v_{\text{B}} B $; hence, $d_{\text{mps}} \sim \rho_{\text{B}} $, where
\begin{align} \label{rho-B}
\rho_{\text{B}} = \frac{v_{\text{B}}}{\Omega} \text{.}
\end{align}
Since $\rho_{\text{B}} \simeq \rho_{\text{s}}$ for $\tau \ll 1$, this is consistent with $d_{\text{mps}} \sim \rho_{\text{s}}$.
When the ion temperature is large, $\tau \gg 1$, the radius of gyration of the ions is larger than $\rho_{\text{B}}$.
The length scale of the magnetic presheath is set by the ion density variation, and therefore must satisfy $d_{\text{mps}} \sim  \rho_{\text{i}} = v_{\text{t,i}} / \Omega $, where $\rho_{\text{i}}$ is the ion gyroradius.
Since $\rho_{\text{i}} \simeq \sqrt{2} \rho_{\text{s}}$ for $\tau \gg 1$, this is consistent with $d_{\text{mps}} \sim \rho_{\text{s}}$.
When $\tau \sim 1$, both arguments are valid, since $\rho_{\text{i}} \sim \rho_{\text{B}} \sim  \rho_{\text{s}}$. 

The assumption of an electron-repelling wall is not valid for any value of $\alpha$ and $\tau$.
We proceed to obtain the condition on $\alpha$ and $\tau$ for this assumption to be valid.
We expect electrons to travel at characteristic velocities equal to their thermal speed,
\begin{align} \label{vte}
v_{\text{t,e}} = \sqrt{\frac{2T_{\text{e}} }{m_{\text{e}}}} \text{,}
\end{align}
where $m_{\text{e}}$ is the electron mass.
The typical electron velocity is so large, $v_{\text{t,e}} \gg v_{\text{B}}$, that electrons are virtually unaffected by the electric field, since they are subject to magnetic forces, $ev_yB \sim ev_{\text{t,e}} B$, much larger than electric forces, $e\phi' \lesssim e v_{\text{B}} B$.
Moreover, electron gyro-orbits are small, $\rho_{\text{e}} \ll \rho_{\text{s}}$.
Hence, averaging over the small-scale gyro-motion, the electrons in the magnetic presheath stream parallel to the magnetic field at a velocity of the order of $v_{\text{t,e}}$.
Conversely, the ion motion close to the wall in the magnetic presheath consists of gyro-orbits distorted by the electric field, and so the ions reach the wall travelling at a velocity of the order of $c_{\text{s}}$.
Considering an ion and an electron initially at a distance $\sim \rho_s$ from the wall, and remembering that the electron motion is constrained to be parallel to the magnetic field, the electron has to travel a longer distance than the ion by a factor of $1/\alpha$.
However, the electron travels this distance at a speed larger than the ion's by a factor $v_{\text{t,e}}/c_s = \sqrt{m_{\text{i}}/\left(m_{\text{e}}(1+\tau)\right)}$.
Hence, the electron reaches the wall in a shorter time than the ion if
\begin{align} \label{alpha-tau-order}
\sqrt{\frac{m_{\text{e}}}{m_{\text{i}}}} \sqrt{\tau + 1} \ll \alpha \text{.}
\end{align}
If condition (\ref{alpha-tau-order}) is satisfied, the wall repels most of the electrons back into the plasma, and the ordering for the magnitude of the ion velocity, equation (\ref{cs}), is self-consistent.

For an electron-repelling wall, the electron distribution function is typically considered to be well-approximated by a Maxwellian.
The reason for this is that the collisional processes outside of the collisionless sheath and presheath drive it to a Maxwellian, and the sheath repels most of the electrons back into the plasma.
Hence, the electron density is assumed to be given by a Boltzmann distribution.

\section{Kinetic ion model} \label{sec-model}

In this section we briefly review the shallow-angle kinetic model presented in detail in \cite{Geraldini-2017, Geraldini-2018}.
In Section \ref{subsec-traj} we use the asymptotic expansion in $\alpha \ll 1$ to write the ion velocity in terms of slowly varying orbit parameters, finding that there are approximately periodic solutions to the ion motion.
In moving across the magnetic presheath, ions conserve two quantities to lowest order in $\alpha$: the total energy $U$ and an adiabatic invariant $\mu$ \citep{Cohen-Ryutov-1998}. 
The adiabatic invariant is directly related to the approximately periodic nature of the ion motion, and coincides with the usual magnetic moment only when the electric field variation over the ion gyroradius scale is small.

When written as a function of $\mu$ and $U$, the distribution function is constant across the magnetic presheath, to lowest order in $\alpha$.
This is used, in Section \ref{subsec-distfunc}, to write an expression for the ion density.
In Section \ref{subsec-modelsummary}, we write the quasineutrality equation and summarize the main equations of the shallow-angle kinetic model.

\subsection{Ion trajectories in terms of slowly changing orbit parameters} \label{subsec-traj}

The equations of motion of an ion in the magnetic presheath are
\begin{align} \label{x-EOM}
\dot{v}_x = - \frac{\Omega \phi'}{B} + \Omega v_y \cos \alpha \text{,}
\end{align}
\begin{align} \label{y-EOM}
\dot{v}_y = -  \Omega v_x \cos \alpha - \Omega v_z \sin \alpha \text{,}
\end{align}
\begin{align} \label{z-EOM}
\dot{v}_z =  \Omega v_y \sin \alpha \text{.}
\end{align}
Expanding equations (\ref{x-EOM})-(\ref{z-EOM}) in $\alpha \ll 1$ and neglecting second order terms, we obtain
\begin{align} \label{x-EOM-alpha}
\dot{v}_x \simeq - \frac{\Omega \phi'}{B} + \Omega v_y  \text{,}
\end{align}
\begin{align} \label{y-EOM-alpha}
\dot{v}_y \simeq -  \Omega v_x  - \Omega v_z  \alpha \text{,}
\end{align}
\begin{align} \label{z-EOM-alpha}
\dot{v}_z \simeq  \Omega v_y  \alpha \text{.}
\end{align}
We introduce three orbit parameters: the orbit position 
\begin{align} \label{xbar-def}
\bar{x} = x + \frac{1}{\Omega} v_y \text{,}
\end{align}
the perpendicular energy 
\begin{align} \label{Uperp-def}
U_{\perp} = \frac{1}{2} v_x^2 + \frac{1}{2} v_y^2  + \frac{\Omega \phi (x)}{B}  \text{,}
\end{align}
and the total energy 
 \begin{align} \label{U-def}
U = \frac{1}{2} v_x^2 + \frac{1}{2} v_y^2  + \frac{1}{2} v_z^2  + \frac{\Omega \phi (x)}{B}  \text{.}
\end{align}
The orbit parameters vary over a timescale which is longer by a factor of $1/\alpha$ than the timescale $1/\Omega$ over which $x$, $v_x$ and $v_y$ vary, $\bar{x}/\dot{\bar{x}} \sim U_{\perp} / \dot{U}_{\perp} \sim 1/\alpha \Omega \gg |\vec{v}| / |\dot{\vec{v}}| \sim 1/\Omega$.
The total energy $U$ is exactly constant, $\dot{U} = 0$.
The instantaneous particle velocities can be expressed in terms of the instantaneous position $x$ and the orbit parameters:
\begin{align} \label{vx-x-xbar-Uperp}
v_x & = \pm V_x \left( x, \bar{x}, U_{\perp} \right) = \pm \sqrt{2\left(U_{\perp} - \chi (x, \bar{x})  \right) }  \text{,}
\end{align}
\begin{align} \label{vy-x-xbar}
v_y = \Omega \left( \bar{x} - x \right) \text{,}
\end{align}
\begin{align} \label{vz-Uperp-U}
v_z = V_{\parallel} \left( U_{\perp}, U \right) = \sqrt{2\left(U - U_{\perp} \right) } \text{,}
\end{align}
where 
\begin{align} \label{chi}
\chi (x, \bar{x}) =  \frac{1}{2} \Omega^2 \left( x - \bar{x} \right)^2 + \frac{\Omega \phi(x)}{B}  
\end{align}
is an effective potential function.
In equation (\ref{vz-Uperp-U}) we assumed $v_z >0$ because all ions enter the magnetic presheath with $v_z > 0$, are accelerated to larger values of $v_z$, reach the Debye sheath and are then absorbed by the wall \citep{Geraldini-2018}.
For convenience, in equation (\ref{vx-x-xbar-Uperp}) we introduced the symbol $V_x$ to denote the absolute value of $v_x$ as a function of $x$, $\bar{x}$ and $U_{\perp}$, and in equation (\ref{vz-Uperp-U}) we introduced the symbol $V_{\parallel}$ to denote $v_z$ as a function of $U_{\perp}$ and $U$.

For times comparable to the typical ion gyroperiod, $2\pi/\Omega$, the orbit parameters are constant to lowest order in $\alpha$ and equations (\ref{vx-x-xbar-Uperp})-(\ref{chi}) can be used to infer the approximate particle trajectory.
From equation (\ref{vx-x-xbar-Uperp}), the ion motion is periodic to lowest order in $\alpha$ if, for some $\bar{x}$ and $U_{\perp}$, turning points $x_{\text{b}}$ (bottom) and $x_{\text{t}}$ (top) exist such that: (i) $V_x \left( x_{\text{b}}, \bar{x}, U_{\perp} \right) = V_x \left( x_{\text{t}}, \bar{x}, U_{\perp} \right) = 0$ and (ii) $\chi(x, \bar{x}) \leqslant U_{\perp}$ in the interval $x_{\text{b}} \leqslant x \leqslant x_{\text{t}}$.
Then, the ion will move back and forth between $x_{\text{b}}$ and $x_{\text{t}}$ with period $\sim 2\pi/\Omega$.
In order to satisfy (ii), the turning points must lie on either side of an effective potential minimum $x_{\text{m}}$ which, by definition, satisfies
\begin{align} \label{chi-minimum-1}
\chi'(x_{\text{m}}, \bar{x}) = \Omega^2 \left( x_{\text{m}} - \bar{x} \right) + \frac{\Omega \phi' (x_{\text{m}})}{B} = 0 
\end{align}
and
\begin{align}
 \chi''(x_{\text{m}} ) = \Omega^2 + \frac{\Omega \phi''(x_{\text{m}} )}{B} > 0 \text{.}  \label{chi-minimum-2}
\end{align}
The value of $\chi$ evaluated at the effective potential minimum is, using equations (\ref{chi}) and (\ref{chi-minimum-1}),
\begin{align} \label{chim}
\chi _{\text{m}}(\bar{x}) = \chi ( x_{\text{m}}, \bar{x} ) = \frac{1}{2} \left( \frac{\phi'(x_{\text{m}})}{B} \right)^2  + \frac{\Omega \phi (x_{\text{m}})}{B}  \text{.}
\end{align}


The ion motion in the $x$ direction (normal to the wall) is exactly periodic for $\alpha =0$, with constant orbit parameters (in the $y$ direction, the motion is a sum of an exactly periodic motion and a constant $\vec{E} \times \vec{B}$ drift, as explained in \cite{Geraldini-2017}).
The small angle $\alpha$ perturbs the periodic motion by a small amount, since the orbit parameters become slowly changing in time. 
Thus, the ion velocity can be approximately decomposed into a periodic piece, with period $\sim 2\pi/\Omega$, and a piece that is approximately constant over the timescale of the periodic motion.
Under such circumstances, there is a quantity related to the underlying periodic motion, called an adiabatic invariant, which is a constant of the overall quasi-periodic motion to lowest order in the perturbation parameter\footnote{In fact, adiabatic invariants can usually be corrected at every order in such a way that they are conserved to all orders in the perturbation parameter.}.
The adiabatic invariant in this system is given by
\begin{align} \label{mu-Uperp-xbar}
\mu = \mu_{\text{gk}} ( \bar{x}, U_{\perp} ) =  \frac{1}{\pi} \int_{x_{\text{b}}}^{x_{\text{t}}} V_x \left( x, \bar{x}, U_{\perp} \right) dx \sim \frac{v_{\text{t,i}}^2}{\Omega}   \text{,}
\end{align}
and is constant to lowest order in $\alpha$.
The ordering $\mu \sim v_{\text{t,i}}^2/\Omega$ on the far right is obtained in the following way. 
We define the quantities
\begin{align} \label{rhox-def}
 \rho_x = x - x_{\text{m}} \sim  x_{\text{t}} - x_{\text{b}}   \text{,} 
\end{align} 
and 
\begin{align} \label{wx-def}
w_x = \dot{\rho}_x \sim \sqrt{2\left( U_{\perp} -\chi_{\text{m}} \right)} \text{.}
\end{align}
Note that the size of $w_x$ is the characteristic orbital velocity, and  the size of $\rho_x$ is the characteristic spatial extent of the orbit in the $x$ direction (normal to the wall).
From equation (\ref{mu-Uperp-xbar}) we estimate $\mu \sim w_x \rho_x$.
At the magnetic presheath entrance, where the electric field is very small, the ion gyro-orbit is circular to a good approximation.
Hence, the orbital velocity is of the order of the ion thermal velocity, $w_x \sim v_{\text{t,i}}$, and the ion orbit size is of the order of the ion thermal gyroradius, $\rho_x \sim \rho_{\text{i}} = v_{\text{t,i}}/\Omega$.
The ordering in equation (\ref{mu-Uperp-xbar}) follows because $\mu$ is an adiabatic invariant, and so $\mu$ is conserved to lowest order in $\alpha$ as the ion moves across the magnetic presheath.
Recall that the ion motion must retain an approximate periodicity for $\mu$ to be an adiabatic invariant.

\subsection{Ion density} \label{subsec-distfunc}

Treating the ion motion as periodic to lowest order in some expansion parameter is akin to conventional gyrokinetics \citep{Rutherford-1968, Taylor-1968, Catto-1978, Antonsen-1980,  Frieman-1982}.
At every point, the ion's trajectory can be approximated to lowest order by a periodic orbit whose period is faster than any other timescale of interest.
As in gyrokinetic theory, the ion distribution function can be shown to be independent of the fast timescale to lowest order in $\alpha$. 
Moreover, since $\mu$ and $U$ are both constants of the perturbed motion (at least to lowest order in $\alpha$), the distribution function written in terms of the variables $\mu$ and $U$, $F (\mu, U )$, can be shown to be constant across the magnetic presheath \citep{Cohen-Ryutov-1998, Geraldini-2017}.
Therefore, the function $F(\mu, U)$ is completely determined by ions entering the magnetic presheath at $x \rightarrow \infty$.
In order to write $F(\mu, U)$ from the distribution function at $x\rightarrow \infty$ expressed in terms of $\vec{v}$, denoted $f_{\infty} ( \vec{v})$, we use the equations
\begin{align} \label{mu-infty-vx-vy}
\mu = \frac{v_x^2 + v_y^2 }{2\Omega}  
\end{align}
and
\begin{align} \label{U-infty-vx-vy-vz}
U = \Omega \mu + \frac{1}{2} v_z^2 \text{.}
\end{align}
These equations are obtained by setting $\phi = 0$ in equations (\ref{U-def}) and (\ref{mu-Uperp-xbar}), and are thus valid at $x \rightarrow \infty$.
Note that the self-consistent form of $f_{\infty}(\vec{v})$ should be independent of the gyrophase angle, which at $x\rightarrow \infty$ is $\tan^{-1} (v_x/v_y)$.

The ion density, $n_{\text{i}}$, can be obtained by taking an integral in the velocity space variables $\bar{x}$, $U_{\perp}$ and $U$, as explained in \cite{Geraldini-2017, Geraldini-2018}. 
There are two distinct contributions to the ion density: one due to ions in quasiperiodic orbits
\begin{align} \label{ni-closed}
n_{\text{i,cl}}[\phi](x)  = \int_{\bar{x}_{\text{m}}(x)}^{\infty} \Omega d\bar{x} \int_{\chi(x, \bar{x})}^{\chi_{\text{M}}(\bar{x})} \frac{ 2 dU_{\perp}}{V_x \left( x, \bar{x}, U_{\perp} \right) } \int_{U_{\perp}}^{\infty} \frac{F\left(\mu_{\text{gk}} (\bar{x}, U_{\perp}), U \right) dU }{V_{\parallel} (U_{\perp}, U )} \text{,}
\end{align}
and another due ions that are about to intersect the wall,
\begin{align} \label{ni-open}
n_{\text{i,op}} [\phi](x)  = \int_{\bar{x}_{\text{m,o}}(x)}^{\infty} \Omega d\bar{x}  \int_{\chi_{\text{M}}(\bar{x})}^{\infty} \frac{F\left(\mu_{\text{gk}} (\bar{x}, \chi_{\text{M}}(\bar{x})), U \right) dU }{V_{\parallel} (\chi_{\text{M}}(\bar{x}), U )} \times \nonumber \\ \left[ V_x \left(x, \bar{x}, \chi_{\text{M}}(\bar{x}) + \Delta_{\text{M}} \left( \bar{x}, U \right) \right) - V_x \left(x, \bar{x}, \chi_{\text{M}}(\bar{x}) \right) \right]  \text{.}
\end{align}
The notation $f[\phi](x)$ represents a functional $f$ that depends on the whole function $\phi$, and not just on its value at a particular position $x$. 
In equation (\ref{ni-closed}), the subscript `cl' stands for `closed' and in equation (\ref{ni-open}) the subscript `op' stands for `open', corresponding to ions whose trajectory can be approximated by a closed orbit (i.e. periodic) and an open orbit (i.e. terminating at the wall).
The total ion density is the sum of the closed and open orbit densities of equations (\ref{ni-closed}) and (\ref{ni-open}), respectively,
\begin{align} \label{ni}
n_{\text{i}} (x) = n_{\text{i,cl}}[\phi](x) + n_{\text{i,op}}[\phi](x) \text{.}
\end{align}

In equations (\ref{ni-closed}) and (\ref{ni-open}), we have introduced several quantities which are derived and explained in detail in \cite{Geraldini-2017} and \cite{Geraldini-2018}, and we have assumed that $\phi(x)$, $\phi'(x)$ and $\phi''(x)$ are all monotonic functions of $x$.
The minimum allowed orbit position $\bar{x}_{\text{m}}$ for an ion at position $x$ to be in an orbit that is periodic to lowest order in $\alpha$ is
\begin{align} \label{xbarm-def}
\bar{x}_{\text{m}} \left( x \right) = \min_{s \in \left[0, x \right)}  \frac{1}{2}\left( x + s \right) + \frac{\phi \left( x \right) - \phi \left( s \right)}{\Omega B \left( x - s \right)} \text{.} 
\end{align}
The minimum allowed orbit position $\bar{x}_{\text{m,o}}$ for an ion at position $x$ to be in an orbit that is not periodic to lowest order in $\alpha$ is
\begin{align} \label{xbarm-open}
\bar{x}_{\text{m,o}} (x) = \begin{cases} 
\bar{x}_{\text{c}}  & \text{ for } x < x_{\text{c}} \text{,} \\
\bar{x}_{\text{m}} (x) &  \text{ for } x \geqslant  x_{\text{c}} \text{.}  
\end{cases}
\end{align}
In equation (\ref{xbarm-open}) we have introduced the two quantities $\bar{x}_{\text{c}}$ and $x_{\text{c}}$, defined via
\begin{align} \label{xbarc}
\bar{x}_{\text{c}} = \min_{x \in [0,\infty]} \left( x  + \frac{\phi'( x )}{\Omega B} \right) = x_{\text{c}} + \frac{\phi'(x_{\text{c}})}{\Omega B}  \text{.}
\end{align}
The effective potential maximum $\chi_{\text{M}}(\bar{x})$ is the largest value of $\chi(s, \bar{x})$ for a given value of $\bar{x}$ and for values of $s$ smaller than the position of the effective potential minimum $x_{\text{m}}$,
\begin{align} \label{chiM-def}
\chi_{\text{M}} \left( \bar{x} \right) = \chi(x_{\text{M}}, \bar{x}) = \max_{s \in \left[ 0, x_{\text{m}} \right]} \chi \left(s, \bar{x} \right) \text{.}
\end{align}
The quantity $x_{\text{M}}$ is the position of the effective potential maximum at a given value of $\bar{x}$.
For $\bar{x} = \bar{x}_{\text{c}}$, the values of $\chi_{\text{M}}$ and $\chi_{\text{m}}$ coincide with
\begin{align} \label{chic-def}
\chi_{\text{c}} \equiv \chi \left( x_{\text{c}}, \bar{x}_{\text{c}} \right) \text{.}
\end{align}
For $\bar{x} \geqslant \bar{x}_{\text{m}}(x)$, we are guaranteed to find $\chi_{\text{M}} (\bar{x}) \geqslant \chi(x, \bar{x}) $ and $x\geqslant x_{\text{M}}$, so that there are closed orbit solutions (to lowest order in $\alpha$) passing through $x$.
For $\bar{x} \geqslant \bar{x}_{\text{m,o}}(x)$, we are guaranteed to find $\chi_{\text{M}} (\bar{x}) \geqslant \chi(x, \bar{x}) $, so that there are open orbit solutions with $U_{\perp} \simeq \chi_{\text{M}} (\bar{x})$ passing through $x$.
Finally, the quantity $\Delta_{\text{M}}$ is the range of possible values of $v_x^2/2$ that an ion in an open orbit can have at a given value of $\bar{x}$ and $U$, and is given by
\begin{align} \label{DeltaM-mu}
\Delta_{\text{M}} \left( \bar{x}, U \right) & = 2\pi \alpha  V_{\parallel} \left( \chi_{\text{M}} (\bar{x}) , U \right)  \left. \frac{ d \mu }{ d \bar{x} } \right\rvert_{ \text{open} } 
 \text{,}
\end{align}
where 
\begin{align}
\left. \frac{ d \mu }{ d \bar{x} } \right\rvert_{ \text{open} } = \frac{ d  }{ d \bar{x} }  \left[ \mu_{\text{gk}} \left( \bar{x}, \chi_{\text{M}} (\bar{x}) \right) \right] \text{.} 
\end{align}
Equation (\ref{DeltaM-mu}) is derived in Appendix \ref{app-DeltaM} from the expression for $\Delta_{\text{M}}$ given in \cite{Geraldini-2018}.

\subsection{Quasineutrality and summary of equations} \label{subsec-modelsummary}

The magnetic presheath is quasineutral: the ion charge density is equal and opposite to the electron charge density, and so
\begin{align} \label{quasineutrality-general}
Zn_{\text{i}} (x) = n_{\text{e}} (x) \text{.}
\end{align}
Since the electrons are assumed to be in thermal equilibrium, the electron number density is 
\begin{align} \label{ne}
n_{\text{e}} (x) = Z n_{\infty} \exp \left( \frac{e\phi(x)}{T_{\text{e}}} \right) \text{,}
\end{align}
where $n_{\infty}$ is the ion density for $x\rightarrow \infty$. 
Using equations (\ref{ni}) and (\ref{ne}), the quasineutrality equation of our kinetic model can be written as
\begin{align} \label{quasineutrality}
n_{\text{i,cl}}[\phi](x) + n_{\text{i,op}}[\phi](x) = n_{\infty} \exp \left( \frac{e\phi(x)}{T_{\text{e}}} \right) \text{.}
\end{align}
Equation (\ref{quasineutrality}) is used to determine the self-consistent electrostatic potential $\phi(x)$ across the magnetic presheath.
A condition that must be satisfied in order for equation (\ref{quasineutrality}) to have a solution is \citep{Geraldini-2018}
\begin{align} \label{kinetic-Chodura}
\int \frac{ f_{\infty} \left( \vec{v} \right)}{v_z^2} d^3v \leqslant \frac{n_{\infty}}{v_{\text{B}}^2}   \text{,}
\end{align}
which we refer to as the kinetic Chodura condition.

Once $\phi(x)$ is calculated, we can obtain several interesting quantities.
The component $u_x$ of the ion fluid velocity in the direction normal to the wall is obtained by use of the steady-state ion continuity equation $d/dx \left( n_{\text{i}} u_x  \right) = 0$.
The quasineutrality equation (\ref{quasineutrality-general}) and the expression for the electron density (\ref{ne}) lead to $n_{\text{i}} = n_{\infty} \exp \left( e\phi / T_{\text{e}} \right) $.
Hence, using the boundary conditions $n_{\text{i}}(\infty) = n_{\infty} $ and $u_x(\infty) = u_{x\infty}$, we obtain
\begin{align} \label{ux}
u_x (x) =u_{x\infty}  \exp \left( - \frac{ e\phi(x) }{T_{\text{e}} } \right) \text{.}
\end{align}
The value of $u_{x\infty}$ is obtained from the flow velocity in the direction parallel to the magnetic field at $x \rightarrow \infty$, projected in the direction normal to the wall.
Since to lowest order in $\alpha$ the velocity component $u_{z\infty}$ is equal to the component of the velocity parallel to the wall, we have
\begin{align} \label{ux-cont}
u_{x \infty} = - \alpha u_{z \infty} =  - \frac{ \alpha }{n_{\infty}} \int  v_z f_{\infty} ( \vec{v} ) d^3v  
 \text{.}
\end{align}

The ion distribution function at the Debye sheath entrance, $x=0$, is given by
\begin{align} \label{f0}
f_0 (\vec{v}) = & F(\mu, U ) \hat{\Pi} \left( v_x, -V_x \left( 0, \bar{x}, \chi_{\text{M}}(\bar{x}) + \Delta_M (\bar{x}, U) \right),  -V_x \left( 0, \bar{x}, \chi_{\text{M}}(\bar{x}) \right) \right)
\end{align}
where $\bar{x} = v_y / \Omega $ at $x=0$, and $\hat{\Pi}$ is the top-hat function defined by
\begin{align} \label{tophat}
 \hat{\Pi} \left( y, h_1, h_2 \right) = \begin{cases}
 1 & \text{ if } h_1 < y \leqslant h_2 \text{,} \\
 0 & \text{ else.}
 \end{cases}
\end{align}
In order to study the 3-dimensional distribution function at the Debye sheath entrance of equation (\ref{f0}), we define the distribution of the velocity component normal to the wall,
\begin{align} \label{f0x-def}
f_{0x} (v_x) = &  \int_0^{\infty} dv_y  \int_0^{\infty} f_0 (\vec{v}) d v_z =  \int_{\bar{x}_{\text{c}}}^{\infty} \Omega d\bar{x} \int_{\chi_{\text{M}}(\bar{x})}^{\infty} \frac{F\left(\mu (\bar{x}, \chi_{\text{M}}(\bar{x})), U \right)  dU }{V_{\parallel} (\chi_{\text{M}}(\bar{x}), U )} \nonumber \\ & \times  \hat{\Pi} \left[ v_x, -V_x \left( 0, \bar{x}, \chi_{\text{M}}(\bar{x}) + \Delta_M (\bar{x}, U) \right),  -V_x \left( 0, \bar{x}, \chi_{\text{M}}(\bar{x}) \right) \right] dU \text{,}
\end{align}
and the two-dimensional distribution of the velocity components tangential to the wall,
\begin{align} \label{f0yz-def}
f_{0yz} (v_y, v_z ) = & \int_{0}^{\infty} f_0(\vec{v}) dv_x  \nonumber \\ 
= & F(\mu ( \bar{x}, \chi_{\text{M}}(\bar{x}) ), U)  \left[ V_x \left(x, \bar{x}, \chi_{\text{M}}(\bar{x}) + \Delta_{\text{M}} (\bar{x}, U)  \right) - V_x \left(x, \bar{x}, \chi_{\text{M}} \right) \right]  \text{.}
\end{align}
Equation (\ref{f0x-def}) is obtained by integrating (\ref{f0}) over $\bar{x}$ and $U$, without integrating over $v_x$.
Equation (\ref{f0yz-def}) is obtained by integrating (\ref{f0}) over $v_x$ and re-expressing the distribution as a function of $v_y$ and $v_z$, using $v_y = \Omega \bar{x}$ (valid at $x=0$) and $v_z = \sqrt{2\left(U - \chi_{\text{M}}(\bar{x}) \right) }$.
In \cite{Geraldini-2018} it was shown that the equation
\begin{align} \label{kinetic-Bohm-marginal}
\int \frac{f_{0} (\vec{v})}{v_x^2} d^3 v \equiv \int \frac{f_{0x} (v_x)}{v_x^2} d v_x = \frac{n_{\text{i}}(0)}{v_{\text{B}}^2} \text{,}
\end{align}
which corresponds to the equality form of the well-known kinetic Bohm condition, is satisfied self-consistently by the magnetic presheath solution.
In the review paper \cite{Riemann-review}, it is shown that in most presheath models the Bohm condition is self-consistently satisfied in the equality form, as in (\ref{kinetic-Bohm-marginal}).

\section{The limits of small and large ion temperature} 
\label{sec-limits}

In order to study the effect of ion temperature in a kinetic model of the magnetic presheath, it is essential to check that the model is consistent with expected results in appropriate limits of $\tau$.
Here, we study the limits $\alpha^{1/3} \ll \tau \ll 1$ (cold ions) and $\tau \gg 1/\alpha \gg 1$ (hot ions) of the kinetic model introduced in section \ref{sec-model}.
For cold ions, $\tau \ll 1$ and so the ion distribution function is narrow when compared to the Bohm velocity, as $v_{\text{t,i}} = \sqrt{2\tau} v_{\text{B}} \ll v_{\text{B}}$.
Since we have argued in section \ref{sec-orderings} that the typical ion speed in the magnetic presheath is, for $\tau \ll 1$, the Bohm velocity, the ion distribution function can be taken to be a delta function to lowest order in $\tau$.
The result of this approximation is the fluid theory first presented in \cite{Chodura-1982}.
Conversely, for hot ions, $\tau \gg 1$ and the size of the Bohm velocity is negligible compared to the thermal velocity, $v_{\text{B}} = v_{\text{t,i}} / \sqrt{2\tau}  \ll v_{\text{t,i}} $.
This means that an accurate knowledge of the whole of the ion distribution function is important when studying the limit $\tau \gg 1$.
For the purpose of this paper, we assume that the ion distribution function is a half-Maxwellian when entering the collisionless magnetic presheath, as was done in \cite{Cohen-Ryutov-1998}.

\subsection{Cold ions ($\tau \ll 1$)} \label{sec-cold} 

In this subsection, we argue that our kinetic model is equivalent to Chodura's fluid model, which is valid for $\tau = 0$, in an appropriate limit for $\tau \ll 1$.

In order to compare the fluid and kinetic models with each other, we first briefly recap the fluid analysis.
We start by generalizing to arbitrary values of $\alpha$.
All ions are assumed to have the same velocity at a given position $x$, such that $\vec{v} = \vec{u} (x)$, where $\vec{u}$ is the fluid velocity vector (a function of position only).
The fluid velocity at $x \rightarrow \infty$ is chosen to be
\begin{align} \label{bc-infty}
u_{x\infty}  = - v_{\text{B}}  \sin \alpha \text{, } 
 u_{y\infty} = 0 \text{, and }
 u_{z\infty} = v_{\text{B}} \cos \alpha    \text{.}
\end{align}
The choice (\ref{bc-infty}) corresponds to flow parallel to the magnetic field satisfying the Chodura (or Bohm-Chodura) condition \citep{Chodura-1982} with the equality sign.
Using equation (\ref{ux}) and (\ref{bc-infty}), the ion fluid velocity at every position can be written in terms of the electrostatic potential at that position,
\begin{align} \label{ux-phi-exact}
u_x = -v_{\text{B}}  \exp\left( - \frac{e\phi}{T_{\text{e}}} \right)  \sin \alpha  \text{.}
\end{align}
As shown in Appendix \ref{app-fluid-exact}, from the momentum equations and equation (\ref{ux-phi-exact}), one obtains a first order differential equation for the electrostatic potential,
\begin{align} \label{udiff-Riemann}
\left( \sin^2 \alpha \exp\left( -  \frac{2e\phi}{T_{\text{e}}} \right) - 1 \right)^2  \frac{v_{\text{B}}^2}{\Omega^2 \cos^2 \alpha } \left( \frac{e\phi'}{T_{\text{e}}} \right)^2  = 1 - \sin^2 \alpha  \exp \left( - \frac{2e\phi}{T_{\text{e}}} \right)  - \frac{2e\phi}{T_{\text{e}}} \nonumber \\ - \frac{1}{\cos^2 \alpha } \left[ 2 -  \exp \left( \frac{e\phi}{T_{\text{e}}} \right)  -   \exp \left( - \frac{e\phi}{T_{\text{e}}}  \right) \sin^2 \alpha   \right]^2 \text{.}
\end{align}
Equation (\ref{udiff-Riemann}) was originally derived in \cite{Chodura-1982} (and later in \cite{Riemann-1994}), in terms of $u_x$ instead of $\phi$, and is valid for all values of $\alpha$ (provided that $\alpha \gg \sqrt{m_{\text{e}}/m_{\text{i}}}$ as discussed in section \ref{sec-orderings}).

For $\alpha \ll 1$, the relationship between electrostatic potential and fluid velocity, equation (\ref{ux-phi-exact}), simplifies to
\begin{align} \label{ux-phi}
u_x \simeq - \alpha v_{\text{B}}  \exp\left( - \frac{e\phi}{T_{\text{e}}} \right)   \text{.}
\end{align}
In the fluid model, the equality form of the Bohm condition is $u_x(0) = - v_{\text{B}}$, which, from equation (\ref{ux-phi}), leads to equation \begin{align} \label{phidrop-tau0}
\frac{e\phi(0)}{T_{\text{e}}} \simeq \ln \alpha \text{.}
\end{align} 
In Appendix \ref{app-expanded}, we expand equation (\ref{udiff-Riemann}) for $\alpha \ll 1$, thus obtaining an equation for the electrostatic potential for $\tau = 0$ and $\alpha \ll 1$,
\begin{align} \label{phi-uniform}
x \simeq  \rho_{\text{B}} \int_{\ln \alpha + \frac{1}{2} - \frac{1}{2} \alpha^2}^{e\phi/T_{\text{e}}  + \frac{1}{2} \alpha^2 \left[ \exp\left( - 2e\phi /T_{\text{e}} \right) - 1 \right]} \frac{dp}{\sqrt{- 3 - 2p + 4  e^{p}   -  e^{2p} } }  \text{.}
\end{align}

In order to obtain a correspondence between our kinetic model and Chodura's fluid model, we define a new expansion parameter, $\epsilon \equiv 1 / |\ln \alpha|$, and take the ordering 
\begin{align} \label{ordering-logtau}
\frac{1}{\epsilon} \equiv |\ln \alpha | \sim |\ln \tau | \sim |\ln \alpha| - 3 |\ln \tau| \gg | \ln \epsilon | \sim 1  \text{.}
\end{align} 
The ordering (\ref{ordering-logtau}) restricts $\alpha$ (and $\tau$) to be \emph{exponentially} small, $\alpha = \exp(-1/\epsilon)$. 
However, with the ordering (\ref{ordering-logtau}), the kinetic model is asymptotically equivalent to the fluid model with an exponentially small error in $\epsilon$ (or, equivalently, a small error in $\alpha$ and $\tau$).
Hence, since the error is so small, in practice $\alpha$ need not be excessively small (we require $\alpha < 0.1$).

Before analyzing the kinetic model for $\tau \ll 1$ using the ordering (\ref{ordering-logtau}), it is instructive to solve equation (\ref{phi-uniform}) explicitly for $\epsilon \ll 1$.
Using the boundary condition (\ref{phidrop-tau0}), we order $e\phi/T_{\text{e}} \sim 1/\epsilon$ in the magnetic presheath.
Then, we expand equation (\ref{phi-uniform}) in $\epsilon  \ll 1$ to obtain
\begin{align} \label{phi-uniform-0}
x \simeq  \rho_{\text{B}} \int_{\ln \alpha}^{e\phi/T_{\text{e}}} \frac{ dp }{\sqrt{- 2p}}   \text{.}
\end{align}
Carrying out the integral in (\ref{phi-uniform-0}), the electrostatic potential in the magnetic presheath is
\begin{align} \label{phi-parabola-0}
\frac{ e\phi(x) }{T_{\text{e}}} \simeq \begin{cases}  -  \frac{1}{2} \left(x/\rho_{\text{B}} - \sqrt{2/\epsilon}\right)^2  &  \text{ for } x/\rho_{\text{B}}< \sqrt{2/\epsilon} \text{,}  \\
 0 & \text{ for } x/\rho_{\text{B}} \geqslant \sqrt{2/\epsilon} \text{.}
\end{cases}
\end{align}
From equation (\ref{phi-parabola-0}), the length scale of the magnetic presheath is $\sim \rho_{\text{B}} / \sqrt{\epsilon}$.

We will find that, in the ordering (\ref{ordering-logtau}), the magnetic presheath can be divided into three regions where different types of ion trajectories are dominant:
\begin{itemize}
\item a region far from the wall,
\begin{align} \label{region-far}
\frac{x}{\rho_{\text{B}}} > \sqrt{\frac{2}{\epsilon}}  - \sqrt{4|\ln\tau|}  \text{,}
\end{align}
where all ions are in small approximately periodic orbits (closed orbits);
\item a region close to the wall,
\begin{align} \label{region-near}
\frac{x}{\rho_{\text{B}}}  < \sqrt{\frac{2}{\epsilon}}  - \sqrt{\frac{2}{\epsilon} - 2 |\ln\tau|}  \text{,}
\end{align}
 where all ions are in open orbits; 
\item an intermediate region,
\begin{align} \label{region-int}
 \sqrt{\epsilon}  \ll \frac{x}{\rho_{\text{B}}} < \sqrt{\frac{2}{\epsilon}}  \text{,}
\end{align}
where ions moving towards the wall transition from small closed orbits to larger, distorted closed orbits, and finally to open orbits. 
\end{itemize}
In subsections \ref{subsec-cold-closed}-\ref{subsec-cold-intermediate}, we study the three regions in the order listed above.
Instead of taking the limit $\tau \ll 1$ of equations (\ref{ni-closed}) and (\ref{ni-open}) directly, which we leave to Appendix~\ref{app-smalltau}, in subsections \ref{subsec-cold-closed} and \ref{subsec-cold-open} we derive the flow velocity of ions in closed and open orbits.
For ions in closed orbits, the flow velocity is much smaller than the particle velocity, as most of the particle velocity is periodic in time and gives no contribution to the flow (because the periodic motion is averaged over);
hence, the flow velocity is equal to the drift velocity of the ion gyro-orbits.
However, for ions in open orbits sufficiently close to the wall the motion has no periodic piece, and so the flow velocity is equal to the individual particle velocity.
From the flow velocity $u_x$ and equation (\ref{ux-phi}), we derive equations for the electrostatic potential $\phi$ in the regions (\ref{region-far}) and (\ref{region-near}).
To lowest order in $\alpha$ and $\tau$, the solution for the electrostatic potential in the part of these two regions that overlaps with the intermediate region (\ref{region-int}) is a parabola.
Therefore, we assume that the lowest order solution for $\phi(x)$ in the whole intermediate region (\ref{region-int}) is a parabola, and use this to write an approximate kinetic quasineutrality equation for the region (\ref{region-int}). 
Finally, in subsection \ref{subsec-cold-uniform} we write an approximate differential equation for the electrostatic potential, whose solution is (\ref{phi-uniform}), and show that it is equivalent to the equations describing the electrostatic potential in the three regions.

\subsubsection{Far from the wall} \label{subsec-cold-closed}

 \begin{figure}
\centering
\includegraphics[width = 0.6\textwidth]{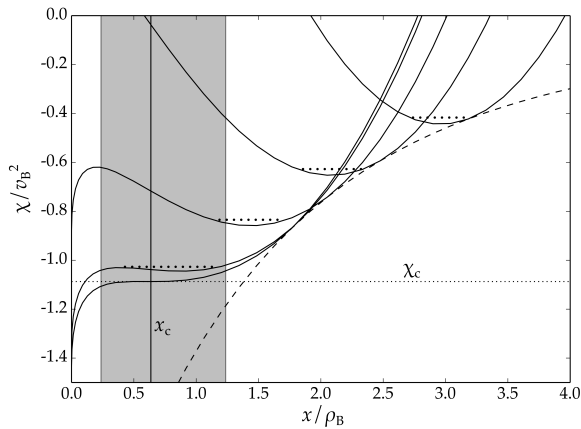}
\caption[Effective potential curves of small-temperature ions]{Effective potential curves $\chi(x, \bar{x})$ (solid lines), corresponding to the electrostatic potential profile $\phi(x)$ (dashed line) given by the approximation (\ref{phi-uniform}) (valid for $\tau = 0$) with $\alpha = 0.05$, shown for five different values of $\bar{x}$.
From the equation $\chi(\bar{x}, \bar{x}) = \phi(\bar{x})$, the values of $\bar{x}$ are where the dashed line intersects the solid lines. 
For the different values of $\bar{x}$, the values of $U_{\perp}$ (horizontal dotted lines) of an ion with $\mu \ll v_{\text{B}}^2 / \Omega $ are $U_{\perp} \simeq \chi_{\text{m}} (\bar{x})$.
When the difference between $\chi_{\text{m}} (\bar{x})$ (local minimum) and $\chi_{\text{M}} (\bar{x})$ (local maximum) becomes so small that $U_{\perp} \simeq \chi_{\text{M}} (\bar{x})$ (shaded region around the solid vertical line, $x = x_{\text{c}}$), the ion gyro-orbit is distorted and enlarged. 
}
\label{fig-chiChodura}
\end{figure}

From equation (\ref{phi-parabola-0}), the characteristic size of the magnetic presheath is $ \rho_{\text{B}} / \sqrt{\epsilon} $. 
Hence, sufficiently far away from the wall, all ions are in closed orbits with a radius of gyration, $\rho_{\text{i}}$, that is small compared with the size of the magnetic presheath, $ \rho_{\text{B}} / \sqrt{\epsilon} $.
The motion of the ions is thus drift-kinetic.
As shown in figure \ref{fig-chiChodura}, for $\chi''(x) \neq 0$ the effective potential $\chi$ looks like a parabola locally near the minimum,
\begin{align} \label{chi-nearmin}
\chi(x, \bar{x}) - \chi_{\text{m}} (\bar{x}) = \frac{1}{2} \chi'' (x_{\text{m}}) \left( x - x_{\text{m}} \right)^2 \left( 1 + O\left( \frac{\rho_x}{l} \right) \right)  \text{,}
\end{align}
where, in the error term, we have introduced the characteristic length scale over which the second derivative of the effective potential, $\chi''$, changes,
\begin{align} \label{l-def}
l = \left| \frac{\chi''(x)}{\chi'''(x)} \right| \text{.}
\end{align}
Consider an ion moving in an effective potential given by (\ref{chi-nearmin}). 
The turning points $x_{\text{b}}$ and $x_{\text{t}}$ are solutions of the equation $U_{\perp} = \chi(x, \bar{x})$, and so
\begin{align} \label{chi-nearmin-Uperp}
U_{\perp} - \chi_{\text{m}} (\bar{x}) = \frac{1}{8} \chi'' (x_{\text{m}}) \left( x_{\text{t}} - x_{\text{b}} \right)^2  \left( 1 + O\left( \frac{\rho_x}{l} \right) \right)   \text{.}
\end{align}
Recalling the definitions and orderings in (\ref{rhox-def})-(\ref{wx-def}), equation (\ref{chi-nearmin-Uperp}) corresponds to $w_x^2 \sim \chi''(x) \rho_x^2 $.
Thus, we obtain the ordering $\rho_x \sim w_x/\sqrt{\chi''(x)}$ relating the typical spatial extent in the $x$ direction (normal to the wall) of the ion orbit, $\rho_x$, to the typical orbital velocity component in the same direction, $w_x$.
Note that $w_x / \rho_x \sim \sqrt{\chi''(x_{\text{m}})}$ is the characteristic gyrofrequency of the approximately periodic motion of the ion.
This is consistent with the elliptical gyro-orbits studied in the Appendix of \cite{Geraldini-2017}.
Moreover, from equation (\ref{mu-Uperp-xbar}) we have the relationship $\mu \sim w_x \rho_x \sim v_{\text{t,i}}^2 /\Omega \sim  \tau v_{\text{B}}^2 / \Omega$, from which we obtain the estimates
\begin{align} \label{wx-est}
w_x \sim \left( \frac{\chi''(x_{\text{m}})}{\Omega^2} \right)^{1/4} \sqrt{\tau} v_{\text{B}}
\end{align}
 and 
\begin{align}  \label{rhox-est}
 \rho_x \sim  \left( \frac{\chi''(x_{\text{m}})}{\Omega^2} \right)^{-1/4} \sqrt{\tau} \rho_{\text{B}} \text{.}
 \end{align}

\begin{figure} 
\centering
\includegraphics[width = 0.8\textwidth]{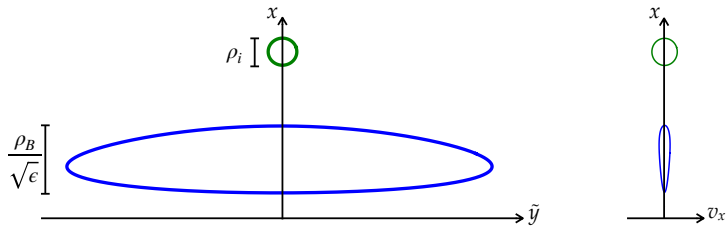} 
\caption{An example of an ion orbit shown at two different positions: far from the wall (green), and in the intermediate region (blue). 
On the left, the approximate trajectory is shown in the coordinates $(\tilde{y}, x)$, where $\tilde{y}$ is a $y$-coordinate in a frame of reference that is moving with the average $v_y$ of the ion.
On the right, the trajectory is shown in phase space co-ordinates $(v_x, x)$.
The invariance of $\mu$ ensures that the area of the closed orbits on the right is constant. } 
\label{fig-orbitgrowth}
\end{figure}

The electrostatic potential $\phi$ given in equation (\ref{phi-parabola-0}) has a discontinuous second derivative: for $x/\rho_{\text{B}} \geqslant \sqrt{2/\epsilon}$, we have $\phi''(x) \simeq 0$ and $\chi''(x) \simeq \Omega^2$, while for $x/\rho_{\text{B}} < \sqrt{2/\epsilon}$ we have $\phi''(x) \simeq -\Omega^2$ and $\chi''(x) \simeq 0$.
Hence, to lowest order in $\epsilon$, the second derivative of the electrostatic potential is not determined (note that equation (\ref{udiff-Riemann}) does not specify $\phi''(x)$).
However, the abrupt jump in the value of $\phi''(x)$ and $\chi''(x)$ occurring at $x/\rho_{\text{B}}=\sqrt{2/\epsilon}$ is a reflection of a decrease of $\phi''(x)$ and $\chi''(x)$ in going from $x \rightarrow \infty$ to $x/\rho_{\text{B}} < \sqrt{2/\epsilon}$.
From equation (\ref{rhox-est}), the size of ion orbits is $\rho_x \sim \sqrt{\tau} \rho_{\text{B}} \sim \rho_{\text{i}}$ when $\chi''(x_{\text{m}}) \simeq \Omega^2$.
Conversely,  when $\chi''(x_{\text{m}}) \ll \Omega^2$, the spatial extent of the ion orbits is larger, $\rho_x \gg \rho_{\text{i}}$.
The growth of the ion orbit as it approaches the wall in the magnetic presheath is shown in figure \ref{fig-orbitgrowth}.
Note that as $\rho_x$ becomes larger, the typical orbital velocity $w_x \simeq v_x$ becomes smaller (see equation (\ref{wx-est})).
When $\rho_x$ grows so large that $\rho_x \sim l$, equations (\ref{chi-nearmin}) and (\ref{chi-nearmin-Uperp})-(\ref{rhox-est}) cease to be valid as the effective potential can no longer be Taylor expanded near its minimum.
This happens when the ion reaches the shaded region in figure \ref{fig-chiChodura}.

We proceed to solve for the ion motion by assuming that $\rho_x$ is small,
\begin{align} \label{x-gc}
x = x_{\text{m}} + \rho_x \simeq x_{\text{m}} \text{.}
\end{align}
From equation (\ref{chi-minimum-1}) and (\ref{x-gc}), we obtain the value of $v_y = \Omega(\bar{x} - x)$,
\begin{align} \label{vy-ExB}
v_y =  \frac{\phi'(x_{\text{m}})}{B} - \Omega \rho_x \simeq \frac{\phi'(x_{\text{m}})}{B}  \text{.} 
\end{align}
Indeed, since the orbital velocity is small, by equation (\ref{wx-est}), and the angle between the magnetic field and the wall is shallow, the motion of the ion is approximately parallel to the wall.
Thus, the magnetic force away from the wall, $Z e B v_y $, is approximately equal to the electric force towards the wall, $  Ze \phi'(x) \simeq Ze \phi'(x_{\text{m}}) $.
Using equations (\ref{Uperp-def}) and (\ref{vy-ExB}), the perpendicular energy of an ion at a position $x \simeq x_{\text{m}}$ is given by 
\begin{align} \label{Uperp-cold}
U_{\perp} = \frac{1}{2} \left( \frac{\phi'(x_{\text{m}})}{B} \right)^2 + \frac{\Omega \phi (x_{\text{m}}) }{B} + O\left(  \chi''(x_{\text{m}}) \rho_x^2,  w_x^2 \right)  \text{.}
\end{align}
The first error in (\ref{Uperp-cold}) is a combination of the orbital component of $\frac{1}{2} v_y^2$, $\Omega^2 \rho_x^2/2$, and the quadratic term, $\Omega \phi''(x_{\text{m}})\rho_x^2 /2B$, of the Taylor expansion of $\Omega \phi(x)/B$ near $x = x_{\text{m}}$.
Note that the term $O\left( \phi'(x_{\text{m}}) \Omega \rho_x / B \right)$ arising from taking the square of equation (\ref{vy-ExB}) has cancelled with the linear term of the Taylor expansion of $\Omega \phi(x)/B$. 
The second error in (\ref{Uperp-cold}) comes from neglecting $\frac{1}{2} v_x^2$.
From equations (\ref{wx-est}) and (\ref{rhox-est}), the two errors have the same size.
From equation (\ref{bc-infty}), the total energy of an ion at $x \rightarrow \infty$ is given by $U = v_{\text{B}}^2 / 2$.
Hence, the $z$-component of the ion velocity is, using equations (\ref{vz-Uperp-U}) and (\ref{Uperp-cold}) with $U = v_{\text{B}}^2 / 2$,
\begin{align} \label{Vz}
v_z  = \sqrt{ v_{\text{B}}^2 -  \left( \frac{\phi'(x_{\text{m}})}{B} \right)^2 - \frac{2 \Omega \phi (x_{\text{m}}) }{B} } + O\left(  \frac{w_x^2}{ v_{\text{B}}} \right)   \text{.}
\end{align}

In order to obtain the ion fluid velocity $u_x$, we do not require the exact velocity of an ion, $v_x \simeq w_x$, as most of this velocity gives a quasi-periodic motion at the small length scale $\rho_x \ll \rho_{\text{B}}$. 
Instead, we require a drift velocity, denoted $v_{\text{d}}$, defined as
\begin{align}
v_{\text{d}} = \dot{x}_{\text{m}} \text{.}
\end{align}
Using this definition for the drift velocity and equations (\ref{wx-def}) and (\ref{x-gc}), the ion velocity $\dot{x} = v_x$ can be split into two distinct contributions,
\begin{align} \label{vx-split}
v_x = w_x + v_{\text{d}} \text{.}
\end{align}
As mentioned at the end of section \ref{subsec-traj}, it is important that the ion motion be approximately periodic for our shallow-angle kinetic model to be valid.
The ion motion can be considered as an approximately periodic orbit only if the characteristic period $\sim \rho_x / w_x$ is much smaller than the characteristic time it takes for the ion orbit to drift (at speed $v_{\text{d}}$) by a distance $l$ such that the electrostatic potential has changed by a significant amount, $\sim l/v_{\text{d}}$.
Thus, the criterion for approximate periodicity is
\begin{align} \label{vd-ordering}
\frac{v_{\text{d}}}{l} \ll \frac{w_x}{\rho_x} \text{.}
\end{align}

We proceed to obtain an expression for $v_{\text{d}}$.
From equation (\ref{y-EOM-alpha}), we obtain
\begin{align} \label{vx-implicit-before}
v_x = - \alpha v_z  - \frac{ \dot{v}_{y}  }{ \Omega }   \text{.}
\end{align} 
Taking the derivative of (\ref{vy-ExB}), we obtain 
\begin{align} \label{vy-dot}
\dot{v}_y  =  v_{\text{d}}  \frac{ \phi''(x_{\text{m}})}{  B} -  \Omega w_x    \text{.} 
\end{align}
Inserting equations (\ref{Vz}), (\ref{vx-split}), and (\ref{vy-dot}) into (\ref{vx-implicit-before}), the terms proportional to $w_x$ on the left and right hand side cancel and we obtain an implicit equation for $v_{\text{d}}$,
\begin{align} \label{vx-implicit}
v_{\text{d}} = - \alpha \sqrt{ v_{\text{B}}^2 -  \left( \frac{\phi'(x_{\text{m}})}{B} \right)^2 - \frac{2 \Omega \phi (x_{\text{m}}) }{B} }  - \frac{v_{\text{d}} \phi''(x_{\text{m}})}{ \Omega B} + O\left( \alpha \frac{w_x^2}{v_{\text{B}}} \right)  \text{.}
\end{align} 
The right hand side of equation (\ref{vx-implicit}) consists of the small component of parallel streaming in the $x$ direction, approximately given by $- \alpha v_z $, a polarization drift, approximately given by $-v_{\text{d}} \phi'' / \Omega B $, and the error term coming from the error in $v_z$. 
By manipulating (\ref{vx-implicit}), we obtain
\begin{align} \label{vx-drift}
v_{\text{d}} (x_{\text{m}}) = \frac{ - \alpha \sqrt{ v_{\text{B}}^2 -  \left( \phi'(x_{\text{m}})/B \right)^2 - 2 \Omega \phi (x_{\text{m}}) / B }  }{1 + \phi''(x_{\text{m}})/\Omega B} \left( 1 + O\left( \frac{w_x^2}{v_{\text{B}}^2}  \right) \right) \text{.}
\end{align}
An alternative procedure to derive equation (\ref{vx-drift}) is to obtain the time derivative of $x_{\text{m}}$ by using the chain rule, $v_{\text{d}} = \dot{x}_{\text{m}} = \dot{\bar{x}} dx_{\text{m}}/d\bar{x}$, as shown in Appendix~\ref{app-xmdot}.
Equation (\ref{vx-drift}) is divergent for $ \phi''(x_{\text{m}}) = -\Omega B$, but approximating the ion motion as a closed orbit becomes invalid close to the divergence, as it requires $v_{\text{d}} $ to be small by equation (\ref{vd-ordering}).

The ion fluid velocity $u_x$ is the average value of $v_x$ at a fixed position $x$, not at a fixed guiding center position $x_{\text{m}}$.
The orbital velocity $w_x$ averages to zero provided that the motion is approximately periodic (condition (\ref{vd-ordering})). 
Moreover, writing $v_{\text{d}}(x_{\text{m}}) \simeq v_{\text{d}}(x) - v_{\text{d}}'(x) \rho_x + O(v_{\text{d}} \rho_x^2 / l^2)$ and using the fact that the linear piece in $\rho_x$ averages to zero for approximately periodic motion (condition (\ref{vd-ordering})), we obtain
\begin{align} \label{ux-far}
u_x (x) = \frac{ - \alpha  \sqrt{ v_{\text{B}}^2 -  \left( \phi'(x)/B \right)^2 + 2 \Omega \phi (x) /B}   }{1 + \phi''(x)/\Omega B} \left( 1 + O\left( \frac{\rho_x^2 }{l^2}  \right) \right)  \text{.}
\end{align}
The $ O( w_x^2 / v_{\text{B}}^2 ) $ error in equation (\ref{vx-drift}) is neglected in (\ref{ux-far}) as it is smaller than the $O\left( \rho_x^2  / l^2  \right)$ error.

The assumption that $w_x $ averages to zero also implies that we have neglected the contribution from the open orbits, $n_{\text{i,op}}(x) \simeq 0$, so that $n_{\text{i}}(x) \simeq n_{\text{i,cl}}(x)$.
The closed orbit density can then be obtained from (\ref{ux-far}) and the fact that $n_{\text{i}} u_x = - \alpha n_{\infty} v_{\text{B}}$,
\begin{align} \label{ni-closed-cold}
n_{\text{i,cl}}(x) = \frac{n_{\infty} v_{\text{B}}  \left( 1 + \phi''(x) / \Omega B \right) }{\sqrt{v_{\text{B}}^2 - \left( \phi'(x) / B \right)^2 - 2\Omega \phi (x)/B}} \left( 1 + O\left( \frac{ \tau \rho_{\text{B}}^2 }{l^2} \sqrt{ \frac{ \Omega^2 }{ \left| \chi''(x)\right|} }   \right) \right)  \text{,}
\end{align}
where the error has been rewritten using the ordering (\ref{rhox-est}).
This result can also be derived by taking the limit $\tau \ll 1 $ in equation (\ref{ni-closed}), which is a more direct though perhaps less intuitive approach (see Appendix \ref{app-smalltau}).
We can substitute either of (\ref{ux-far}) or (\ref{ni-closed-cold}) into (\ref{ux-phi}) or (\ref{quasineutrality}), respectively, to obtain a differential equation for the electrostatic potential,
\begin{align} \label{quasi-cold}
 \frac{ v_{\text{B}} \left( 1 + \phi'' / \Omega B  \right) }{\sqrt{v_{\text{B}}^2 - \left(  \phi' / B \right)^2 - 2\Omega \phi / B}} \left( 1 + O\left( \frac{\rho_{\text{B}}^2}{l^2} \left(\frac{ \left| \chi''\right|}{\Omega^2 }\right)^{-1/2} \tau   \right) \right) = \exp \left( \frac{e\phi  }{T_{\text{e}}} \right)  \text{.}
\end{align}
Recalling that $1+\Omega \phi''(x)/B = \chi''(x)/\Omega^2$, the ordering that results from equation (\ref{quasi-cold}) is $\chi''(x)/\Omega^2 \sim \exp \left( e\phi  / T_{\text{e}} \right)$, which also leads to the ordering $1/l \sim  \chi'''/ \chi''  \sim e\phi'/T_{\text{e}}$.
Moreover, using the fact that balancing the terms in the denominator of equation (\ref{quasi-cold}) gives $ \rho_{\text{B}}^2 \left( e\phi'/T_{\text{e}} \right)^2 \sim e\phi/T_{\text{e}}$, we obtain the ordering 
\begin{align}  \label{l-scaling}
\frac{\rho_{\text{B}}^2}{l^2} \sim \frac{e\phi}{T_{\text{e}}} \text{.}
\end{align}
Then, upon rearranging equation (\ref{quasi-cold}) and re-expressing the error, we obtain
\begin{align} \label{quasi-cold-1}
 \frac{ v_{\text{B}} \left( 1 + \phi'' / \Omega B  \right) }{\sqrt{v_{\text{B}}^2 - \left(  \phi' / B \right)^2 - 2\Omega \phi / B}} = \exp \left( \frac{e\phi  }{T_{\text{e}}} \right) + O\left( \tau \frac{e\phi}{T_{\text{e}}}  \exp \left( \frac{e\phi}{ 2T_{\text{e}}}\right)   \right) \text{.}
\end{align}
Multiplying equation (\ref{quasi-cold-1}) by $e\phi'/T_{\text{e}}$, integrating once and using the boundary condition $\phi = \phi' = 0$ at $x \rightarrow \infty$, gives
\begin{align} \label{quasi-cold-2}
 2 - \sqrt{ 1 - \rho_{\text{B}}^2 \left( \frac{e\phi' }{T_{\text{e}}} \right)^2 - \frac{2 e \phi }{T_{\text{e}}} }  = &  \exp \left( \frac{e\phi }{T_{\text{e}}} \right)   + O\left(  \tau    , \tau  \frac{e\phi}{T_{\text{e}}} \exp \left( \frac{e\phi}{2T_{\text{e}}} \right)  \right)  \text{.}
\end{align}
Upon integrating the error on the right hand side of (\ref{quasi-cold-1}), we obtained two distinct contributions to the error in (\ref{quasi-cold-2}): one 
is $O\left( \tau  \right) $ and the other is $O\left( \tau \left(  e\phi / T_{\text{e}} \right) \exp \left( e\phi / 2T_{\text{e}} \right) \right)$.
Note that both these error terms are exactly equal to zero for $\phi = 0$ and they are comparable in size for $- e\phi / T_{\text{e}}  \lesssim 1$, while for $- e\phi / T_{\text{e}}  \sim 1/\epsilon \gg 1$ the term $O\left(   \tau \right) $ is larger. 
However, this larger term tends to a constant, which as we will see does not affect the functional form of the solution $\phi (x)$, but only shifts the value of the constant of integration by a small amount. 
Equation (\ref{quasi-cold-2}) can be rearranged to obtain
\begin{align} \label{udiff-far}
\rho_{\text{B}}^2 \left( \frac{e\phi'}{T_{\text{e}}} \right)^2  + 3  + \frac{2e\phi}{T_{\text{e}}} = &  4  \exp \left( \frac{e\phi}{T_{\text{e}}} \right)  -  \exp \left( \frac{2e\phi}{T_{\text{e}}} \right) +  O\left(  \tau  , \tau  \frac{e\phi}{T_{\text{e}}} \exp \left( \frac{e\phi}{2T_{\text{e}}} \right)  \right)    \text{.}
\end{align}
Finally, equation (\ref{udiff-far}) can be integrated to obtain the electrostatic potential far away from the wall, although a boundary condition in the intermediate region, which we have not yet specified, is required to carry out the integration.

For $-e\phi/T_{\text{e}}   \gg 1$, all the terms on the right hand side of equation (\ref{udiff-far}) become small except for the $O(\tau )$ term which approaches a constant, and the solution approaches the parabola
\begin{align} \label{phi-parabola}
\frac{e\phi(x)}{T_{\text{e}}} \simeq \frac{e\phi_{\text{p}}(x)}{T_{\text{e}}}  = - \frac{3}{2} + \kappa \tau  - \frac{1}{2} \frac{ (x-C )^2 }{ \rho_{\text{B}}^2 }  \text{.}
\end{align}
Here, $C$ is a constant determined by boundary conditions, and we denoted the constant $O(\tau )$ error coming from the right hand side of (\ref{udiff-far}) as $\kappa \tau$, where $\kappa$ is an unknown constant of order unity.
The electrostatic potential at the wall is large, $-e\phi(0)/T_{\text{e}}  = |\ln \alpha | = 1/\epsilon \gg 1$, and so we expect equation (\ref{phi-parabola}) to become valid closer to the wall. 
If we assume that (\ref{phi-parabola}) is valid at $x=0$ to lowest order in $\epsilon$ and impose $-e\phi(0)/T_{\text{e}}  = |\ln \alpha | = 1/\epsilon$, we obtain 
\begin{align} \label{C-lowest}
C \simeq \rho_{\text{B}} \sqrt{\frac{2}{\epsilon}} \text{.}
\end{align}
To lowest order in $\epsilon$, equation (\ref{phi-parabola}), with $C$ given by (\ref{C-lowest}), is equivalent to equation (\ref{phi-parabola-0}), which was obtained from the fluid model.
However, note that the non-constant piece of the error in equation (\ref{udiff-far}) becomes comparable to the first term on the right hand side when $\exp \left( e\phi/T_{\text{e}} \right) \sim \tau \left(e\phi/T_{\text{e}} \right) \exp \left( e\phi/2T_{\text{e}} \right)$.
Hence, equation (\ref{udiff-far}) fails to correctly determine the potential when $\exp \left( e\phi/T_{\text{e}} \right) \sim \tau^2 |\ln \tau |^2 \sim \tau^2 / \epsilon^2$.
From equation (\ref{rhox-est}), with the ordering $\chi'' / \Omega^2 \sim \exp ( e\phi / T_{\text{e}} )$, and equation (\ref{l-scaling}), this value of $\exp \left( e\phi / T_{\text{e}} \right)$ corresponds to $\rho_x \sim l \sim \sqrt{\epsilon} \rho_{\text{B}}$, which is the point at which the approximation in (\ref{chi-nearmin}) ceases to be valid.
The validity of (\ref{udiff-far}) is thus restricted to
\begin{align} \label{ineq-far}
\frac{\tau^2}{\epsilon^2} \ll \exp \left( \frac{e\phi}{T_{\text{e}}} \right)  \text{.}
\end{align}
Note that, from (\ref{phi-parabola}), (\ref{C-lowest}) and (\ref{ineq-far}), the validity region is given by (\ref{region-far}) to lowest order in $\epsilon$.
Ion gyro-orbits grow in size as they approach the wall, as shown in figure \ref{fig-orbitgrowth}, making the treatment of this section invalid for $x/\rho_{\text{B}} \leqslant \sqrt{2/\epsilon} - \sqrt{4|\ln\tau|}$, where $\rho_x$ is no longer small.

\subsubsection{Near the wall} \label{subsec-cold-open}

When $U_{\perp} \simeq \chi_{\text{M}}$, ions transition to open orbits and thereafter reach the wall in a timescale of the order of a gyroperiod, $ \rho_{x} / w_{x} \sim 1 / \sqrt{\chi''(x_{\text{m}})}$.
Previously, we saw that the spatial extent of a closed ion orbit enlarges as the ion approaches the wall; for the moment, we take $\rho_x \sim \rho_{\text{B}}$ for ions transitioning from closed to open orbits, ignoring any potential scaling with $\epsilon$.
For such transitioning ions, we expect that $\chi_{\text{M}} - \chi_{\text{m}} \sim w_x^2 \sim  \tau^2  v_{\text{B}}^2 / \rho_x^2$, since $\mu \sim \rho_x w_x \sim \tau v_{\text{B}}^2$.
Thus, $\chi_{\text{M}} - \chi_{\text{m}}$ is small in $\tau$.
Recall, from equation (\ref{chic-def}), that $\chi_{\text{c}}$ is defined to be the value of the effective potential at $\bar{x} = \bar{x}_{\text{c}}$ such that $\chi_{\text{c}} = \chi_{\text{M}} (\bar{x})  = \chi_{\text{m}} (\bar{x}) $.
Hence, it follows that $ \chi_{\text{M}} (\bar{x})  \simeq \chi_{\text{c}} $ and $\bar{x} \simeq \bar{x}_{\text{c}}$ for all ions in open orbits, as can be seen in figure \ref{fig-chiChodura}.
The error in approximating $ \chi_{\text{M}}(\bar{x}) - \chi(x, \bar{x}) \simeq \chi_{\text{c}}  - \chi(x, \bar{x}_{\text{c}} )$ can be obtained by calculating
\begin{align}
\left. \frac{d}{d\mu} \right\rvert_{\text{open}} \left( \chi_{\text{M}} - \chi(x, \bar{x}) \right) = \left( \frac{d\mu}{d\bar{x}} \right)_{\text{open}}^{-1}   \Omega^2 \left( x - x_{\text{M}} \right) \sim \frac{\tau  \Omega \rho_{\text{B}}^2}{\rho_x^2}  \text{,}
\end{align}
where we used $ d \left( \chi_{\text{M}} - \chi(x, \bar{x}) \right) / d\bar{x} = \Omega^2 \left( x - x_{\text{M}} \right) \sim \Omega^2 \rho_x  $ and estimated $\left( d\mu / d\bar{x} \right)_{\text{open}} \sim \Omega^2 \rho_x^2 / w_x \sim \Omega \rho_x^3 / \tau \rho_{\text{B}}^2$ from equation (\ref{dmudxbar-open}).
Since typical ion orbits have values of $\mu$ differing by $O( \tau v_{\text{B}}^2 / \Omega)$, the values of $ \chi_{\text{M}}(\bar{x}) - \chi(x, \bar{x}) $ of such orbits change by $O(\tau^2 v_{\text{B}}^2 \rho_{\text{B}}^2 / \rho_{x}^2)$.
Recall that an open orbit has $U_{\perp} - \chi(x, \bar{x}) = \chi_{\text{M}}(\bar{x}) - \chi(x, \bar{x}) + O(\Delta_{\text{M}})$: from equation (\ref{DeltaM-mu}) and the previous estimate for $\left( d\mu / d\bar{x} \right)_{\text{open}}$, we obtain the scaling $ \Delta_{\text{M}} \sim \alpha \Omega v_{\text{B}} \rho_x^3 / \tau \rho_{\text{B}}^2$.
If the condition 
\begin{align} \label{validity-cond}
\frac{\alpha}{ \tau^3 } \sim \alpha^A = \exp \left( - \frac{A}{\epsilon} \right) \ll 1 
\end{align}
is satisfied, with $A$ a positive constant, the $O(\Delta_{\text{M}})$ error term is small in $\alpha$ compared to the $O(\tau^2 v_{\text{B}}^2 \rho_{\text{B}}^2 / \rho_{x}^2)$ error.
Then, using equation (\ref{vx-x-xbar-Uperp}) with $ U_{\perp} - \chi(x, \bar{x}) \simeq \chi_{\text{c}}  - \chi(x, \bar{x}_{\text{c}} ) + O(\tau^2 v_{\text{B}}^2 \rho_{\text{B}}^2 / \rho_{x}^2)$, the velocity of an ion in an open orbit near the wall is
\begin{align} \label{vx-open-smalltau}
v_x = - \sqrt{ 2\left( \chi_{\text{c}} - \chi (x, \bar{x}_{\text{c}}) \right) + O\left( \tau^2 v_{\text{B}}^2 \rho_{\text{B}}^2 / \rho_{x}^2 \right)  } \text{.}
\end{align}
As we will see at the end of this section, condition (\ref{validity-cond}) must be satisfied for our kinetic model to be valid; otherwise, the velocity of all ions transitioning from closed to open orbits is not known to lowest order in $\alpha$, making the ion density incorrect in a large region. 
The ordering (\ref{ordering-logtau}) includes the validity condition (\ref{validity-cond}).

Assuming that $x$ is sufficiently close to the wall that most ions are in open orbits, $n_{\text{i}}(x) \simeq n_{\text{i,op}}(x)$, the ion fluid velocity is
\begin{align} \label{ux-open}
u_x (x) = - \sqrt{ 2\left( \chi_{\text{c}}  - \chi (x, \bar{x}_{\text{c}}) \right) + O\left(  \tau^2 v_{\text{B}}^2 \rho_{\text{B}}^2 / \rho_{x}^2 \right) } \text{.}
\end{align}
Then, from equation (\ref{ux-open}) and the continuity equation $n_{\text{i,op}}(x)u_x(x) = -\alpha n_{\infty} v_{\text{B}}$, we obtain an expression for the open orbit density,
\begin{align} \label{ni-open-cold}
n_{\text{i,op}} (x) = \frac{  \alpha  n_{\infty} v_{\text{B}} }{\sqrt{ 2\left( \chi_{\text{c}} - \frac{1}{2} \Omega^2 ( x - \bar{x}_{\text{c}} )^2 - \Omega \phi(x)/B \right) + O \left(  \tau^2 v_{\text{B}}^2 \rho_{\text{B}}^2 / \rho_{x}^2  \right) }} \text{.}
\end{align}
In Appendix \ref{subapp-cold-open}, we derive equation (\ref{ni-open-cold}) by expanding $n_{\text{i,op}}(x)$ in equation (\ref{ni-open}) to lowest order in $\tau \ll 1$.
In order for equation (\ref{ni-open-cold}) to be valid, we require the error term to be small, implying $n_{\text{i,op}} \ll \left( \alpha n_{\infty} / \tau \right) \left( \rho_x / \rho_{\text{B}} \right)$.
Moreover, since $\rho_x$ here quantifies the characteristic size of closed orbits while the ion is transitioning from a closed to an open orbit, the ion density changes from $n_{\text{i,cl}} \gg n_{\infty} \tau^2 / \epsilon^2 $ (recall equation (\ref{ineq-far}) and the fact that $n_{\text{i,cl}} \sim n_{\infty} \exp \left( e\phi / T_{\text{e}} \right)$) to $n_{\text{i,op}} \ll  \left( \alpha n_{\infty} / \tau \right) \left( \rho_x / \rho_{\text{B}} \right) $ over a length scale of $\rho_x$.
If the validity condition (\ref{validity-cond}) is satisfied, this drop in density corresponds to a decrease of $ \ln \left( \tau^3 /  \alpha  \right) \sim  1/\epsilon $ in the normalized electrostatic potential $e  \phi / T_{\text{e}}$.
In order to be consistent with the lowest order electric field obtained from (\ref{phi-parabola}) and (\ref{C-lowest}), $e\phi'/T_{\text{e}} = - (x-C)/\rho_{\text{B}}^2 \sim 1/\sqrt{\epsilon} \rho_{\text{B}} $, transitioning ions must have $\rho_x \sim \rho_{\text{B}} /\sqrt{\epsilon}$.

Inserting equation (\ref{ni-open-cold}) into (\ref{quasineutrality}) with $n_{\text{i,cl}}(x) = 0$, or inserting equation (\ref{ux-open}) into (\ref{ux-phi}),
we obtain
\begin{align} \label{phi-near-kinetic}
   \frac{e \phi}{T_{\text{e}}} + \frac{1}{2} \alpha^2   \exp\left(-\frac{ 2e\phi}{T_{\text{e}}}\right)  = K - \frac{3}{2}  - \frac{ ( x - \bar{x}_{\text{c}} )^2 }{2\rho_{\text{B}}^2} + O\left( \epsilon \tau^2  \right) \text{.}
\end{align}
The constants
\begin{align} \label{K-def}
K = \frac{\chi_{\text{c}}}{v_{\text{B}}^2} + \frac{3}{2}
\end{align}
and $\bar{x}_{\text{c}}$ are to be determined; they are related by the boundary condition $e\phi(0)/T_{\rm{e}} = \ln \alpha$, giving 
\begin{align} \label{xbarc-K-rel}
\bar{x}_{\text{c}} \simeq \rho_{\text{B}} \sqrt{- 2\ln \alpha  - 4 + 2K} \sim \frac{\rho_{\text{B}}}{ \sqrt{ \epsilon} } \text{.}
\end{align}
Note that far from the wall, equation (\ref{phi-near-kinetic}) gives
\begin{align} \label{phi-near-kinetic-far}
\frac{e \phi}{T_{\text{e}}} \simeq K - \frac{3}{2} - \frac{ ( x - \bar{x}_{\text{c}} )^2 }{2\rho_{\text{B}}^2}  \text{.}
\end{align}
For equation (\ref{phi-near-kinetic-far}) to be valid, we require $\alpha^2   \exp\left(- 2e\phi / T_{\text{e}}\right) \sim \alpha^2   \exp\left( ( x - \bar{x}_{\text{c}} )^2 / \rho_{\text{B}}^2\right) \sim \exp \left( - x\bar{x}_{\text{c}} / \rho_{\text{B}}^2 + x^2 / 2\rho_{\text{B}}^2 \right) \ll 1$, where we have used $\bar{x}_{\text{c}} \simeq  \rho_{\text{B}}  \sqrt{2/ \epsilon} $; hence, the electrostatic potential becomes well-approximated by (\ref{phi-near-kinetic-far}) for $x \gg \rho_{\text{B}} \sqrt{\epsilon}$.
The derivation of equation (\ref{phi-near-kinetic}) fails when  $n_{\text{i,op}} \sim \alpha n_{\infty} / \sqrt{\epsilon} \tau$ (recall equation (\ref{ni-open-cold}) with $\rho_x \sim \rho_{\text{B}} / \sqrt{\epsilon}$), and so the validity of (\ref{phi-near-kinetic}) is restricted to
\begin{align} \label{ineq-near}
 \exp \left( \frac{e\phi}{T_{\text{e}}} \right) \ll \frac{\alpha}{\sqrt{\epsilon} \tau} \text{.}
\end{align}
From equations (\ref{phi-near-kinetic-far}) and (\ref{ineq-near}), we obtain the estimate (\ref{region-near}) for the region where (\ref{phi-near-kinetic}) is valid.
Outside of the validity region (\ref{region-near}), the velocity of a typical ion is of the order of the gyration velocity, $ \sqrt{ 2\left( \chi_{\text{M}}(\bar{x}) - \chi(x, \bar{x}) \right)} \sim w_x$, and so the assumption that all ions are in open orbits is invalid.

\subsubsection{Intermediate region} \label{subsec-cold-intermediate}

With the ordering (\ref{validity-cond}), there is a finite region where equations (\ref{udiff-far}) and (\ref{phi-near-kinetic}) are not valid: from equations (\ref{region-far}) and (\ref{region-near}), this region is
\begin{align} \label{region-transition}
 \sqrt{\frac{2}{\epsilon}} - \sqrt{\frac{2}{\epsilon} - 2 |\ln \tau|} \leqslant \frac{x}{\rho_{\text{B}}} \leqslant \sqrt{\frac{2}{\epsilon}} - \sqrt{4 |\ln \tau|} \text{.}
\end{align}
However, the solution of equation (\ref{udiff-far}) tends to (\ref{phi-parabola}) for $\sqrt{2/\epsilon} - x/ \rho_{\text{B}} \gg 1$ and (\ref{phi-near-kinetic}) tends to (\ref{phi-near-kinetic-far}) for $x / \rho_{\text{B}} \gg \sqrt{\epsilon}$.
Hence, we proceed by assuming that in the intermediate region (\ref{region-int}), which includes the region (\ref{region-transition}), the electrostatic potential is simultaneously given by the parabolas in equations (\ref{phi-parabola}) and (\ref{phi-near-kinetic-far}) to lowest order in $\alpha$ and $\tau$.
This provides the value of $K$, $K= \kappa \tau  \ll 1$, and the equality $C = \bar{x}_{\text{c}}$.
Using equation (\ref{xbarc-K-rel}) with $K\simeq 0$, the value of $C$ and $\bar{x}_{\text{c}}$ is
\begin{align} \label{C-xbarc-matched}
C = \bar{x}_{\text{c}} \simeq \rho_{\text{B}} \sqrt{-2\ln\alpha - 4}  \text{.}
\end{align}
The neglected term $\kappa \tau$ causes a small constant correction to the value of $C$, as we had claimed in the discussion following equation (\ref{quasi-cold-2}).

From equation (\ref{chi}), the effective potential curves associated with the parabolic electrostatic potential of equation (\ref{phi-parabola}) are a set of straight lines,
\begin{align}\label{chi-inter}
\chi (x; \bar{x} ) \simeq  - \frac{3}{2} v_{\text{B}}^2  +  \frac{1}{2} \Omega^2 ( \bar{x}^2  - C^2 )  -  \Omega^2 \left( \bar{x} - C \right) x 
   \text{.}
\end{align}
In figure \ref{fig-chiChodura}, a family of effective potential curves $\chi(x;\bar{x})$ are plotted for different values of the orbit position $\bar{x}$ for $\alpha = 0.05$: the curves shown are indeed close to straight lines in the shaded region, as equation (\ref{chi-inter}) suggests.
Since straight lines do not have a local minimum --- which is necessary to approximate the ion motion as a periodic orbit --- the small non-parabolic piece of the electrostatic potential,
\begin{align} \label{phinp-intermediate}
\phi_{\text{np}}(x) = \phi(x) - \phi_{\text{p}}(x)  \text{,}
\end{align}
must be retained.
In equation (\ref{phinp-intermediate}), $\phi(x)$ is the solution to the quasineutrality equation (\ref{quasineutrality}) for a given value of $\tau$ and $\alpha$.
In the intermediate region we take $\phi(x) \simeq \phi_{\text{p}}(x)$ and calculate $\phi_{\text{np}}(x) $ as a higher order asymptotic correction from the following equation:
\begin{align} \label{phinp-kinetic}
n_{\text{i,cl}} \left[ \phi_{\text{p}} + \phi_{\text{np}}  \right] (x) + n_{\text{i,op}} \left[ \phi_{\text{p}} + \phi_{\text{np}}   \right] (x)  = n_{\infty} \exp \left( \frac{e\phi_{\text{p}} (x) }{T_{\text{e}}} \right) \left( 1 + O\left(\frac{e\phi_{\text{np}}}{T_{\text{e}}} \right) \right) \text{.}
\end{align}
On the right hand side of equation (\ref{phinp-kinetic}), we neglected terms small in $e\phi_{\text{np}} / T_{\text{e}} \ll 1$ to simplify the expression for the electron density.
On the left hand side, we included the non-parabolic piece $\phi_{\text{np}} $ because no effective potential minima exist --- and so no closed or open ion orbits can be solved for --- when $\phi = \phi_{\text{p}}$.

At the beginning of this section, we noted that equations (\ref{udiff-far}) and (\ref{phi-near-kinetic}) do not have a common region of validity.
Therefore, it is crucial that equations (\ref{udiff-far}) and (\ref{phinp-kinetic}) be simultaneously valid in some overlap region of finite size; the same has to be true for equations (\ref{phi-near-kinetic}) and (\ref{phinp-kinetic}).
Equation (\ref{udiff-far}) is valid in region (\ref{region-far}), and equation (\ref{phinp-kinetic}) is valid in the region (\ref{region-int}).
Hence, the overlap region in which both equations are valid is 
\begin{align}
 1 \ll  \sqrt{\frac{2}{\epsilon}} - \frac{x}{\rho_{\text{B}}} < \sqrt{4 |\ln \tau|}  \text{,}
\end{align}
where we re-expressed the lowest order inequality $ x / \rho_{\text{B}} <  \sqrt{2 / \epsilon} $ to the more precise form $ 1 \ll  \sqrt{2 / \epsilon} - x / \rho_{\text{B}} $ in order to emphasize the necessity of the ordering $|\ln \tau | \sim 1/\epsilon \gg 1$.
We proceed to calculate $\phi_{\text{np}}(x)$ in this region.
Inserting $\phi = \phi_{\text{p}} + \phi_{\text{np}}$ in (\ref{quasi-cold}) and rearranging the error term, we obtain
\begin{align}  \label{eqB}
\frac{\phi''_{\text{np}} (x) }{2\Omega B  }    =  \exp \left( - \frac{3}{2} - \frac{(x-C)^2}{2\rho_{\text{B}}^2} \right) + O \left( \frac{\tau \rho_{\text{B}}}{l^2}  \sqrt{\frac{e\phi_{\text{np}}''}{T_{\text{e}}}} \right) \text{,}
\end{align}
where we have used $\phi \simeq \phi_{\text{p}}$ in the denominator to get $ \sqrt{v_{\text{B}}^2 - \left( \phi'(x) / B \right)^2 - 2\Omega \phi (x) / B} \simeq 2v_{\text{B}}  $.
From the definition of $l$ in (\ref{l-def}) and using $\chi'' = \Omega \phi_{\text{np}}'' / B$ with equation (\ref{eqB}), we obtain $l \sim \rho_{\text{B}}^2 / ( C - x ) \sim \rho_{\text{B}} \sqrt{\epsilon} $, consistent with our previous estimate for $l$ in this region (before equation (\ref{ineq-far})).
Integrating (\ref{eqB}) twice and imposing $\phi'_{\text{np}} (x)  = \phi_{\text{np}} (x) = 0$ at $(C - x)/\rho_{\text{B}} \rightarrow \infty$ (where the electrostatic potential becomes more parabolic) gives
\begin{align} \label{phinp-far}
\frac{ e \phi_{\text{np}} (x) }{T_{\text{e}}}  = &  2\exp\left(-\frac{3}{2} - \frac{(x-C)^2}{2\rho_{\text{B}}^2} \right)  - \sqrt{2\pi} \left(C-x\right)  \exp \left(-\frac{3}{2} \right)  \left(1-\text{erf}\left( \frac{C-x}{\sqrt{2}\rho_{\text{B}}} \right)  \right) \nonumber \\ & + O \left( \frac{\tau }{\sqrt{\epsilon}}  \sqrt{\frac{e\phi_{\text{np}}}{T_{\text{e}}}} \right) \text{,}
\end{align}
where we have used that the double integral of the term $O\left( ( \tau \rho_{\text{B}}/ l^2 ) \sqrt{ e\phi_{\text{np}}'' / T_{\text{e}}} \right)$ is $O \left( (\tau \rho_{\text{B}}/ l )  \sqrt{e\phi_{\text{np}} / T_{\text{e}}} \right)$.
We proceed to consider the part of the intermediate region that is closest to the wall. 
Equation (\ref{phi-near-kinetic}) is valid in the region (\ref{region-near}) near the wall and equation (\ref{phinp-kinetic}) is valid in the intermediate region (\ref{region-int}).
Hence, the region in which these two equations are both valid is the overlap of (\ref{region-near}) and (\ref{region-int}),
\begin{align}
 \sqrt{\epsilon} \ll \frac{x}{\rho_{\text{B}}} < \sqrt{\frac{2}{\epsilon}} - \sqrt{\frac{2}{\epsilon} - 2 |\ln \tau|} \text{.}
\end{align}
From equation (\ref{phi-near-kinetic}) and using $e\phi_{\text{np}} / T_{\text{e}} \ll 1$ (the assumption behind equation (\ref{phinp-kinetic})), we extract
\begin{align} \label{phinp-near}
\frac{e\phi_{\text{np}}(x)}{T_{\text{e}}}  = - \frac{1}{2} \alpha^2  \exp\left( 3 + \frac{ ( x - C )^2 }{\rho_{\text{B}}^2} \right) +  O\left(   \tau^2 \epsilon  \right) \text{.}
\end{align}

In the region (\ref{region-transition}), equation (\ref{phinp-kinetic}) cannot be simplified further.  
Ion orbits are large, $\rho_{\text{B}}  \sqrt{\epsilon} \lesssim \rho_x \lesssim \rho_{\text{B}} / \sqrt{\epsilon}$, and the non-parabolic piece of the electrostatic potential is small, $- \tau^2 \epsilon \lesssim e\phi_{\text{np}}(x)/T_{\text{e}} \lesssim  \tau^2 / \epsilon$ (consistent with the errors in (\ref{phinp-far}) and (\ref{phinp-near}), noting that $\phi_{\text{np}}$ is positive in (\ref{phinp-far}) and negative in (\ref{phinp-near})). 
From the previous paragraph we deduce that $l \sim \sqrt{\epsilon} \rho_{\text{B}}$, and so the double derivative of the effective potential is small, $- \tau^2  \lesssim \chi''(x)/\Omega^2 = \rho_{\text{B}}^2 e\phi_{\text{np}}''/T_{\text{e}} \lesssim \tau^2 / \epsilon^2$.
The characteristic size of the periodic piece of the ion velocity is obtained from the relation $\mu \sim \rho_x w_x \sim \tau v_{\text{B}}^2 / \Omega$ (which holds provided the adiabatic invariant is still conserved), giving $ \tau \sqrt{\epsilon} \lesssim w_x / v_{\text{B}} \lesssim \tau / \sqrt{\epsilon}$.
The characteristic size of the drift of closed ion orbits is obtained from equation (\ref{vx-drift}), giving $ \alpha \epsilon^2 / \tau^2  \lesssim v_{\text{d}} / v_{\text{B}} \lesssim \alpha  / \tau^2 $.
Ignoring the dependences on $\epsilon$ and focusing only on how quantities scale with $\alpha$ and $\tau$ (which, from equation (\ref{ordering-logtau}), are both exponentially small in $\epsilon$), we obtain $ \rho_x \sim l \sim \rho_{\text{B}} $, $w_x \sim \tau v_{\text{B}}$ and $v_{\text{d}} \sim \alpha v_{\text{B}} / \tau^2$.
Hence, the condition $v_{\text{d}} / l \ll w_x / \rho_x$ (equation (\ref{vd-ordering})), which is necessary for the ion motion to be approximately periodic, implies that $v_{\text{d}} / w_x \sim \alpha  / \tau^3 \ll 1 $.
In equation (\ref{validity-cond}) $\alpha / \tau^3$ is required to be small in $\alpha$ to ensure that the motion remains periodic to lowest order in $\alpha$. 
The scaling with $\tau^3$ in (\ref{validity-cond}) implies that the kinetic model is not valid for relatively large values of $\tau \ll 1$.
This unfortunate scaling arises because of the growth of the ion orbits: if small ion orbits reached $x=0$, we would expect $l \sim \rho_x \sim \rho_{\text{i}}$, $\chi''(x) \sim \Omega^2$, $w_x \sim \sqrt{\tau} v_{\text{B}}$, $v_{\text{d}} \sim \alpha v_{\text{B}}$, and so the condition (\ref{vd-ordering}) would give the weaker requirement $\alpha \ll \sqrt{\tau}$ for the model to be valid near $x=0$.
Hence, the orbit growth and the associated large polarization drift have a strong negative effect on the condition for validity of the model, multiplying the power by which $\tau$ is raised by a factor of six.
It is for this reason that, as we will see in section \ref{sec-finite}, we do not obtain numerical solutions of equation (\ref{quasineutrality}) for values of $\tau$ lower than $\tau = 0.2$.

\subsubsection{Uniformly valid solution} \label{subsec-cold-uniform}

We proceed to obtain an expression for $\phi(x)$, equation (\ref{phi-uniform}), that is uniformly valid across the whole magnetic presheath to lowest order in $\alpha$ and $\tau$.
In order to do this, we first make a change of variables: guided by the form of (\ref{phi-near-kinetic}), we introduce the function
\begin{align} \label{psi-def}
\psi (x) = \frac{e\phi(x)}{T_{\text{e}}} + \frac{1}{2} \alpha^2 \left( \exp\left( -\frac{2e\phi(x)}{T_{\text{e}}} \right) - 1\right) \text{.}
\end{align}
The term $-\alpha^2 / 2$ is small but is included in the definition (\ref{psi-def}) in order to have the desirable exact property that $\phi = 0$ when $\psi = 0$: then, far from the wall, where $-e\phi/T_{\text{e}} \ll 1/\epsilon$, the relation $\psi = \left( e\phi/T_{\text{e}} \right) \left( 1 + O(\alpha^2) \right)$ is satisfied.

We proceed to show that the equation
\begin{align} \label{psidiff-uniform}
\rho_{\text{B}}^2 \psi'^2  + 3 + 2\psi = 4  \exp(\psi)   -  \exp(2\psi) + O\left(\tau , \tau \psi  \exp \left( \frac{1}{2} \psi \right),  \tau^2 \right)  \text{} 
\end{align}
is equivalent to the equations describing the electrostatic potential in the three regions of the magnetic presheath.
All the errors on the right hand side of (\ref{psidiff-uniform}) are exactly equal to zero at $\psi = 0$.
Moreover, the $O(\tau )$ error tends to a constant for $-\psi \gg 1$, and this constant does not have an effect on the functional form of $\psi (x)$.
Hence, we retain the smaller errors $O( \tau \psi  \exp \left( \psi/2 \right),  \tau^2 )$.
First, we compare equation (\ref{psidiff-uniform}) with equation (\ref{udiff-far}), valid in the region (\ref{region-far}).  
Since, from (\ref{ineq-far}), $\alpha^2 \exp ( - 2e\phi/T_{\text{e}} ) \ll \alpha^2 /\tau^4$ in this region, it follows that $e\phi/T_{\text{e}} = \psi + O(\alpha^2/\tau^4)$.
Hence, equation (\ref{psidiff-uniform}) directly follows from (\ref{udiff-far}), after noting that the $O(\alpha^2/\tau^4)$ error term is smaller than the $O(\tau^2  )$ error term (the smallest in (\ref{psidiff-uniform})) because of the validity condition (\ref{validity-cond}).
Next, we compare the solution to equation (\ref{psidiff-uniform}) with equation (\ref{phi-near-kinetic}) (recall that $K=\kappa \tau$, where $\kappa$ is an unknown constant of order unity, and $\bar{x}_{\text{c}} = C$), valid in the region (\ref{region-near}) close to the wall.
From (\ref{psi-def}), in this region $-e\phi/T_{\text{e}} = -\psi + O(1) \sim 1/\epsilon$ and so, from (\ref{ineq-near}), $ \exp(\psi / 2) \ll \sqrt{\alpha/\tau} \ll \tau$.
Therefore, all terms on the right hand side of (\ref{psidiff-uniform}), except for the constant $O(\tau )$ term, become smaller than the $O (  \tau^2 )$ term.
Re-expressing the constant $O(\tau ) $ error term as $\kappa \tau$ and integrating gives
\begin{align} \label{psi-near}
\psi = -\frac{3}{2} +\kappa \tau - \frac{ \left( x-C \right)^2}{2\rho_{\text{B}}^2} + O\left(\epsilon \tau \psi  \exp \left( \frac{1}{2} \psi \right), \epsilon \tau^2 \right) \text{.}
\end{align}
The size of the error terms in (\ref{psi-near}) is obtained as follows.
From $\psi \sim 1/\epsilon$ and $l \sim \psi'' / \psi''' \sim \rho_{\text{B}}^2/(x-C) \sim \sqrt{\epsilon} \rho_{\text{B}}$, the error terms in the expression for $\psi'$ are smaller by a factor of $\sqrt{\epsilon}$ compared with the error terms in the expression for $\psi'^2$, equation (\ref{psidiff-uniform}).
Upon integrating the expression for $\psi'$ to obtain $\psi$, we multiply these error terms by a factor of $l\sim \sqrt{\epsilon} \rho_{\text{B}}$.
Using (\ref{psi-def}), and remembering that in the region (\ref{region-near}) the size of the largest error term in (\ref{psi-near}) is $O(\epsilon \tau^2 )$, observe that equation (\ref{psi-near}) is equivalent to (\ref{phi-near-kinetic}).

It only remains to be shown that (\ref{psidiff-uniform}) is valid in the part of the intermediate region (\ref{region-transition}) where neither (\ref{udiff-far}) (the equation determining $\phi$ far from the wall) nor (\ref{phi-near-kinetic}) (the equation determining $\phi$ close to the wall) are valid. 
From the equation $\phi = \phi_{\text{p}}  + \phi_{\text{np}} $, equation (\ref{phi-parabola}) for $\phi_{\text{p}}$, and the fact that $e|\phi_{\text{np}}|/T_{\text{e}} \lesssim \tau^2 / \epsilon$, the electrostatic potential in this region is given by $e\phi / T_{\text{e}} = - 3/2 + \kappa \tau - \left( x-C \right)^2 / 2\rho_{\text{B}}^2 + O(\tau^2 / \epsilon ) $.
From equation (\ref{ineq-far}) for the size of the electrostatic potential in the region far from the wall, and equation (\ref{ineq-near}) for the size of the electrostatic potential in the region close to the wall, we obtain the ordering $\alpha / \sqrt{\epsilon} \tau \lesssim \exp(e\phi / T_{\text{e}} ) \lesssim \tau^2 / \epsilon^2 $ for the size of the electrostatic potential in this region.
Hence, the first error term in equation (\ref{psi-near}) becomes $O(\tau^2 / \epsilon )$. 
Moreover, from $\epsilon^4\alpha^2 /\tau^4 \lesssim \alpha^2 \exp ( - 2e\phi/T_{\text{e}} ) \lesssim \epsilon \tau^2 $ we obtain $e\phi / T_{\text{e}} = \psi + O(\epsilon \tau^2)$.
Hence, from equation (\ref{psi-near}) and the associated $O(\tau^2 / \epsilon )$ error in this region, we obtain $ e\phi / T_{\text{e}}  = -3/2 + \kappa \tau -  \left( x-C \right)^2 / 2\rho_{\text{B}}^2 + O\left( \tau^2 / \epsilon \right)$. 
Equation (\ref{psidiff-uniform}) is thus a good approximation also in the region (\ref{region-transition}).

Using the definition (\ref{psi-def}), and equation (\ref{phidrop-tau0}), the boundary condition at the wall is $\psi(0) = \ln \alpha + 1/2 -  \alpha^2/2 $. This can be used to integrate equation (\ref{psidiff-uniform}) (neglecting the error terms) and obtain the approximate electrostatic potential solution, as in equation (\ref{phi-uniform}).

\subsection{Hot ions ($\tau \gg 1/\alpha$)} \label{sec-hot}

In the limit of very hot ions, $\tau \gg 1/\alpha$, we assume that the ion distribution function is a half-Maxwellian at the magnetic presheath entrance,
\begin{align} \label{f-infty-hot}
f_{\infty} \left( \vec{v} \right) =  \frac{ 2 n_{\infty} }{  \pi^{3/2} v_{\text{t,i}}^3} \exp \left( - \frac{\left| \vec{v} \right|^2 }{v_{\text{t,i}}^2} \right)\Theta \left( v_z \right)  \text{,}
\end{align}
where we introduced the Heaviside step function
 \begin{align} \label{Heaviside}
\Theta \left( s \right) = \begin{cases} 1 \text{ for } s\geqslant 0 \text{,} \\
0 \text{ for } s< 0 \text{.} \end{cases}
\end{align}
Since $\phi(\infty) = 0$, $U = |\vec{v}|^2/2$ and we re-express (\ref{f-infty-hot}) to
\begin{align} \label{F-hot}
F = 2 n_{\infty} \left( \frac{ m_{\text{i}} }{2\pi T_{\text{i}}} \right)^{3/2} \exp\left( -\frac{ m_{\text{i}} U }{ T_{\text{i}}} \right)  \text{.}
\end{align}
Equation (\ref{f-infty-hot}) is one of many choices that could be made.
The reason we choose this distribution function is that it was also used in \cite{Cohen-Ryutov-1998} in the equivalent limit of small electron temperature. 
We consider the limit $\tau \ll \alpha^2 m_{\text{i}} / m_{\text{e}} $ in order to be consistent with condition (\ref{alpha-tau-order}) for an electron repelling sheath.

For $\tau \rightarrow \infty$, ion orbits are undistorted by the presheath potential drop necessary to repel the electrons. 
We expect $e\phi(x)/T_{\text{e}} \sim 1$, and therefore the ion flow and density can be computed using $Ze\phi(x)/T_{\text{i}} = (1/\tau) e\phi(x)/T_{\text{e}} \simeq 0$ across the magnetic presheath. 
The effective potential is a parabola with its minimum at $x_{\text{m}} = \bar{x}$,
\begin{align}
\chi(x, \bar{x}) = \frac{1}{2} \Omega^2 \left(x-\bar{x} \right)^2 \text{.}
\end{align}
This is an effective potential whose maximum for $x<x_{\text{m}}$ is given by
\begin{align}
\chi_{\text{M}} (\bar{x} ) = \chi (0, \bar{x} ) = \frac{1}{2} \Omega^2 \bar{x}^2 \text{.}
\end{align}
The minimum value of $\bar{x}$ necessary for an ion at position $x$ to be in a closed orbit or an open orbit is, using equations (\ref{xbarm-def}) and (\ref{xbarm-open}) with $\phi(x)=0$,
\begin{align} \label{xbarm-open-flatphi}
\bar{x}_{\text{m,o}} \left( x \right) = \bar{x}_{\text{m}} \left( x \right) = \frac{1}{2} x \text{.}
\end{align}
Moreover, the adiabatic invariant is $ \mu = U_{\perp}/\Omega $.

Inserting the distribution function (\ref{F-hot}) into equation (\ref{ni-closed}), the closed orbit density is 
\begin{align}
 n_{\text{i,cl}}(x)  = &  2 n_{\infty} \left( \frac{ m_{\text{i}} }{2\pi T_{\text{i}}} \right)^{3/2} \nonumber \\ & \times  \int_{x/2}^{\infty} \Omega d\bar{x} \int_{\frac{1}{2} \Omega^2 \left(x-\bar{x} \right)^2}^{\frac{1}{2}\Omega^2 \bar{x}^2 } \frac{ 2 dU_{\perp} }{\sqrt{2\left(U_{\perp} - \chi (x, \bar{x}) \right)}} \int_{U_{\perp}}^{\infty}  \frac{ \exp \left(  - m_{\text{i}} U / T_{\text{i}}   \right) dU }{\sqrt{2\left( U - U_{\perp}\right)}}  \text{.}
\end{align}
Changing variables to $\tilde{ v }_y =  \left(  \bar{x} - x \right)/\rho_{\text{i}} $, $\tilde{U}_{\perp} = m_{\text{i}} \left(  U_{\perp} - \frac{1}{2} \Omega^2 (x-\bar{x})^2 \right) / T_{\text{i}}$ and $\tilde{U} = m_{\text{i}} \left(U - U_{\perp} \right) / T_{\text{i}}$ gives
\begin{align}
n_{\text{i,cl}}(x) = \frac{ n_{\infty} }{ \pi^{3/2} }  \int_{ - \frac{x}{2\rho_{\text{i}}} }^{\infty} d\tilde{v}_y \exp (-  \tilde{ v }_y^2)  \int_{0}^{ \frac{x}{\rho_{\text{i}}} \left( 2\tilde{v}_y + \frac{x}{\rho_{\text{i}}} \right) }  \tilde{U}_{\perp}^{-1/2} \exp (-\tilde{U}_{\perp}) d\tilde{U}_{\perp} \nonumber \\
\times  \int_{0}^{\infty}  \tilde{U}^{-1/2} \exp(-\tilde{U}) d\tilde{U}  \text{.}
\end{align}
Evaluating the integral over $\tilde{U}$ and the integral over $\tilde{U}_{\perp}$ leads to
\begin{align} \label{ni-closed-coldelectrons}
n_{\text{i,cl}}(x) = \frac{n_{\infty}}{\sqrt{\pi}}  \int_{-\frac{x}{2\rho_{\text{i}}}}^{\infty}  \exp \left(- \tilde{v}_y^2 \right) \text{erf}  \left( \sqrt{  \frac{x}{\rho_{\text{i}}} \left( 2\tilde{ v }_y + \frac{x}{\rho_{\text{i}}}\right) } \right) d\tilde{v}_y \text{,}
\end{align}
where we introduced the error function,
\begin{align}
\text{erf}(s) = \frac{2}{\sqrt{\pi}} \int_0^s \exp \left( - s'^2 \right) ds' \text{.}
\end{align}
For $x\ll \rho_{\text{i}}$, the integral in (\ref{ni-closed-coldelectrons}) simplifies in the following ways: (i) the lower limit of integration can be set to $\tilde{v}_y = 0$, since the contribution to the integral from the integration range $[0,\infty]$ is dominant; (ii) the factor $\sqrt{ \left( x/\rho_{\text{i}} \right) \left( 2\tilde{ v }_y + x / \rho_{\text{i}}\right) }$ in the argument of the error function can be replaced by $\sqrt{ 2\tilde{ v }_y x / \rho_{\text{i}} }$, since this replacement is accurate in most of the integration range except where $\tilde{v}_y \sim x/\rho_{\text{i}} \ll 1$; (iii) the error function can be approximated by $\text{erf}\left( \sqrt{ 2\tilde{ v }_y x / \rho_{\text{i}} }\right) \simeq 2\sqrt{ 2\tilde{ v }_y x / \rho_{\text{i}} }/\sqrt{\pi}$ for $x/\rho_{\text{i}} \ll 1/\tilde{v}_y$, which holds everywhere except in the region $\tilde{v}_y \gtrsim \rho_{\text{i}} / x \gg 1$, where the integrand is exponentially small.
Hence, equation (\ref{ni-closed-coldelectrons}) becomes, for $x \ll \rho_{\text{i}}$,
\begin{align} \label{ni-closed-coldelectrons-nearwall-1}
n_{\text{i,cl}}(x) \simeq \frac{2n_{\infty}}{\pi} \sqrt{  \frac{2 x}{\rho_{\text{i}}} }  \int_0^{\infty} \sqrt{\tilde{ v }_y}  \exp \left(- \tilde{v}_y^2 \right)   d\tilde{v}_y \text{.}
\end{align}
Re-expressing the integral over $\tilde{v}_y$ in terms of the standard Gamma function $\Gamma$,
\begin{align}
\int_0^{\infty} \sqrt{\tilde{ v }_y}  \exp \left(- \tilde{v}_y^2 \right)   d\tilde{v}_y = \frac{1}{2} \int_0^{\infty} \frac{\exp \left( - \xi \right)}{\xi^{1/4}} d\xi = \frac{1}{2} \Gamma \left( \frac{3}{4} \right) \text{,}
\end{align}
we simplify (\ref{ni-closed-coldelectrons-nearwall-1}) to 
\begin{align} \label{ni-closed-coldelectrons-nearwall}
n_{\text{i,cl}}(x) \simeq \frac{\sqrt{2}}{\pi}  \Gamma \left( \frac{3}{4} \right) \sqrt{  \frac{ x}{\rho_{\text{i}}} } n_{\infty} \text{.}
\end{align}

The density of open orbits is given by
\begin{align} \label{ni-open-coldelectrons-1}
n_{\text{i,op}} (x)  = & \int_{\frac{1}{2}x }^{\infty}  \Omega d\bar{x}  \int_{\frac{1}{2} \Omega^2 \bar{x}^2 }^{\infty} \frac{ F\left( \Omega^2\bar{x}^2/2, U \right) }{\sqrt{2\left( U - \chi_{\text{M}} (\bar{x})  \right) }} \nonumber \\
 & \times \left( \sqrt{2\left( \chi_{\text{M}}(\bar{x})  - \chi \left( x, \bar{x} \right) + \Delta_{\text{M}}(\bar{x}, U) \right) } - \sqrt{2\left( \chi_{\text{M}}(\bar{x})  - \chi \left( x, \bar{x} \right)  \right) } \right) dU \text{.}
\end{align}
Note that, in equation (\ref{ni-open-coldelectrons-1}), we have used 
\begin{align}
\mu =\frac{1}{2} \Omega \bar{x}^2 \text{}
\end{align}
for the adiabatic invariant of ions with $U_{\perp} = \chi_{\text{M}} (\bar{x}) = \chi ( 0, \bar{x} )= \Omega^2 \bar{x}^2 / 2 $.
Using equation (\ref{DeltaM-mu}), we obtain 
 \begin{align}
\Delta_{\text{M}} = 2 \alpha \pi \Omega \bar{x} \sqrt{2\left( U - \frac{1}{2} \Omega^2 \bar{x}^2   \right) }  \text{.}
\end{align}
Then, using the dimensionless integration variables $\tilde{v}_z = \sqrt{ m_{\text{i}} \left( U - \Omega^2 \bar{x}^2/2 \right) / T_{\text{i}} }$ and $\tilde{\bar{x}} = \bar{x}/\rho_{\text{i}}$, equation (\ref{ni-open-coldelectrons-1}) reduces to
\begin{align} \label{ni-open-coldelectrons}
n_{\text{i,op}} (x) = & \frac{2n_{\infty}}{\pi^{3/2}} \int_{\frac{x}{2\rho_{\text{i}}} }^{\infty} d\tilde{\bar{x}}  \exp \left( - \tilde{\bar{x}}^2 \right)  \int_{ 0 }^{\infty} \exp\left(-\tilde{v}_{z}^2 \right)  \nonumber \\
 & \times  \left( \sqrt{ \frac{x}{\rho_{\text{i}}} \left( 2\tilde{ \bar{x} } - \frac{x}{\rho_{\text{i}}} \right)  + 4 \alpha \pi \tilde{ \bar{x} } \tilde{v}_z  } - \sqrt{ \frac{x}{\rho_{\text{i}}} \left( 2\tilde{\bar{x}} - \frac{x}{\rho_{\text{i}}} \right) } \right) d\tilde{v}_z   \text{.}
\end{align}
Equation (\ref{ni-open-coldelectrons}) does not simplify further for general values of $x$, but can be simplified for $x \ll \alpha \rho_{\text{i}}$.
Evaluating the ion density at $x=0$ using equation (\ref{ni-open-coldelectrons}), we obtain
\begin{align} \label{ni-open-coldelectrons-x=0}
n_{\text{i,op}} (0)  
= \frac{ 1 }{\pi }  \Gamma^2 \left( \frac{3}{4} \right) \sqrt{\alpha } n_{\infty}  \text{.}
\end{align}
We then proceed to evaluate $n_{\text{i,op}} (x) - n_{\text{i,op}} (0)$ for $x\ll \alpha \rho_{\text{i}}$.
In this ordering, $\left( x/ \rho_{\text{i}} \right) \left( 2\tilde{\bar{x}} - x / \rho_{\text{i}} \right) \ll 4\alpha \pi \tilde{\bar{x}} \tilde{v}_z$ and so
\begin{align} \label{ni-open-coldelectrons-nearwall-1}
n_{\text{i,op}} (x) - n_{\text{i,op}} (0) \simeq & - \frac{2n_{\infty}}{\pi^{3/2}} \int_{\frac{x}{2\rho_{\text{i}}} }^{\infty} d\tilde{\bar{x}}  \exp \left( - \tilde{\bar{x}}^2 \right)  \sqrt{ \frac{x}{\rho_{\text{i}}} \left( 2\tilde{\bar{x}} - \frac{x}{\rho_{\text{i}}} \right) } \int_{ 0 }^{\infty} \exp\left(-\tilde{v}_{z}^2 \right) d\tilde{v}_z   \text{.}
\end{align}
Note that there is a small integration region, $\tilde{v}_z \lesssim x / \alpha \rho_{\text{i}}$, where $\left( x/ \rho_{\text{i}} \right) \left( 2\tilde{\bar{x}} - x / \rho_{\text{i}} \right) \gtrsim 4\alpha \pi \tilde{\bar{x}} \tilde{v}_z$, but the contribution to the integral from this region is higher order in $x/\alpha \rho_{\text{i}} \ll 1$.
In equation (\ref{ni-open-coldelectrons-nearwall-1}) we take $\sqrt{ \left( x/\rho_{\text{i}} \right) \left( 2\tilde{ \bar{x} } + x / \rho_{\text{i}}\right) } \simeq \sqrt{ 2\tilde{ \bar{x} } x / \rho_{\text{i}} }$, since this is accurate everywhere except where $\tilde{\bar{x}} \sim x/\rho_{\text{i}} \ll \alpha$, and evaluate the integrals over $\tilde{v}_z$ and $\tilde{\bar{x}}$ to obtain
\begin{align} \label{ni-open-coldelectrons-nearwall}
n_{\text{i,op}} (x) - n_{\text{i,op}} (0) \simeq & - \frac{\sqrt{2}}{2\pi}  \Gamma \left( \frac{3}{4} \right) \sqrt{  \frac{ x}{\rho_{\text{i}}} } n_{\infty} \text{.}  
\end{align}
From equation (\ref{ni-open-coldelectrons-nearwall}), the open orbit density near $x=0$ decreases proportionally to $\sqrt{x}$, but the increase of the closed orbit density in equation (\ref{ni-closed-coldelectrons-nearwall}) is faster by a factor of $2$, leading to the total ion density increasing proportionally to $\sqrt{x}$,
\begin{align} \label{ni-coldelectrons-nearwall}
n_{\text{i}} (x) - n_{\text{i}} (0) \simeq &  \frac{\sqrt{2}}{2\pi}  \Gamma \left( \frac{3}{4} \right) \sqrt{  \frac{ x}{\rho_{\text{i}}} } n_{\infty} \text{,}  
\end{align}
for $x \ll \alpha \rho_{\text{i}}$.

The ion density profile for $\tau \rightarrow \infty$ is, according to (\ref{ni}), the sum of equations (\ref{ni-closed-coldelectrons}) and (\ref{ni-open-coldelectrons}).
The potential profile is obtained by imposing quasineutrality and inverting the Boltzmann relation for the electron density, to find
\begin{align} \label{phi-hot}
\frac{ e\phi (x) }{T_{\text{e}}} =  \ln \left( \frac{ n_{\text{i}}(x) }{n_{\infty}} \right) \text{.}
\end{align}
The potential drop across the magnetic presheath can be calculated by using $n_{\text{i,cl}}(0) = 0$ (from equation (\ref{ni-closed-coldelectrons})) and equation (\ref{ni-open-coldelectrons-x=0}),
\begin{align} \label{phidrop-tauinf}
\frac{ e\phi (0) }{T_{\text{e}}} =  \ln \left(  \frac{  \Gamma^2 \left( 3/4 \right) }{\pi } \sqrt{\alpha }  \right) \simeq  \ln \left( 0.48 \sqrt{\alpha }  \right) \text{.}
\end{align}
Inserting the distribution function (\ref{F-hot}) and the value of $\bar{x}_{\text{m,o}}$ in (\ref{xbarm-open-flatphi}) into equation (\ref{f0x-def}), the distribution of the ion velocity component perpendicular to the wall at $x=0$ is 
\begin{align} \label{f0x-hot}
f_{0x}(v_x ) = \frac{n_{\infty}}{v_{\text{t,i}}\pi} \Theta (-v_x) \int_0^{\infty} \exp \left( - \tilde{\bar{x}}^2 \right) \left[ 1 - \text{erf} \left( \frac{v_x^2 }{4\pi \alpha \tilde{\bar{x}} v_{\text{t,i}}^2} \right) \right] d\tilde{\bar{x}}  \text{.}
\end{align}
Inserting the distribution function (\ref{F-hot}) into equation (\ref{f0yz-def}), the distribution of the ion velocity components parallel to the wall at $x=0$ is 
\begin{align} \label{f0yz-hot}
f_{0yz}(v_y, v_z ) = \frac{4\sqrt{\alpha} n_{\infty}}{\pi} \frac{  \sqrt{v_y v_z } }{v_{\text{t,i}}^3} \exp \left( - \frac{ v_y^2 + v_z^2  }{v_{\text{t,i}}^2} \right)  \Theta (v_y) \Theta(v_z) \text{.}
\end{align}

To conclude, we briefly point out and resolve an apparent contradiction in the validity of our kinetic model when $\tau \gg 1/\alpha$.
In reference \cite{Geraldini-2018}, we found that the self-consistent electrostatic potential prohibits the presence of ions entering the Debye sheath with zero velocity normal to the wall.
This is in apparent contradiction with the situation described in this section: when undistorted circular orbits reach the wall, there are ion trajectories tangential to the wall and thus there is a finite number of ions which have a normal component of the velocity equal to zero.
This is reflected in the fact that, from equation (\ref{f0x-hot}), $f_{0x}(0) \neq 0$.
In reality, there is a small region near $x=0$ in which the electric field distorts ion orbits just before they reach the wall, so that $\chi_{\text{M}}(\bar{x}) = \chi(x_{\text{M}}, \bar{x})$ with $x_{\text{M}} \ll \rho_{\text{i}}$.
The quasi-tangential ions (with $v_x \simeq 0$) must be accelerated to values of $v_x$ such that the Bohm condition (\ref{kinetic-Bohm-marginal}) is satisfied with the equality sign.
If these very slow ions do not accelerate to large enough values of $v_x$, the integral on the left hand side of (\ref{kinetic-Bohm-marginal}) becomes too large and the Bohm condition cannot be satisfied.
Conversely, if these ions are accelerated too much towards the wall, the Bohm condition cannot be satisfied with the equality sign, as in (\ref{kinetic-Bohm-marginal}), which is in contradiction with our theory.
Thus, one can think of the real distribution function as the distribution function in (\ref{f0x-hot}) (which is plotted as a dashed line in the bottom-left panel of figure \ref{fig-f0x}), but shifted in such a way that the peak of the distribution function is at $v_x = - \bar{v}$ instead of $v_x = 0$, and the distribution function is effectively equal to zero for $|v_x | < \bar{v} \ll \sqrt{\alpha} v_{\text{t,i}}$\footnote{In \cite{Geraldini-2018} we found that the distribution decreases to zero exponentially fast as $v_x \rightarrow 0$.}.
Since the width of the distribution function, $\sqrt{\alpha} v_{\text{t,i}}$, is much larger than $v_{\text{B}}$ for $\tau \gg 1/ \alpha$, the Bohm integral on the right hand side evaluates approximately to $f_{0x}(\bar{v})/\bar{v}$ for the real distribution function.
Then, approximating $f_{0x}(\bar{v}) \sim n_{\text{i}}(0) / \sqrt{\alpha} v_{\text{t,i}}$, we obtain the estimate $\bar{v} \sim v_{\text{B}} / \sqrt{\alpha \tau}$ to satisfy the Bohm condition (\ref{kinetic-Bohm-marginal}).
Hence, the final piece of the electrostatic potential drop, which is responsible for distorting the ion orbits enough to satisfy the kinetic Bohm condition, is smaller than the total electrostatic potential drop by a factor of $m_{\text{i}} \bar{v}^2 / T_{\text{e}} \sim 1 / \alpha \tau \ll 1$.
Note that the pair of conditions (\ref{alpha-tau-order}) and $1/\alpha\tau \ll 1$ require $m_{\text{i}} / m_{\text{e}} \gg \tau^3 $ to be satisfied.
The size $h$ of the region near $x=0$ where this final potential drop occurs is obtained by balancing the electric force, $Ze\phi' \sim ZT_{\text{e}}/ h \alpha \tau$, with the magnetic force $Zev_yB \sim m_{\text{i}} \Omega v_{\text{t,i}} $, giving $h /  \rho_{\text{i}} \sim 1 /\alpha  \tau^2 \ll 1$. 
The spatial resolution necessary to resolve this region can be prohibitively high even for $1/\alpha \tau \sim 1$, since $\tau \gg 1$, and it is for this reason that, as we will see in section \ref{sec-finite}, we do not obtain numerical solutions for values of $\tau$ larger than $\tau = 10$.

\section{Numerical results} \label{sec-finite}

In this section, we study the magnetic presheath at finite values of $\tau$ using numerical simulations.
First, in section \ref{subsec-finite-bc}, we parameterize a set of magnetic presheath entrance distribution functions using $\tau$ in a way that is consistent with the limits of small ($\tau \ll 1$) and large ($\tau \gg 1/\alpha$) ion temperature studied in the previous section.
Then, in section \ref{subsec-finite-numsol}, we present numerical solutions of the electrostatic potential profile and of the ion distribution function at the Debye sheath entrance.

\subsection{Boundary conditions}
\label{subsec-finite-bc}

The ion distribution function, $f_{\infty}(\vec{v})$, that enters the magnetic presheath is determined by a kinetic solution of the bulk plasma or of the collisional presheath.
Without such a solution, there is an infinite possible number of distribution functions we could choose as boundary conditions.  
We proceed to parameterize a set of such distribution functions using $\tau = T_{\text{i}} / ZT_{\text{e}}$.
We design them to recover the two limits studied in section \ref{sec-limits}.

We proceed to make a number of observations about the properties that an appropriate set of distribution functions must satisfy.
Considering the strong resemblance of the kinetic Chodura condition (\ref{kinetic-Chodura}) with the kinetic Bohm condition, whose equality form is equation (\ref{kinetic-Bohm-marginal}), we choose that (\ref{kinetic-Chodura}) be satisfied with the equality sign,
\begin{align} \label{kinetic-Chodura-marginal}
\int \frac{ f_{\infty} \left( \vec{v} \right)}{v_z^2} d^3v = \frac{n_{\infty}}{v_{\text{B}}^2}   \text{.}
\end{align}
The assumption behind equation (\ref{kinetic-Chodura-marginal}) is that, just as the magnetic presheath solution self-consistentely satisfies the kinetic Bohm condition with the equality sign, the collisional presheath will self-consistentely satisfy the kinetic Chodura condition with the equality sign.
In order to be consistent with the models in section \ref{sec-limits} in the limits $\tau \rightarrow 0$ and $\tau \rightarrow\infty$, we also choose a set of distribution functions that:
\begin{itemize}
\item for $\tau \rightarrow 0$ is a Maxwellian that peaks at $v_z =  v_{\text{B}}$;
\item for $\tau \rightarrow \infty$ is a half Maxwellian that peaks at $v_z=0$.
\end{itemize}

A set of distribution functions that has all the above properties is
\begin{align} \label{f-infty}
f_{\infty} \left( \vec{v} \right) =  \begin{cases}
\mathcal{N}  n_{\infty} \frac{4 v_z^2}{\pi^{3/2} v_{\text{t,i}}^5}   \exp \left( - \frac{ \left| \vec{v} - u v_{\text{t,i}} \hat{\vec{z}} \right|^2 }{v_{\text{t,i}}^2} \right) \Theta \left( v_z \right) & \text{ for } \tau \leqslant 1 \text{,} \\
\mathcal{N}  n_{\infty}  \frac{ 4 v_z^2 }{  \pi^{3/2} v_{\text{t,i}}^3 \left( v_{\text{t,i}}^2 +r v_z^2 \right)} \exp \left( - \frac{\left| \vec{v} \right|^2 }{v_{\text{t,i}}^2} \right)\Theta \left( v_z \right)  & \text{ for } \tau > 1 \text{,}
\end{cases}
\end{align}
where $\Theta $ is the Heaviside step function defined in (\ref{Heaviside}).
The values of $u$ and $r$ in (\ref{f-infty}) are chosen such that condition (\ref{kinetic-Chodura-marginal}) is satisfied.
For $\tau \leqslant 1$, decreasing $\tau $ increases the parameter $u$, which increases the flow velocity of the distribution function.
For $\tau \ll 1$ and $u\gg 1$, the distribution function tends to a shifted Maxwellian with flow velocity given by $u v_{\text{t,i}}$. 
For $\tau > 1$, the parameter $r$ increases from $0$ to $\infty$ for increasing $\tau$. 
For $r > 1$, the distribution function becomes small for values of $v_z$ smaller than $ v_{\text{t,i}}/\sqrt{r}$, and thus the parameter $r$ determines the region of velocity space around $v_z = 0$ where there are almost no particles. 
For $\tau \gg 1$ and $r \gg 1$, the distribution function at the entrance of the magnetic presheath is a half Maxwellian with a very narrow region around $v_z = 0$ where the distribution function vanishes.
The quantity $\mathcal{N}$ is a normalization constant that ensures that 
\begin{align} \label{n-infty}
n_{\infty} = \int f_{\infty} \left( \vec{v} \right) d^3v  \text{.}
\end{align}
Note that, from equations (\ref{mu-infty-vx-vy}), (\ref{U-infty-vx-vy-vz}) and (\ref{f-infty}), we can write the distribution function in the form $F(\mu, U)$,
\begin{align} \label{F-infty}
F \left( \mu, U \right) =  \begin{cases}
\mathcal{N}  n_{\infty} \frac{8\left( U - \Omega \mu \right) }{\pi^{3/2} v_{\text{t,i}}^5}  \exp \left[ - \frac{2}{v_{\text{t,i}}^2}  \left(  \Omega \mu + \left( \sqrt{2\left(U - \Omega \mu \right)}  - u v_{\text{t,i}} \right)^2 \right)  \right]   & \text{ for } \tau \leqslant 1 \text{,} \\
\mathcal{N}  n_{\infty}  \frac{ 8\left( U - \Omega \mu \right) }{  \pi^{3/2} v_{\text{t,i}}^3 \left( v_{\text{t,i}}^2 + 2r \left(U - \Omega \mu \right) \right)} \exp \left( - \frac{2U}{v_{\text{t,i}}^2} \right)   & \text{ for } \tau > 1 \text{.}
\end{cases}
\end{align} 

\begin{figure} 
\centering
\includegraphics[width=0.47\textwidth]{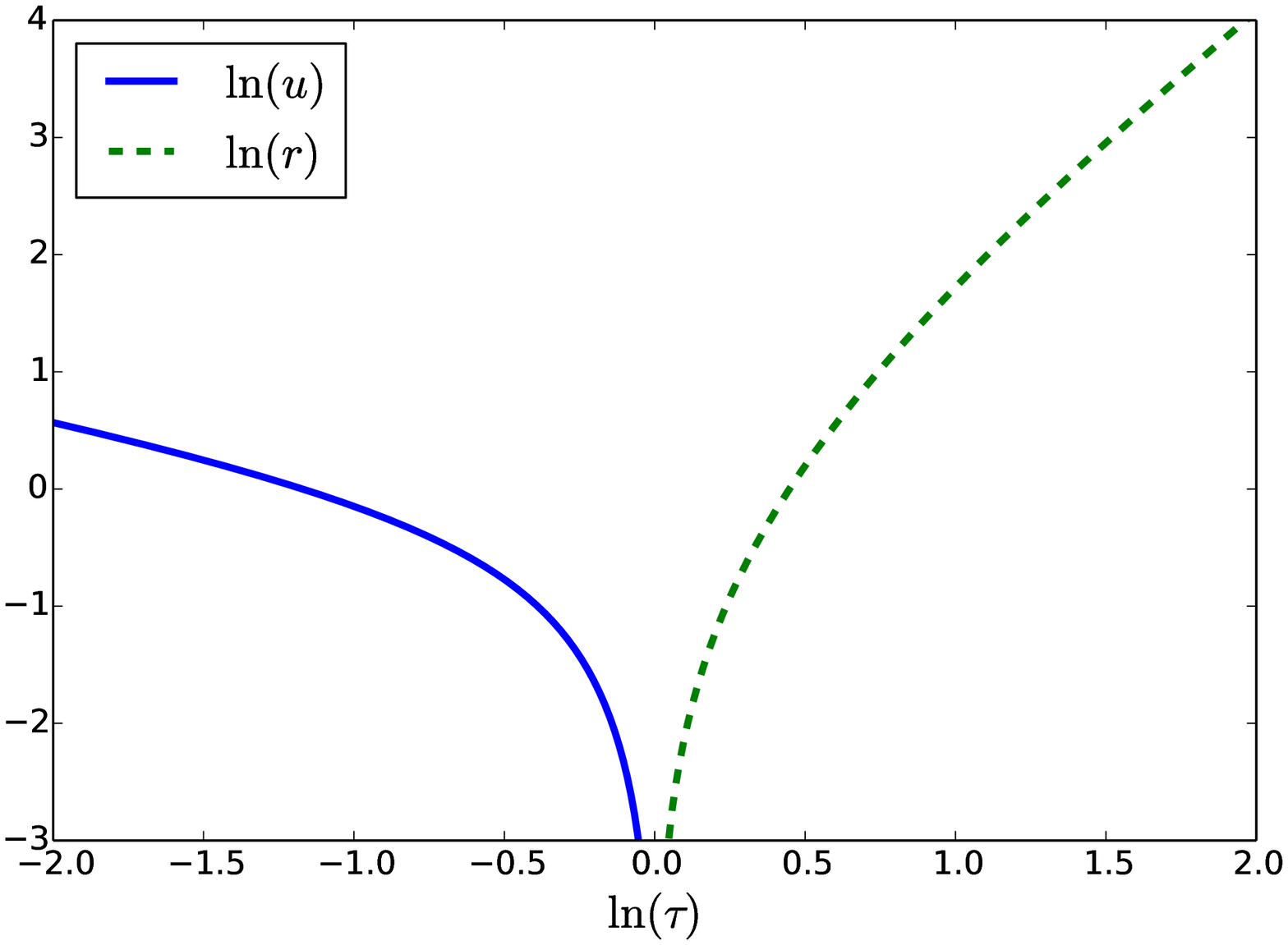}
\includegraphics[width=0.47\textwidth]{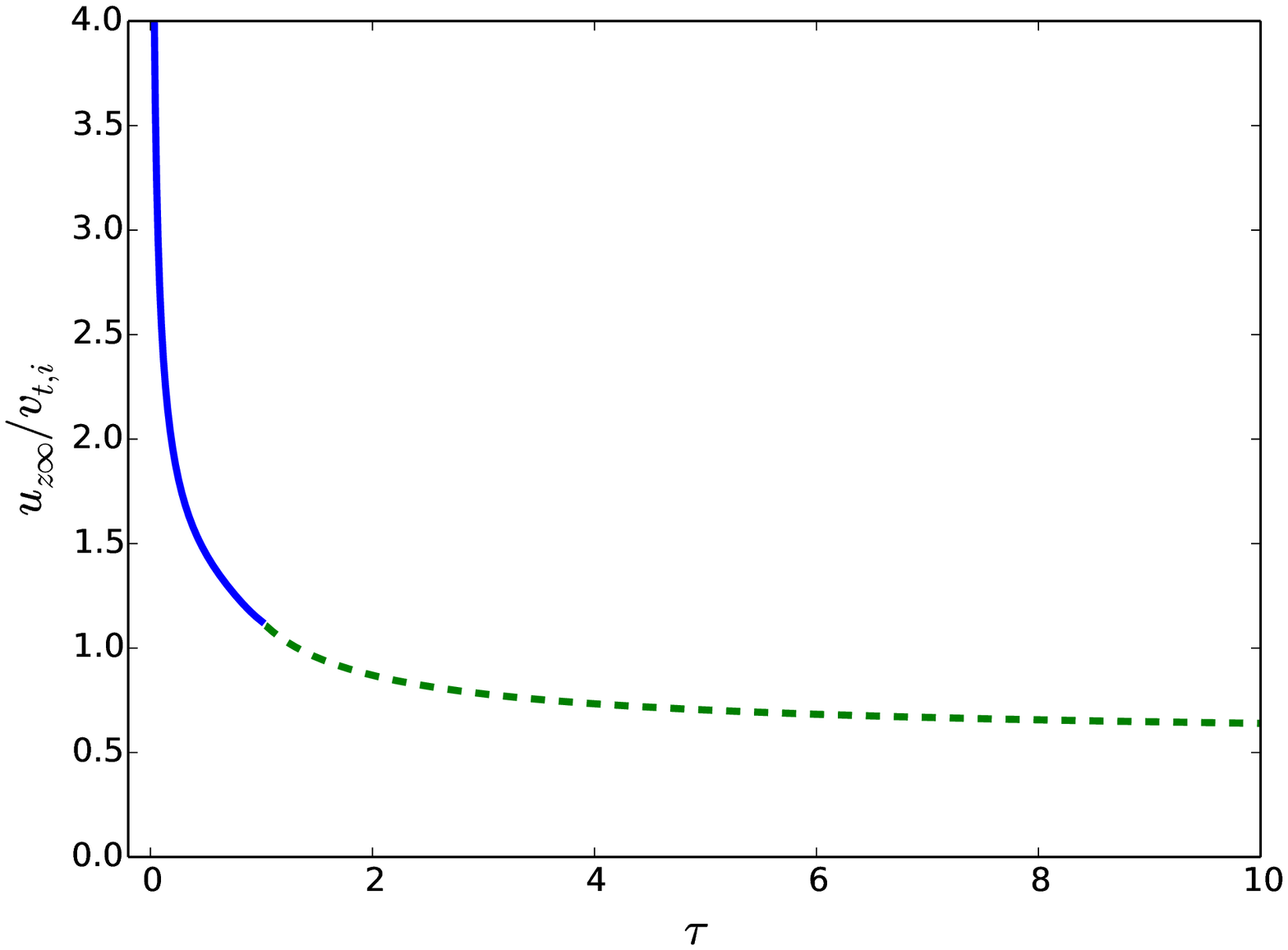}
\caption{On the left, the numbers $\ln r$ and $\ln u$ as a function of the parameter $\ln \tau$. On the right, the flow velocity at the magnetic presheath entrance, $u_{z\infty}$, as a function of the parameter $\tau$. The dashed line corresponds to $\tau >1$, where $r$ (instead of $u$) is used to parameterize the distribution functions in equation (\ref{f-infty}). Note that $u_{z\infty}/v_{\text{t,i}} \rightarrow 1/\sqrt{2\tau}$ for $\tau \rightarrow 0 $, $u_{z\infty}/v_{\text{t,i}} = 2/\sqrt{\pi}\approx 1.13$ for $\tau = 1$, and $u_{z\infty}/v_{\text{t,i}} \rightarrow 1/\sqrt{\pi} \approx 0.56$ for $\tau \rightarrow \infty$.}
\label{fig-u-r}
\end{figure}

The value of the normalization constant $\mathcal{N}$ is, from equation (\ref{n-infty}),
\begin{align} \label{N-infty}
\mathcal{N}  =  \begin{cases}
\left[  \left( 1 + 2u^2 \right) \left( 1 + \text{erf}(u) \right) + \frac{2u}{\sqrt{\pi}}  \exp(-u^2) \right]^{-1}  & \text{ for } \tau \leqslant 1 \text{,} \\
r^{3/2} \left[ 2\sqrt{r} - 2\sqrt{\pi} \exp\left(\frac{1}{r}\right) \left( 1 - \text{erf} \left( \frac{1}{\sqrt{r}} \right) \right) \right]^{-1}  & \text{ for } \tau > 1 \text{.}
\end{cases}
\end{align}
The values of $u$ and $r$ are, from equation (\ref{kinetic-Chodura-marginal}), given by 
\begin{align} \label{u-def}
 1 + \text{erf}(u)  = \tau \left[ \left( 1 + 2u^2 \right) \left( 1 + \text{erf} (u) \right) + \frac{2u}{\sqrt{\pi}} \exp(-u) \right]  \text{,}
\end{align}
\begin{align} \label{r-def}
r \sqrt{\pi} \exp\left(\frac{1}{r}\right) \left( 1 - \text{erf} \left( \frac{1}{\sqrt{r}} \right) \right)  = \tau \left[  2\sqrt{r} - 2\sqrt{\pi} \exp\left(\frac{1}{r}\right) \left( 1 - \text{erf} \left( \frac{1}{\sqrt{r}} \right) \right) \right] \text{,}
\end{align}
and are plotted as functions of $\tau$ in figure~\ref{fig-u-r}.
The fluid velocity in the $z$ direction at the magnetic presheath entrance, $u_{z\infty}$, is given by the equations
\begin{align} \label{flow-u}
\frac{ u_{z\infty} }{v_{\text{t,i}} } =  \frac{  u \left( 3+ 2u^2\right) \left( 1 + \text{erf} \left( u \right) \right) + \frac{2}{\sqrt{\pi}} \exp(-u^2) \left(1+u^2\right) }{  \left( 1 + 2u^2 \right) \left( 1 + \text{erf}(u) \right) + \frac{2}{\sqrt{\pi}} u  \exp(-u^2) }   & \text{ for } \tau \leqslant 1 \text{,}
\end{align}
and
\begin{align} \label{flow-r}
\frac{ u_{z\infty} }{v_{\text{t,i}} } =  \frac{2}{\sqrt{\pi r}} \frac{ r - \exp\left(\frac{1}{r} \right)  E_1 \left( \frac{1}{r} \right) }{  2\sqrt{r} - 2\sqrt{\pi} \exp\left(\frac{1}{r}\right) \left( 1 - \text{erf} \left( \frac{1}{\sqrt{r}} \right) \right)  }   & \text{ for } \tau > 1 \text{.}
\end{align}
In equation (\ref{flow-r}), we have introduced the exponential integral,
\begin{align} \label{E1}
E_1(\xi) = \int_{\xi}^{\infty} \frac{\exp(-\eta)}{\eta} d\eta \text{.}
\end{align}
Using equations (\ref{u-def})-(\ref{flow-r}), in figure \ref{fig-u-r} we plot the value of $u_{z\infty}$ as a function of $\tau$.
Equations (\ref{N-infty})-(\ref{flow-r}) are derived in Appendix \ref{app-integrals-Tdep}.

To conclude this subsection, we verify that the distribution functions have the required properties at $\tau \rightarrow 0$ and $\tau \rightarrow \infty$. 
 From (\ref{u-def}), note that taking the limit $\tau \rightarrow 0$ leads to $u \simeq \sqrt{1/2\tau} \gg 1$, so that the ion distribution function $f_{\infty}$ in equation (\ref{f-infty}) is indeed a Maxwellian that peaks at $v_z = v_{\text{t,i}} /\sqrt{2\tau} = v_{\text{B}}$. 
Moreover, note that taking the limit $\tau \rightarrow \infty$ in (\ref{r-def}) leads to $r \simeq \left(2\tau\right)^2/\pi \gg 1$, so that $f_{\infty}$ is a half Maxwellian that peaks at $v_z = 0$.
In the next subsection, we present the numerical results obtained for finite values of $\tau$.

\subsection{Numerical solutions}
\label{subsec-finite-numsol}

The numerical scheme presented in \cite{Geraldini-2018} is used to obtain numerical solutions to the quasineutrality equation (\ref{quasineutrality}) for values of $\alpha$ and $\tau$ in the range $0.01 \leqslant \alpha \leqslant 0.2$ (roughly corresponding to $0.57^{\circ} \leqslant \alpha \leqslant 11^{\circ}$) and $0.2 \leqslant \tau \leqslant 10$.
We define a quantity
\begin{align}
\tilde{n}(x) = 1 - \frac{Zn_{\text{i}}(x)}{n_{\text{e}}(x)} \text{.}
\end{align}
In the numerical scheme, all quantities are discretized and so $\tilde{n}_{\mu} = \tilde{n} (x_{\mu})$ is a set of values defined on a grid of values of $x_{\mu}$, where $\mu$ is an index running from $0$ to some value $\eta$.
The exact solution to equation (\ref{quasineutrality}) has $\tilde{n}(x)  = 0$ everywhere, but numerically $\tilde{n}_{\mu}$ cannot be made to be arbitrarily small at all grid points.
Hence, we use the following convergence criterion to define what constitutes a valid numerical solution to equation (\ref{quasineutrality}),
\begin{align}
\left( \frac{1}{\eta + 1} \sum_{\mu = 0}^{\eta} \tilde{n}_{\mu}^2 \right)^{1/2} < E \text{,}
\end{align}
where $E$ is a small number.
An iteration scheme, outlined in \cite{Geraldini-2018}, is performed to find the numerical electrostatic potential solution $\phi_{\mu} = \phi (x_{\mu})$ for a given value of $\alpha$ and $\tau$.
The solution numerically satisfies the quasineutrality equation with an error $E = 0.7\%$ for all values of $\tau$ except for $\tau = 0.2$, where $E = 1.2\%$.

\begin{figure}
\centering
\includegraphics[width = 0.9\textwidth]{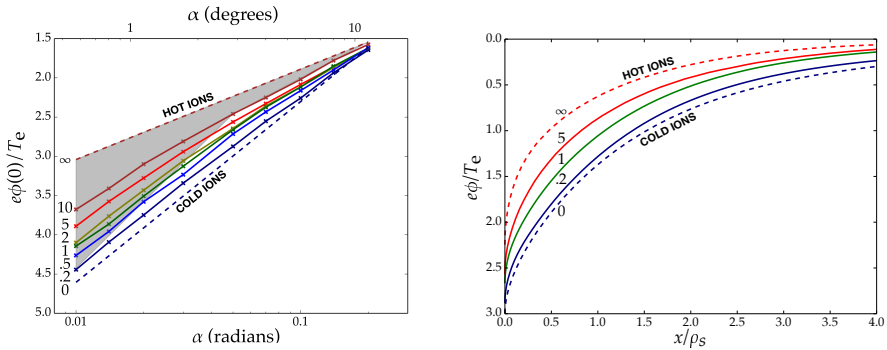} 
\caption{On the left, the electrostatic potential drop across the magnetic presheath $\phi \left( 0 \right)$ is shown as a function of the angle $\alpha$ and the parameter $\tau$.   
The region where $ \alpha \lesssim \sqrt{1+\tau} \sqrt{m_e / m_i  } $, and therefore the ordering (\ref{alpha-tau-order}) breaks down, is shaded. 
On the right, electrostatic potential profiles for $\alpha = 0.05$ at different values of $\tau$, marked on the curves. }
\label{fig-phi-Tdep}
\end{figure}

The electrostatic potential drop across the magnetic presheath is shown on the left in figure \ref{fig-phi-Tdep} as a function of $\alpha$ and $\tau$.
The numerical results approaching $\tau = 0.2$ and $\tau = 10$ are consistent with the results obtained using equation (\ref{phidrop-tau0}) (valid for small $\tau$, $ 3/|\ln\alpha| < 1/|\ln\tau|  \ll 1$) and using equation (\ref{phidrop-tauinf}) (valid for $\alpha \tau \gg 1$), shown with dashed lines.
The shaded region is where we expect the assumption of an electron-repelling wall not to be suitable for Deuterium ions, $\alpha \lesssim \sqrt{1+\tau} \sqrt{m_{\text{e}}/m_{\text{i}}} \sim 0.02 \sqrt{1+\tau} $.
Considering the unshaded region in figure \ref{fig-phi-Tdep}, the potential drop with finite ion temperature is up to $10-15\%$ smaller than the cold ion ($\tau =0$) potential drop. 
For a fixed angle, $\alpha = 0.05 \text{ rad} \approx 3^{\circ}$, the electrostatic potential profiles for different values of $\tau$ are shown on the right in Figure~\ref{fig-phi-Tdep}. 
The blue dashed curve labelled ``0'' in Figure~\ref{fig-phi-Tdep} is obtained from equation (\ref{phi-uniform}), while the red dashed curve marked ``$\infty$'' is obtained from equation (\ref{phi-hot}).
The numerical profiles are consistent with the limits $\tau = 0$ and $\tau = \infty$. 

\begin{figure}
\centering
\includegraphics[width = 1.0\textwidth]{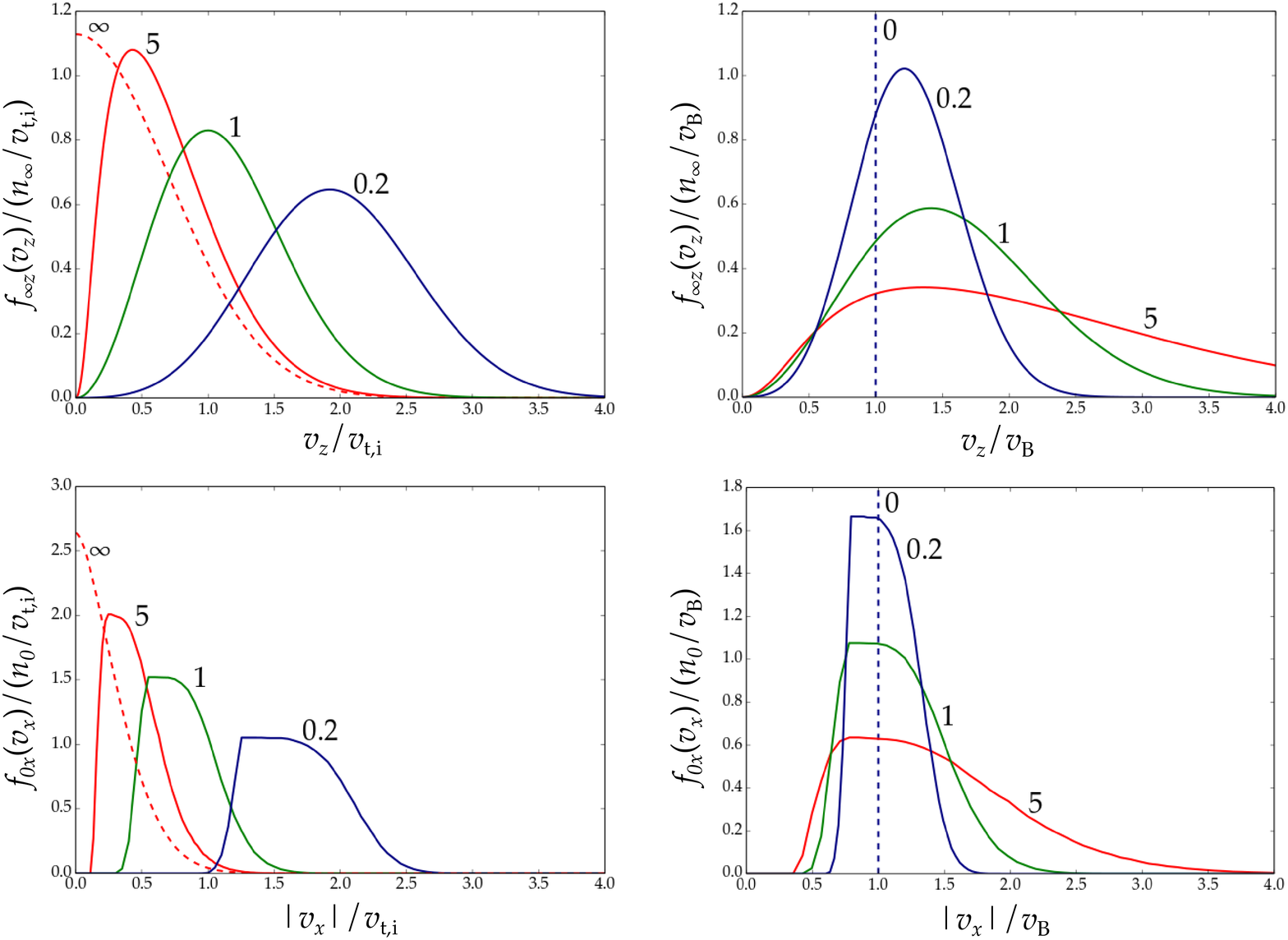} 
\caption{The distributions of the component $v_z$ of the ion velocity at the magnetic presheath entrance $x \rightarrow \infty$ (top) and the component $v_x$ of the velocity at the Debye sheath entrance $x=0$ (bottom) are shown for $\alpha = 0.05$ for three different values of the parameter $\tau$, labelled next to the corresponding curve. 
The velocities are normalized to $v_{\text{t,i}}$ on the left diagrams and to $v_{\text{B}}$ on the right diagrams.
Magnetized ions at the magnetic presheath entrance move parallel to the magnetic field. 
Hence, $v_z$ is responsible for the flow of ions to the wall.
At the Debye sheath entrance, the ion flow towards the wall is determined by $|v_x|$. 
The red dashed lines on the left diagrams are the distribution functions in the limit $\tau \rightarrow \infty$.
The blue vertical dashed lines on the right diagrams are the cold ion distribution functions, $\tau = 0$.
}
\label{fig-f0x}
\end{figure}

\begin{figure}
\centering
\includegraphics[width = 0.5\textwidth]{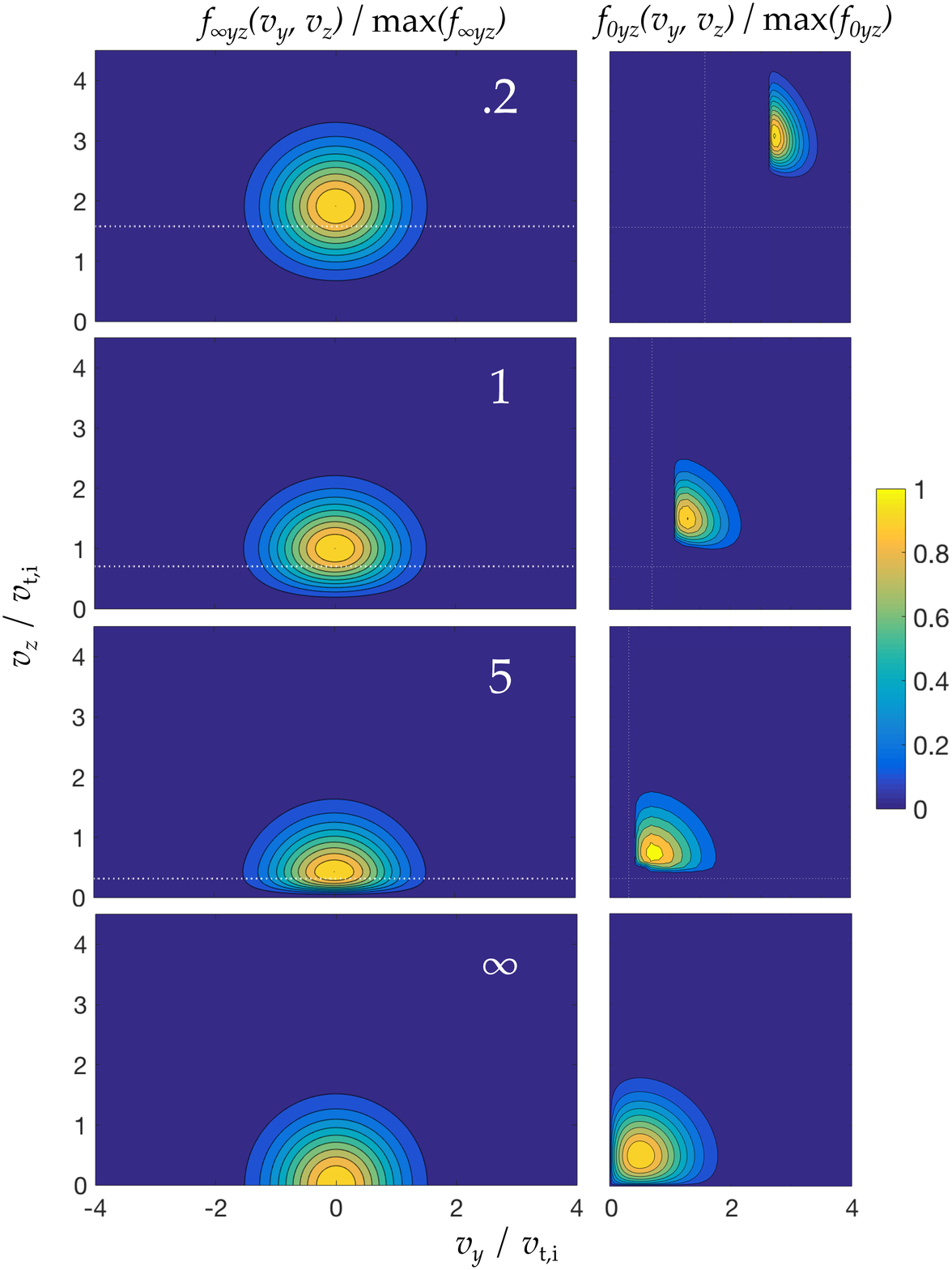} 
\caption[2D distribution function at the Debye and magnetic presheath entrance]{The ion distribution functions $f_{\infty yz}(v_y, v_z)$ (left) and $f_{0yz} (v_y, v_z)$ (right) for $\alpha = 0.05$ and, from top to bottom, for $\tau = 0.2$, $\tau = 1$, $\tau = 5$ and $\tau = \infty$ (see section \ref{sec-hot}). 
The Bohm speed $v_{\text{B}}/v_{\text{t,i}} = 1/ \sqrt{2\tau} $ is marked as a horizontal line in all panels, and also as a vertical line on the right panels. 
}
\label{fig-f0yz-Tdep}
\end{figure}

While the solution to a fluid model can give a good estimate of the electrostatic potential profile in the magnetic presheath at some range of finite temperatures, it provides no information on the velocity distribution of the ions.
The ions hitting the wall can cause sputtered neutral impurities to be thrown back into the plasma, and the sputtering yield is sensitively dependent on the kinetic energy and angle of incidence of the ion on the target.
Hence, it is important to predict the ion distribution function at the wall. 
Since in the Debye sheath ions only undergo an acceleration towards the wall, see e.g. \cite{Riemann-review}, the distribution function of ions at the Debye sheath entrance is expected to be similar in shape to the distribution function at the wall. 
For different values of $\tau$, in figure~\ref{fig-f0x} we plot the distribution function $f_{0x} (v_x ) $ (defined in equation (\ref{f0x-def})) and compare it with the boundary condition $f_{\infty z} (v_z) = \int_{-\infty}^{\infty} \int_{-\infty}^{\infty} f_{\infty } (\vec{v}) dv_y dv_x$.
Equation (\ref{f0x-hot}) is the dashed curve on the bottom-left panel in figure \ref{fig-f0x}.
The equality form of the kinetic Bohm condition (\ref{kinetic-Bohm-marginal}) \citep{Riemann-review} is approximately numerically satisfied for all distribution functions in the parameter range of the presented simulations; recall that (\ref{kinetic-Bohm-marginal}) is an analytical property of the self-consistent solution of equation (\ref{quasineutrality}) \citep{Geraldini-2018}.
Note that at values of $\tau$ larger than $\tau = 10$, it becomes computationally expensive to resolve the sharp gradient of the distribution function near $v_x=0$, as discussed at the end of section \ref{sec-hot}.
In all of our simulations, the distribution $f_{0x} (v_x )$ is found to be both narrower and more centred around $v_{\text{B}}$ than $f_{\infty z} (v_z)$. 
In figure~\ref{fig-f0yz-Tdep}, we plot the functions $f_{\infty yz}(v_y, v_z) = \int_{-\infty}^{\infty} f_{\infty } (\vec{v}) dv_x $ and $f_{0yz}(v_y, v_z)$. 
Equation (\ref{f0yz-hot}) is the bottom right panel in figure \ref{fig-f0yz-Tdep}.
For $\tau \lesssim 1$, the ions have very large tangential velocities at $x=0$ (compared with $x=\infty$) due to the large increase in the $y$-component of the velocity, related to the $\vec{E} \times \vec{B}$ drift acquired by the ion orbit in the magnetic presheath. 

We can summarize the numerical results for the distribution function as follows:
\begin{itemize}
\item for $\tau \gg 1$, the velocity components tangential to the wall, $v_y$ and $v_z$, remain unaffected while the velocity component normal to the wall, $v_x$, becomes of the order of whichever is largest between $\sqrt{\alpha} v_{\text{t,i}}$ and $v_{\text{B}} \sim v_{\text{t,i}} / \sqrt{\tau}$;
\item for $\tau \lesssim 1$, all velocity components are affected by the magnetic presheath electric field and become of order $v_{\text{B}}$ (ignoring factors of $|\ln \alpha|$).
\end{itemize} 
For large ion temperatures, $\tau \gtrsim 5$, the velocity component normal to the wall at the Debye sheath entrance is small because the electrostatic potential necessary to repel electrons barely affects the ions.
In this case, there are two regimes of interest.
Firstly, if $1 \ll \tau \ll 1/\alpha$, most ions are accelerated to $|v_x| \simeq v_{\text{B}} \sim v_{\text{t,i}} / \sqrt{\tau} \ll v_{\text{t,i}} $, as expected if the Bohm condition (\ref{kinetic-Bohm-marginal}) is to be satisfied, and the spread of the ion distribution function in the $x$ direction, $f_{0x}(v_x)$, is $v_{\text{B}}$.
The numerical solution for $\tau = 5$ and $\alpha = 0.05$, where $f_{0x}(v_x)$ is shown in the bottom panels of figure \ref{fig-f0x}, is adequately described by this regime.
Secondly, if $\tau$ is such that $\tau \gg 1/\alpha$, the velocity spread of the distribution function is $\sqrt{\alpha} v_{\text{t,i}}$, satisfying $v_{\text{B}} \ll \sqrt{\alpha} v_{\text{t,i}}  \ll v_{\text{t,i}}$; this regime corresponds to the limit taken in section \ref{sec-hot}, where $f_{0x}(v_x)$ is given in equation (\ref{f0x-hot}) and plotted in the bottom-left panel of figure \ref{fig-f0x} as a red dashed line. 
For $\alpha \sim 1/\tau$, the velocity spread is $\sqrt{\alpha } v_{\text{t,i}} \sim v_{\text{B}} \ll v_{\text{t,i}}$, as both of the estimates above are valid.
The tangential velocity of a typical ion with $\tau \gtrsim 5$ remains roughly of the same size, $v_y \sim v_z \sim v_{\text{t,i}}$, and therefore the angle between the ion trajectory and the wall is shallow at the Debye sheath entrance.
For $\tau \lesssim 1$, the typical size of all the velocity components is $v_{\text{B}}$ and thus the angle between the ion trajectory and the wall is of order unity.
Hence, an ion is expected to impinge on the wall at an angle whose size is small when $\tau \gg 1$ and order unity when $\tau \lesssim 1$.

\section{Conclusion} 
\label{sec-disc}

In this paper we have studied the dependence of a grazing-angle electron-repelling magnetic presheath on ion temperature using the kinetic model in \cite{Geraldini-2017, Geraldini-2018}.
The cold ion limit, $\tau = T_{\text{i}} / ZT_{\text{e}} \ll 1$, is described by Chodura's fluid model, giving the solution (\ref{phi-uniform}) to lowest order in $\alpha$.
In the limit $ 3/|\ln\alpha| < 1/|\ln\tau|  \ll 1$, we have analytically shown that the solution of the shallow-angle kinetic model is asymptotically equivalent to the fluid solution in (\ref{phi-uniform}) to lowest order in $\tau$ and $\alpha$. 
The numerical results for $\tau = 0.2$, shown in figure \ref{fig-phi-Tdep}, confirm that the kinetic solution tends to the fluid solution at small $\tau$. 
We have also shown that, despite the ordering $\rho_{\text{i}} \ll \rho_{\text{B}}$ for $\tau \ll 1$, 
the characteristic spatial extent of ion gyromotion in the direction normal to the wall grows to $\rho_{\text{B}}  \sqrt{| \ln \alpha |}$ as the ion approaches the wall, thus becoming comparable to the size of the magnetic presheath.
The growth of ion gyro-orbits is accompanied by a decrease in the gyration velocity in order to conserve the adiabatic invariant, as can be seen in figure \ref{fig-orbitgrowth}.
Hence, if the ion thermal energy is too small, the gyration velocity of ion orbits becomes comparable to the orbit drift, thus invalidating the gyrokinetic assumption underlying our kinetic model.
For the largest orbits, our kinetic model breaks down if $\tau^3 \lesssim \alpha$.

In the hot ion limit, $\tau \rightarrow \infty$, our model corresponds to a model briefly studied in \cite{Cohen-Ryutov-1998}, which we described in section \ref{sec-hot}.
From the electrostatic potential results shown in figure \ref{fig-phi-Tdep}, the largest values of ion temperature, $\tau = 5$ and $\tau = 10$, are consistent with the large ion temperature limit.
Our results for the distribution function at the Debye sheath entrance (shown in figures \ref{fig-f0x} and \ref{fig-f0yz-Tdep}, for $\alpha = 0.05$) show that the angle between a typical ion trajectory and the wall is smaller at large values of $\tau$.
Correspondingly, ions that have traversed the magnetic presheath tend to have a smaller spread of the normal component of the velocity, $v_x$.
The latter effect, which is also present for $\tau \sim 1$ and $|\ln \alpha | \gg 1$ (to be treated in a future publication), is particularly prominent for $\tau \gg 1$.
For $1\ll \tau \ll 1/\alpha$ ions reach the wall with a range of velocities that is centred at $v_x \approx v_{\text{B}}$ (consistent with the kinetic Bohm condition (\ref{kinetic-Bohm-marginal})) and whose spread is $ v_{\text{B}} \sim v_{\text{t,i}} / \sqrt{\tau}$ (see, for example, $\alpha = 0.05$ and $\tau = 5$ in figure \ref{fig-f0x}).
For $\tau \gg 1/\alpha$, ions reach the wall with a range of velocities that is peaked at $v_x \sim  v_{\text{B}} / \sqrt{\alpha \tau } \ll v_{\text{B}}$ (essentially $v_x \simeq 0$), and whose spread is $\alpha^{1/2} v_{\text{t,i}}$ (see the plot for $\alpha = 0.05$ and $\tau \rightarrow \infty$ in figure \ref{fig-f0x}).

Chodura's fluid model of the magnetic presheath can give electrostatic potential profiles that are qualitatively similar to the ones obtained using our kinetic model for $\tau \lesssim 1$ (see figure \ref{fig-phi-Tdep}).
At larger values of $\tau$, the quantitative difference between the fluid profile and the kinetic profile becomes more evident.
For very large values of $\tau$, the potential drop normalized to electron temperature is up to a factor of $30\%$ smaller than for $\tau = 0$.
However, at such large values of $\tau$ the electrons would not be adiabatic, as was assumed here, since the assumption $\alpha / \sqrt{1+\tau} \gg \sqrt{m_{\text{e}} / m_{\text{i}} }$ would not be satisfied.
In this case, the Debye sheath would not repel most of the electrons back into the magnetic presheath, and a kinetic treatment of both ions and electrons would be necessary.
The ordering $\alpha / \sqrt{1+\tau} \sim \sqrt{m_{\text{e}}/ m_{\text{i}}}$ has mostly been avoided in the literature to date, but is becoming more relevant for fusion devices since $\sqrt{m_{\text{e}}/ m_{\text{i}}  } \sim 0.02 \text{ rad} \approx  1^{\circ} $ for Deuterium plasmas, $\tau \gtrsim 1$ near divertor targets \citep{Mosetto-2015} and $\alpha \sim 2.5^{\circ}$ is expected in ITER \citep{Pitts-2009}. 

~

This work was supported by the US Department of Energy through grant number DE-FG02-93ER-54197.
This work has been carried out within the framework of the EUROfusion Consortium and has received funding from the Euratom research and training programme 2014-2018 and 2019-2020 under grant agreement No 633053. The views and opinions expressed herein do not necessarily reflect those of the European Commission.

\appendix
\newpage

 \section{Glossary of notation} 
 \label{app-symbols}
 
Here, we provide a glossary of some of the notation used in this paper.
For each symbol, we give a brief description and a reference to the equation where the symbol first appears.
 
 \begin{align*}
 \centering
& \text{symbol} & & \text{name or description }                                           & & \text{appears in } \nonumber \\ \hline
& \tau & & \text{ion temperature / } Z\times \text{ electron temperature } & & \text{equation (\ref{tau})} \nonumber \\ 
& \alpha & & \text{angle between magnetic field and target } & & \text{equation (\ref{B-def})} \nonumber  \\ 
& \phi & & \text{electrostatic potential } & & \text{equation (\ref{E-field})} \nonumber \\
& v_{\text{B}} & & \text{Bohm speed } & & \text{equation (\ref{vB})} \nonumber \\
& c_{\text{s}} & & \text{sound speed } & & \text{equation (\ref{cs})} \nonumber \\
& \rho_{\text{s}} & & \text{sound gyroradius } & & \text{equation (\ref{rho-s})} \nonumber \\
& \rho_{\text{B}} & & \text{Bohm gyroradius } & & \text{equation (\ref{rho-B})} \nonumber \\
& \bar{x} & & \text{orbit position } & & \text{equation (\ref{xbar-def})} \nonumber  \\
& U_{\perp} & & \text{perpendicular energy } & & \text{equation (\ref{Uperp-def})}  \nonumber \\
& U  & & \text{total energy } & & \text{equation  (\ref{U-def})} \nonumber \\
& V_x & & \text{absolute value of } v_x \text{ as a function of } x,~\bar{x} \text{ and } U_{\perp} & & \text{equation (\ref{vx-x-xbar-Uperp})} \nonumber \\
& V_{\parallel} & & v_z \text{ as a function of } U_{\perp} \text{ and } U & & \text{equation (\ref{vz-Uperp-U})} \nonumber \\
& \chi & & \text{effective potential appearing in the function } V_x & & \text{equation (\ref{chi})} \nonumber \\
& \chi_{\text{m}} ~ [x_{\text{m}}]  & & \text{effective potential minimum [position]} & & \text{equation (\ref{chim})} \nonumber \\
& \mu~[\mu_{\text{gk}}] & & \text{adiabatic invariant [functional form]} & & \text{equation (\ref{mu-Uperp-xbar})} \nonumber \\
& \rho_x & & \text{periodic piece of ion position for } \tau \ll 1  & & \text{equation (\ref{rhox-def})} \nonumber \\
& w_x  & & \text{periodic piece of ion velocity for } \tau \ll 1  & & \text{equation (\ref{wx-def})} \nonumber \\
& n_{\text{i,cl}} & & \text{number density of ions in closed orbits } & & \text{equation (\ref{ni-closed})} \nonumber \\
& n_{\text{i,op}} & & \text{number density of ions in open orbits }  & & \text{equation (\ref{ni-open})} \nonumber \\
& \bar{x}_{\text{m}} & & \text{minimum allowed } \bar{x} \text{ for closed orbit crossing } x & & \text{equation (\ref{xbarm-def})} \nonumber \\
& \bar{x}_{\text{m,o}} & & \text{mininum allowed } \bar{x} \text{ for open orbit crossing } x & & \text{equation (\ref{xbarm-open})} \nonumber \\
& \bar{x}_{\text{c}}~[x_{\text{c}}] & & \text{minimum allowed } \bar{x} \text{ [stationary pt. of } \chi (\bar{x}_{\text{c}}, x) \text{]} & & \text{equation (\ref{xbarc})} \nonumber  \\
& \chi_{\text{M}} ~[x_{\text{M}}] & & \text{effective potential maximum [position]} & & \text{equation (\ref{chiM-def})} \nonumber \\
& \chi_{\text{c}} & & \chi (\bar{x}_{\text{c}}, x_{\text{c}}) & & \text{equation (\ref{chic-def})} \nonumber \\
& \Delta_{\text{M}} & & \text{spread of values of } v_x^2/2 \text{ of open orbits} & & \text{equation (\ref{DeltaM-mu})} \nonumber \\
& n_{\infty} & & \text{number density of ions at } x\rightarrow \infty & & \text{equation (\ref{ne})} \nonumber \\
& u_x & & \text{fluid velocity component normal to target} & & \text{equation (\ref{ux})} \nonumber \\
& u_{x\infty}, u_{z\infty}  & & \text{fluid velocity components at } x\rightarrow \infty & & \text{equation (\ref{ux-cont})} \nonumber \\
& f_0  & & \text{ion distribution function at } x=0 & & \text{equation (\ref{f0})} \nonumber \\
& \hat{\Pi} & & \text{top hat function} & & \text{equation (\ref{tophat})} \nonumber \\
& \epsilon & & 1 / |\ln \alpha | & & \text{equation (\ref{ordering-logtau})} \nonumber \\
& l & & \text{length scale of } \phi''(x) & & \text{equation (\ref{l-def})} \nonumber \\
& v_{\text{d}} & & x\text{-component of ion drift velocity} =\dot{x}_{\text{m}}  & & \text{equation (\ref{vx-drift})} \nonumber \\
& C ~[\kappa ] & & \text{constant parameters of parabolic piece of } \phi  & & \text{equation (\ref{phi-parabola})} \nonumber \\
& K & & \text{dimensionless constant related to } \chi_{\text{c}} & & \text{equation (\ref{K-def})} \nonumber \\
& \phi_{\text{np}} & & \text{non-parabolic piece of electrostatic potential} & & \text{equation (\ref{phinp-kinetic})} \nonumber \\
& \psi & & \text{function related to } \phi & & \text{equation (\ref{psi-def})} \nonumber \\
& \Theta & & \text{Heaviside step function } & & \text{equation (\ref{Heaviside})} \nonumber \\
& \tilde{v}_y,~\tilde{U}_{\perp},~\tilde{U} & & \text{dimensionless } v_y,~U_{\perp},~U \text{ (integration variables)} & & \text{equation (\ref{ni-closed-coldelectrons})} \nonumber \\
& \tilde{\bar{x}},~\tilde{v}_z & & \text{dimensionless } \bar{x},~v_z \text{ (integration variables)} & &\text{equation (\ref{ni-open-coldelectrons-1})} \nonumber \\
& f_{\infty} & & \text{ion distribution function at } x\rightarrow \infty & & \text{equation (\ref{f-infty})} \nonumber \\
& F & & \text{ion distribution function in magnetic presheath} & & \text{equation (\ref{F-infty})} \nonumber \\
& \mathcal{N} & & \text{normalization of } f_{\infty} & & \text{equation (\ref{N-infty})}  \nonumber \\
& u & & \text{parameter of } f_{\infty} \text{ for } \tau \leqslant 1 & & \text{equation (\ref{u-def})} \nonumber \\
& r & & \text{parameter of } f_{\infty} \text{ for } \tau > 1 & & \text{equation (\ref{r-def})} \nonumber  \\
 \end{align*}

 \section{Derivation of equation (\ref{DeltaM-mu})} 
 \label{app-DeltaM}

In \cite{Geraldini-2017} the quantity $\Delta_{\text{M}} $ appearing in the open orbit density (\ref{ni-open}) was expressed as
\begin{align} \label{DeltaM}
\Delta_{\text{M}} (\bar{x}, U) = 2 \alpha \Omega^2 V_{\parallel} \left( \chi_{\text{M}}(\bar{x}) , U \right)  \int_{x_{\text{M}}}^{x_{\text{t,M}}} \frac{ x- x_{\text{M}} }{V_x \left(x, \bar{x}, \chi_{\text{M}}(\bar{x}) \right)} dx \text{.}
\end{align}
We proceed to show that equations (\ref{DeltaM}) and (\ref{DeltaM-mu}) for $\Delta_{\text{M}} $ are equivalent.

Open orbits have $U_{\perp} = \chi_{\text{M}} (\bar{x})$ to lowest order.
Hence, their orbit position $\bar{x}$ determines the perpendicular energy $U_{\perp}$. 
Every ion in an open orbit must have come from a closed orbit which had an adiabatic invariant equal to $ \mu = \mu_{\text{gk}} (\bar{x}, \chi_{\text{M}} (\bar{x})) $, where $\mu_{\text{gk}}$ is defined in equation (\ref{mu-Uperp-xbar}).
Taking the total derivative of $\mu$ with respect to $\bar{x}$ leads to 
\begin{align} \label{dmudxbar-int}
\left. \frac{ d \mu }{ d \bar{x} } \right\rvert_{ \text{open} } = \frac{\partial \mu_{\text{gk}}}{\partial U_{\perp} } (\bar{x}, \chi_{\text{M}} )  \frac{d \chi_{\text{M}} }{d \bar{x}} + \frac{\partial \mu_{\text{gk}}}{\partial \bar{x} }  (\bar{x}, \chi_{\text{M}} )  \text{.}
\end{align}
Using equation (\ref{vx-x-xbar-Uperp}), we obtain the partial derivatives $\partial V_x / \partial U_{\perp} = 1 /  V_x $, $\partial V_x / \partial \bar{x} = \Omega^2 \left( x - \bar{x} \right) /  V_x $. 
Then, differentiating equation (\ref{mu-Uperp-xbar}) under the integral sign (which is possible because the limits of integration are points where the integrand vanishes), we get
\begin{align} \label{dmudUperp}
\frac{\partial \mu_{\text{gk}}}{\partial U_{\perp} } (\bar{x}, U_{\perp} ) =  \frac{1}{\pi} \int_{x_{\text{b}}}^{x_{\text{t}}} \frac{1}{V_x \left( x, \bar{x}, U_{\perp} \right)} dx \text{,}
\end{align}
and 
\begin{align} \label{dmudxbar}
\frac{\partial \mu_{\text{gk}}}{\partial \bar{x}} (\bar{x}, U_{\perp} ) =  \frac{1}{\pi} \int_{x_{\text{b}}}^{x_{\text{t}}} \frac{\Omega^2 \left( x - \bar{x} \right)}{V_x \left( x, \bar{x}, U_{\perp} \right)} dx \text{.}
\end{align}
To obtain $  d \chi_{\text{M}}  / d \bar{x} $, we first write 
\begin{align} \label{chiM-exp}
 \chi_{\text{M}}(\bar{x}) =  \chi(x_{\text{M}} , \bar{x}) = \frac{1}{2} \Omega^2 \left( x_{\text{M}} - \bar{x} \right)^2 + \frac{\Omega \phi(x_{\text{M}} )}{B} \text{.}
\end{align} 
As was argued in \cite{Geraldini-2018}, one of the two terms in $ d \chi_{\text{M}}  / d \bar{x} $ is $\chi'(x_{\text{M}}, \bar{x} ) d x_{\text{M}}/ d \bar{x} = 0$, because $\chi'(x_{\text{M}}, \bar{x} ) = 0$ if the maximum is a stationary point of $\chi$, and $d x_{\text{M}}/d \bar{x} =0$ if the maximum is the non-stationary point $x_{\text{M}} = 0$.
Hence, only one term is left when differentiating equation (\ref{chiM-exp}),
\begin{align} \label{dxMdxbar}
\frac{d \chi_{\text{M}} }{d \bar{x}}  =  \Omega^2 \left( \bar{x} - x_{\text{M}} \right) \text{.}
\end{align}
Inserting (\ref{dmudUperp}), (\ref{dmudxbar}) and (\ref{dxMdxbar}) into (\ref{dmudxbar-int}), we obtain
 \begin{align} \label{dmudxbar-open}
\left. \frac{ d \mu }{ d \bar{x} } \right\rvert_{ \text{open} } =  \frac{\Omega^2}{\pi} \int_{x_{\text{M}}}^{x_{\text{t,M}}}  \frac{ x- x_{\text{M}} }{V_x \left(x, \bar{x}, \chi_{\text{M}} \right)} dx \text{.}
\end{align}
Then, equation (\ref{DeltaM-mu}) follows from (\ref{dmudxbar-open}) and (\ref{DeltaM}).

\section{Chodura's fluid model}

In this appendix, we first recap Chodura's fluid model, valid for any angle $\alpha$, and derive the differential equation (\ref{udiff-Riemann}).
We then proceed to expand the fluid model to lowest order in $\alpha$ using the ordering $\alpha \ll 1$.
We thus derive equation (\ref{phi-uniform}), which coincides with the solution of the kinetic model in the ordering (\ref{ordering-logtau}) to lowest order in $\alpha$ and $\tau$.

\subsection{General oblique angles: derivation of equation (\ref{udiff-Riemann})} \label{app-fluid-exact}

In this appendix subsection, we consider general oblique angles, $\alpha \sim 1$ (in radians).
For $\tau = T_{\text{i}} / T_{\text{e}} = 0$, all ions have the same velocity, the ion fluid velocity $\vec{u} = (u_x, u_y, u_z)$, and thus the ion equations of motion (\ref{x-EOM})-(\ref{z-EOM}) reduce to
\begin{align} \label{x-momentum}
u_x u_x' = - \frac{\Omega \phi'}{B} + \Omega u_y \cos \alpha \text{,}
\end{align}
\begin{align} \label{y-momentum}
u_x u_y' = -  \Omega u_x \cos \alpha - \Omega u_z \sin \alpha \text{,}
\end{align}
\begin{align} \label{z-momentum}
u_x u_z' =  \Omega u_y \sin \alpha \text{.}
\end{align}
Here, $'$ indicates differentiation with respect to $x$.
The fluid equations (\ref{x-momentum})-(\ref{z-momentum}) follow from the particle equations of motion (\ref{x-EOM})-(\ref{z-EOM}) by setting $\vec{v}=\vec{u}$ and using $u_x = \dot{x}$ to write $\dot{\vec{u}} = u_x \vec{u}'$ (thus changing the time derivative of every velocity component to a spatial derivative).

Adding equations (\ref{x-momentum})-(\ref{z-momentum}) multiplied by $u_x$, $u_y$ and $u_z$ respectively, dividing by $u_x$ and integrating leads to
\begin{align} \label{energy-equation}
\frac{1}{2} v_{\text{B}}^2 = \frac{1}{2} u_x^2  + \frac{1}{2} u_y^2 + \frac{1}{2} u_z^2 + \frac{\Omega \phi}{B} \text{,}
\end{align}
where we used $\phi (\infty) =0$ and the boundary condition (\ref{bc-infty}). 
We proceed to obtain a differential equation for $\phi(x)$ from equation (\ref{energy-equation}), following the derivation in \cite{Riemann-1994}\footnote{In \cite{Riemann-1994} (originally in \cite{Chodura-1982}) the corresponding differential equation for $u_x(x)$ was derived.}.
Differentiating (\ref{ux-phi-exact}) gives
\begin{align} \label{ux'-phi}
u_x' = \frac{e \phi' }{T_{\text{e}}} v_{\text{B}}  \exp\left( - \frac{e\phi}{T_{\text{e}}} \right)  \sin \alpha  \text{.}
\end{align}
Inserting (\ref{ux'-phi}) in (\ref{x-momentum}) and re-arranging gives
\begin{align} \label{uy-phi}
 u_y = \frac{ \phi'}{B  \cos \alpha  } \left( 1 - \exp \left( - \frac{ 2e\phi}{T_{\text{e}}} \right) \sin^2 \alpha \right)   \text{.}
\end{align}
Equations (\ref{ux-phi-exact}) and (\ref{uy-phi}) are substituted in equation (\ref{z-momentum}) to obtain
\begin{align} \label{uz-inter}
u_z' =- \frac{v_{\text{B}}}{\cos\alpha} \left( \exp \left( \frac{e\phi}{T_{\text{e}}} \right) - \exp \left( - \frac{e\phi}{T_{\text{e}}} \right) \sin^2 \alpha \right) \frac{e\phi'}{T_{\text{e}}} \text{.}
\end{align}
Using the boundary conditions in (\ref{bc-infty}), equation (\ref{uz-inter}) integrates to
\begin{align} \label{uz-phi}
u_z = \frac{v_{\text{B}}}{\cos\alpha} \left[ 2 -  \exp \left( \frac{e\phi}{T_{\text{e}}} \right)  -   \exp \left( - \frac{e\phi}{T_{\text{e}}}  \right) \sin^2 \alpha \right] \text{.}
\end{align}
Substituting equations (\ref{ux-phi-exact}), (\ref{uy-phi}) and (\ref{uz-phi}) into the energy equation (\ref{energy-equation}) results in equation (\ref{udiff-Riemann}), which is solved by imposing a boundary condition at $x=0$, the Debye sheath entrance. 

We proceed to discuss this boundary condition.
First, we note that equation (\ref{udiff-Riemann}) has a singularity at $|u_x|/v_{\text{B}} =  \sin\alpha \exp \left( e\phi / T_{\text{e}}  \right) = 1$ and that our boundary condition at $x\rightarrow \infty$ imposed $|u_x|/v_{\text{B}} =  \sin\alpha \exp \left( e\phi / T_{\text{e}}  \right) =  \sin\alpha  < 1$.
Since a crossing of the singularity in equation (\ref{udiff-Riemann}) would not be physical, it follows that the quantity $|u_x|/v_{\text{B}} =  \sin\alpha \exp \left( e\phi / T_{\text{e}}  \right)$ should stay below unity or reach unity at $x=0$, $|u_x(0)|/v_{\text{B}} \leq 1$. 
However, the Bohm condition for a stationary Debye sheath requires that $|u_x(0)|/v_{\text{B}} \geq 1$.
Therefore, the only way to match the magnetic presheath with the Debye sheath 
is by using the boundary condition $u_x(0) /v_{\text{B}} =  \sin\alpha \exp \left( e\phi(0) / T_{\text{e}}  \right) = 1$.
The electrostatic potential profile in the magnetic presheath can then be obtained by numerically integrating equation (\ref{udiff-Riemann}) using $e\phi(0) / T_{\text{e}} = \ln \left( \sin \alpha \right)$ as a boundary condition. 

\subsection{Shallow angles: derivation of equation (\ref{phi-uniform})} \label{app-expanded}

\begin{figure}
\centering
\includegraphics[width = 0.9\textwidth]{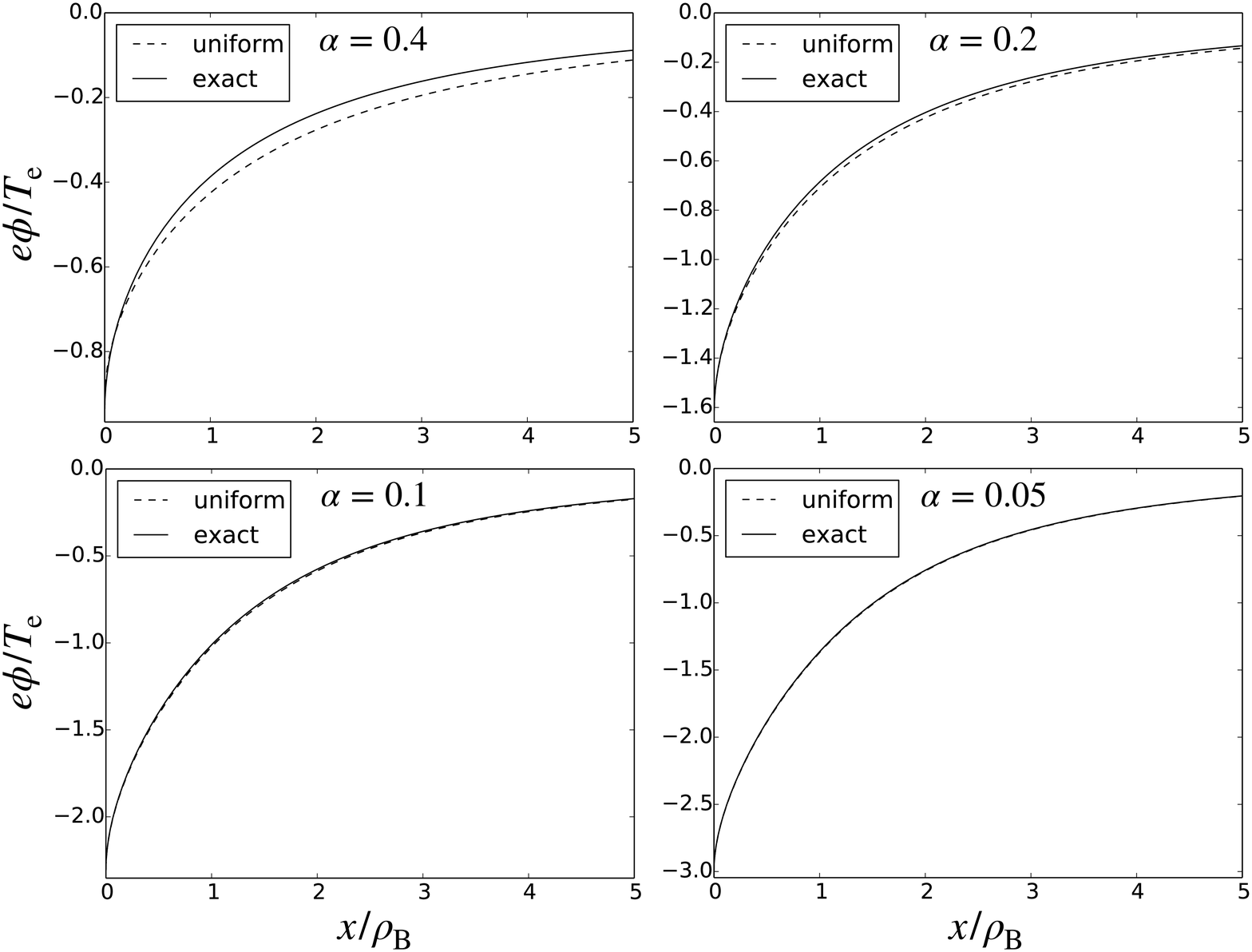}
\caption{Electrostatic potential $\phi(x)$ for four different values of $\alpha$.
The solid line results from solving the exact equation (\ref{udiff-Riemann}), while the dashed line results from the approximation (\ref{phi-uniform}).}
\label{fig-phiunif}
\end{figure}

We proceed to expand equation (\ref{udiff-Riemann}) for $\alpha \ll 1$, with the aim of obtaining $e\phi(x)/T_{\text{e}}$ correct excluding terms that are small in $\alpha$.
The electrostatic potential $\phi$ in equation (\ref{udiff-Riemann}) changes from $\phi(\infty) = 0$ to $e \phi(0) / T_{\text{e}} \simeq  \ln \left( \alpha \right) $ at $x=0$. 
Neglecting terms that are small in $\alpha$ over the entire range of values of $\phi$, equation (\ref{udiff-Riemann}) becomes
\begin{align} \label{udiff-Riemann-1}
\left( \alpha^2 \exp\left( -  \frac{2e\phi}{T_{\text{e}}} \right) - 1 \right)^2  \frac{v_{\text{B}}^2}{\Omega^2 } \left( \frac{e\phi'}{T_{\text{e}}} \right)^2  = -3 -  \alpha^2  \exp \left( - \frac{2e\phi}{T_{\text{e}}} \right)  - \frac{2e\phi}{T_{\text{e}}} \nonumber \\ + 4 \exp \left( \frac{e\phi}{T_{\text{e}}} \right) - \exp \left( \frac{2e\phi}{T_{\text{e}}} \right)  + O\left(  \alpha^2 \exp \left( - \frac{e\phi}{T_{\text{e}}}  \right) \right)  \text{.}
\end{align}
By substituting the definition of $\psi$ in equation (\ref{psi-def}), equation (\ref{udiff-Riemann-1}) becomes 
\begin{align} \label{psidiff-fluid}
\rho_{\text{B}}^2 \psi'^2  = - 3 - 2\psi + 4  \exp( \psi )   -  \exp(2\psi) + O(\alpha^2 \exp(-\psi)) \text{.}
\end{align}
Notice that equation (\ref{psidiff-fluid}) satisfies $\psi' = 0$ for $\psi = 0$ and therefore also satisfies $\phi' = 0$ for $\phi = 0$, a condition which is satisfied by the exact equation (\ref{udiff-Riemann})\footnote{As \cite{Riemann-1994} showed, the derivative of the right hand side of equation (\ref{udiff-Riemann}) evaluated at $\phi = 0$ is equal to zero, and the second derivative is equal to zero when the Chodura condition is marginally satisfied (which is the case we consider). Equation (\ref{psidiff-fluid}) has both of these properties, while equation (\ref{udiff-Riemann-1}) has neither of them.} but is not exactly satisfied by equation (\ref{udiff-Riemann-1}).
There are two terms of equal size that give rise to the error in (\ref{psidiff-fluid}).
One is the error in (\ref{udiff-Riemann-1}), and the other arises from the non-equivalence of $\psi$ and $e\phi/T_{\text{e}}$, giving
\begin{align}
 \exp( \psi ) - \exp \left( \frac{e\phi}{T_{\text{e}}} \right) \sim \alpha^2 \exp \left( - \psi \right) \lesssim \alpha \text{.}
\end{align}
Hence, equation (\ref{phi-uniform}) gives the uniformly valid magnetic presheath electrostatic potential in Chodura's fluid model to lowest order in $\alpha$.

In figure \ref{fig-phiunif}, we plot the electrostatic potential, $\phi(x)$, that results from solving (\ref{udiff-Riemann}) (exact) and (\ref{phi-uniform}) (approximate) for four different values of $\alpha$: the approximate solution is different from the exact solution for $\alpha = 0.4$, is very close to the exact solution for $\alpha = 0.2$ and almost overlaps with the exact solution for $\alpha = 0.1$ and $\alpha = 0.05$.
 
 \section{Alternative derivation of drift velocity of closed ion orbits into the wall} \label{app-xmdot}
 
The drift velocity $v_{\text{d}} = \dot{x}_{\text{m}}$ can be obtained using the relation
\begin{align}
\dot{x}_{\text{m}} = \frac{dx_{\text{m}}}{d\bar{x}} \dot{ \bar{x} } \text{.}
\end{align}
From equation (\ref{chi-minimum-1}) we have
\begin{align}
\bar{x} = x_{\text{m}} + \frac{\phi'(x_{\text{m}})}{B\Omega} \text{,}
\end{align}
which can be differentiated to obtain 
\begin{align} \label{dxbardxm}
\frac{d\bar{x}}{dx_{\text{m}}} = \frac{\chi''(x_{\text{m}})}{\Omega^2} \text{.}
\end{align}
Therefore, the drift velocity is
\begin{align} \label{xmdot-appf}
v_{\text{d}} = \frac{\Omega^2 \dot{\bar{x}}}{\chi''(x_{\text{m}})} \text{.}
\end{align}
Inserting $\dot{x} = v_x$ and equation (\ref{y-EOM-alpha}) into $\dot{\bar{x}} = \dot{x} + \dot{v}_y/\Omega$, we obtain the relation $\dot{\bar{x}} = - \alpha v_z$.
As a final step, we insert $\dot{\bar{x}} = - \alpha v_z$ into equation (\ref{xmdot-appf}), and we use equation (\ref{Vz}) for $v_z$ to recover equation (\ref{vx-drift}).

\section{Alternative derivation of closed and open orbit ion density for small $\tau$}  
\label{app-smalltau}

For $\tau \rightarrow 0$, the ion distribution function $F$ tends to
\begin{align} \label{F-cold}
F (\mu, U ) =  \frac{n_{\infty} v_{\text{B}}}{2\pi \Omega} \delta (  \mu ) \delta  \left( U  - \frac{1}{2} v_{\text{B}}^2 \right) \text{.}
 \end{align}
We proceed to use this distribution function to derive equations (\ref{ni-closed-cold}) and (\ref{ni-open-cold}).

\subsection{Closed orbit density} 
\label{subapp-cold-closed}

Using equation (\ref{chi-nearmin}), the adiabatic invariant of an ion in a closed orbit is given by
\begin{align}  \label{mu-cold-1}
\mu  \simeq  \frac{\sqrt{2\left( U_{\perp} - \chi_{\text{m}} (\bar{x}) \right) }}{\pi} \int_{x_{\text{b}}}^{ x_{\text{t}} } \sqrt{  1 - \frac{\chi''(x_{\text{m}})  \left( x - x_{\text{m}} \right)^2 }{ 2\left( U_{\perp} - \chi_{\text{m}} (\bar{x})  \right) }  } dx \left( 1 + O\left( \frac{\rho_x^2}{l^2} \right) \right) \text{,}
\end{align}
with $x_{\text{b}} = x_{\text{m}} - \sqrt{ 2\left( U_{\perp} - \chi_{\text{m}} (\bar{x}) \right) / \chi''(x_{\text{m}}) }$ and $x_{\text{t}} = x_{\text{m}} + \sqrt{ 2\left( U_{\perp} - \chi_{\text{m}} (\bar{x}) \right) / \chi''(x_{\text{m}}) }$.
In equation (\ref{mu-cold-1}), the $O\left( \rho_x^2 / l^2 \right)$ error comes from the fourth order term of the Taylor expansion of $\chi$ around $x_{\text{m}}$, since the third order term integrates to zero.
Equation (\ref{mu-cold-1}) thus reduces to
\begin{align} \label{mu-Uperp-cold}
\mu \simeq \frac{U_{\perp} -\chi_{\text{m}}(\bar{x})}{\sqrt{ \chi''(x_{\text{m}}) } }  \left( 1 + O\left( \frac{\rho_x^2}{l^2} \right) \right)  \text{.}
\end{align}
Inserting the distribution function of equation (\ref{F-cold}) into the closed orbit integral (\ref{ni-closed}) and changing from $U_{\perp}$ to $\mu$ using equation (\ref{mu-Uperp-cold}) gives
\begin{align} \label{ni-closed-cold-1}
n_{\text{i,cl}}(x) = \frac{n_{\infty} v_{\text{B}}}{2\pi}  \int_{\bar{x}_{\text{m}}(x)}^{\infty} \Omega d\bar{x} \int_0^{ \infty } \frac{  2  \sqrt{ \chi''(x_{\text{m}}) } \delta (  \mu )  d\mu}{\sqrt{2\left(\sqrt{ \chi''(x_{\text{m}}) } \mu + \chi_{\text{m}}(\bar{x}) - \chi (x, \bar{x}) \right)}} \nonumber \\ \times \int_{\Omega \mu}^{\infty} \frac{ \delta   \left( U  -  v_{\text{B}}^2 / 2 \right) dU  }{\sqrt{2\left( U - \chi_{\text{m}}(\bar{x}) - \chi''(x_{\text{m}}) \mu \right)}}  \left( 1 + O\left( \frac{\rho_x^2}{l^2} \right) \right)   \text{.}
\end{align}
In equation (\ref{ni-closed-cold-1}), the upper limit of integration in $\mu$ was extended to $\infty$ because $\delta (\mu)$ is zero for $\mu \neq 0$ (in practice, $F (\mu, U)$ is small for orbits with $\mu \gg \tau v_{\text{B}}^2 / \Omega $).

To calculate the integral in equation (\ref{ni-closed-cold-1}), we change variable from $\bar{x}$ to $x_{\text{m}}$ and change the order of integration so that the integral over $x_{\text{m}}$ is carried out first.
By using the relation (\ref{dxbardxm}) for $d\bar{x}/dx_{\text{m}}$, and taking $\chi''(x_{\text{m}}) = \chi''(x) \left( 1 - \rho_x / l  + O \left(  \rho_x^2 / l^2 \right) \right) $, equation (\ref{ni-closed-cold-1}) becomes
\begin{align} \label{ni-closed-cold-2}
n_{\text{i,cl}}(x) = \frac{n_{\infty} v_{\text{B}} \chi''(x) }{2\pi \Omega^2} \int_0^{\infty}  \delta  (  \mu )  d\mu   \int_{\Omega\mu}^{\infty} \frac{ \delta   \left( U  -  v_{\text{B}}^2 / 2 \right)  dU  }{\sqrt{2\left( U - \frac{1}{2} \left(  \phi'(x) / B \right)^2 - \Omega \phi (x) / B \right) }}   \nonumber \\  \times    \int_{x-\frac{ \sqrt{2\mu}}{(\chi''(x))^{1/4}}}^{x+\frac{ \sqrt{2\mu} }{(\chi''(x))^{1/4}}}  \frac{2  \sqrt{ \chi''(x) } dx_{\text{m}}}{\sqrt{2 \sqrt{ \chi''(x) } \mu - \chi'' (x) \left( x - x_{\text{m}} \right)^2  }}  \left( 1 + O\left( \frac{\rho_x^2}{l^2} \right) \right)  \text{.}
\end{align}
Note that, when Taylor expanding the integrand, the terms linear in $\rho_x = x- x_{\text{m}}$ coming from the correction to $\chi''(x_{\text{m}}) \simeq \chi''(x)$ integrate to zero.
Hence, the size of the relative error has remained $O(\rho_x^2/l^2)$. 
The rightmost integral evaluates to $2\pi$, and thus equation (\ref{ni-closed-cold-2}) becomes
\begin{align} \label{ni-closed-cold-3}
n_{\text{i,cl}}(x) = \frac{n_{\infty} v_{\text{B}}\chi''(x) }{\Omega^2} \int_0^{\infty} \delta   \left( \mu \right)   d\mu  \int_{\Omega\mu}^{\infty} \frac{ \delta   \left( U  - v_{\text{B}}^2 / 2 \right)  \left( 1 + O\left( \rho_x^2 / l^2 \right) \right)   dU  }{\sqrt{2\left( U - \frac{1}{2} \left(  \phi'(x) / B \right)^2 - \Omega \phi (x) / B \right)}}  \text{.}
\end{align}
The straightforward integrals over Dirac delta functions give the density of closed orbits in (\ref{ni-closed-cold}).

\subsection{Open orbit density} 
\label{subapp-cold-open}

Expanding the integrand in equation (\ref{ni-open}) gives
\begin{align} \label{vxopen-expanded}
\sqrt{2\left( \Delta_{\text{M}}(\bar{x}, U) + \chi_{\text{M}}(\bar{x}) - \chi(x, \bar{x}) \right) }  - \sqrt{2\left( \chi_{\text{M}}(\bar{x}) - \chi(x, \bar{x}) \right) }   \simeq \frac{ \Delta_{\text{M}}(\bar{x}, U)  }{\sqrt{ 2\left( \chi_{\text{M}}(\bar{x}) -  \chi(x, \bar{x}) \right) } } \text{.}
\end{align}
By changing variable from $\bar{x}$ to $\mu$, substituting (\ref{vxopen-expanded}) and inserting $\chi_{\text{M}}(\bar{x}) - \chi(x, \bar{x}) =  \chi_{\text{c}} -  \chi(x, \bar{x}_{\text{c}}) + O(\tau^2 \epsilon v_{\text{B}}^2) $ (recall the discussion preceding equation (\ref{vx-open-smalltau})), where $\chi_{\text{c}} = \chi (x_{\text{c}}, \bar{x}_{\text{c}} ) $, the integral (\ref{ni-open}) simplifies to
\begin{align} \label{ni-op-int}
n_{\text{i,op}} = \frac{1}{\sqrt{ 2\left( \chi_{\text{c}} -  \chi( x, \bar{x}_{\text{c}} )  \right) +  O(\tau^2 \epsilon v_{\text{B}}^2)  }   }    \int_{0}^{\infty}  \left. \frac{ d\mu}{d\bar{x}}  \right\rvert_{\text{open}}^{-1}   \Omega d\mu  \nonumber \\ \times \int_{ \Omega \mu }^{\infty} \frac{ F(\mu, U ) \Delta_{\text{M}} (\bar{x}_{\text{c}}, U)  dU}{\sqrt{2\left( U - \chi_{\text{c}} \right) }} 
\text{.}  
\end{align}
Inserting the relation (\ref{DeltaM-mu}) into (\ref{ni-op-int}) gives
\begin{align} \label{ni-op-int-2}
n_{\text{i,op}} = \frac{2\pi\alpha}{\sqrt{ 2\left( \chi_{\text{c}} -  \chi( x, \bar{x}_{\text{c}} )  \right) +  O(\tau^2 \epsilon v_{\text{B}}^2 )  }   }    \int_{0}^{\infty}   \Omega d\mu  \int_{ \Omega \mu }^{\infty} F(\mu, U ) dU
\text{.}  
\end{align}
Using (\ref{F-cold}) for the distribution function, the density of open orbits becomes (\ref{ni-open-cold}).

\section{Integrals of distribution functions (\ref{f-infty})}
\label{app-integrals-Tdep}

We proceed to derive equations (\ref{N-infty})-(\ref{flow-r}) for the values of $\mathcal{N}$, $u$, $r$ and $u_{z\infty}$ associated with the distribution functions in (\ref{f-infty}).
Integrating (\ref{f-infty}) over $v_y$ and $v_z$, we obtain the functions
\begin{align} \label{f-inftyz}
f_{\infty z} (v_z) = \int f_{\infty}( \vec{v}) dv_x dv_y =  \begin{cases}
\mathcal{N}  n_{\infty} \frac{4 v_z^2}{\sqrt{\pi} v_{\text{t,i}}^3}   \exp \left( - \frac{ \left( v_z - u v_{\text{t,i}}  \right)^2 }{v_{\text{t,i}}^2} \right) \Theta \left( v_z \right) & \text{ for } \tau \leqslant 1 \text{,} \\
\mathcal{N}  n_{\infty}  \frac{ 4 v_z^2 }{  \sqrt{\pi} v_{\text{t,i}} \left( v_{\text{t,i}}^2 +r v_z^2 \right)} \exp \left( - \frac{v_z^2 }{ v_{\text{t,i}}^2 } \right)\Theta \left( v_z \right)  & \text{ for } \tau > 1 \text{.}
\end{cases} 
\end{align}
All the integrals in this appendix are carried out using the dimensionless variables $\tilde{w}_z = v_z / v_{\text{t,i}} - u$ and $\tilde{v}_z = v_z / v_{\text{t,i}} $.

Using (\ref{f-inftyz}), the normalization condition (\ref{N-infty}) is
\begin{align} \label{normcondz}
n_{\infty} = \int_0^{\infty} f_{\infty z} (v_z ) dv_z \text{.}
\end{align}
Evaluating equation (\ref{normcondz}) for $\tau \leqslant 1$, and changing integration variable to $\tilde{w}_z $ gives 
\begin{align}
n_{\infty} 
& =   \mathcal{N}  n_{\infty} \frac{4 }{\sqrt{\pi} } \int_{-u}^{\infty}  \left( \tilde{w}_z + u \right)^2  \exp \left( -  \tilde{w}_z^2  \right) d\tilde{w}_z \text{.}
\end{align}
Thus,
\begin{align} \label{N-u}
 \frac{4 \mathcal{N}  }{\sqrt{\pi} } \int_{-u}^{\infty}  \left( \tilde{w}_z^2  + 2\tilde{w}_z u + u^2 \right)   \exp \left( -  \tilde{w}_z^2  \right) d\tilde{w}_z = 1 \text{,}
\end{align}
The integral in equation (\ref{N-u}) evaluates to
\begin{align}
\int_{-u}^{\infty}  \left( \tilde{w}_z^2  + 2\tilde{w}_z u + u^2 \right)   \exp \left( -  \tilde{w}_z^2  \right)  d\tilde{w}_z = \frac{\sqrt{\pi}}{4} \left( 1 + 2u^2 \right) \left( 1 + \text{erf} (u) \right) + \frac{u}{2}  \exp(-u^2)    \text{.}
\end{align}
Hence, equation (\ref{N-infty}) for $\tau \leqslant 1$ follows.

Evaluating equation (\ref{normcondz}) for $\tau >1$, one finds, after changing the integration variable to $\tilde{v}_z$,
\begin{align} \label{N-r-1}
n_{\infty} 
=  \frac{ 4 \mathcal{N}  n_{\infty} }{  \sqrt{\pi}  } \int_0^{\infty}  \frac{ \tilde{v}_z^2  \exp \left( - \tilde{v}_z^2   \right) }{   1 +r \tilde{v}_z^2  }    \text{.}
\end{align}
The last integral in equation (\ref{N-r-1}) is calculated in the following way.
First, one can obtain the integral of the function $\exp(- \tilde{v}_z^2)/(1+r\tilde{v}_z^2)$ (which will be useful when imposing the kinetic Chodura condition (\ref{kinetic-Chodura-marginal}) in the next paragraph).
Re-expressing $1/(1+r\tilde{v}_z^2) = \int_0^{\infty} \exp\left( -\eta \left( 1+r\tilde{v}_z^2 \right) \right) d\eta$, one has
\begin{align}
\int_0^{\infty} \frac{\exp(-\tilde{v}_z^2) }{1+r\tilde{v}_z^2} dx & = \int_0^{\infty} d\eta \exp(-\eta)  \int_0^{\infty} \exp \left( -\left( 1 + \eta r \right) \tilde{v}_z^2 \right) d\tilde{v}_z  \nonumber \\
& =  \frac{\sqrt{\pi}}{2} \int_0^{\infty} \frac{ \exp(-\eta) }{ \sqrt{\eta r + 1} } d\eta \text{.}
\end{align}
Changing the integration variable to $\xi = \sqrt{ \eta + 1 / r}$ gives
\begin{align} \label{fancy-int-0}
\int_0^{\infty} \frac{\exp(-\tilde{v}_z^2) }{1+r\tilde{v}_z^2} dx & =  \sqrt{\frac{\pi}{r}} \exp \left( \frac{1}{r} \right) \int_{1/\sqrt{r}}^{\infty} \exp(-\xi^2) d\xi   \nonumber \\
&  =  \frac{\pi}{2\sqrt{r}} \exp \left( \frac{1}{r} \right) \left[ 1 - \text{erf} \left( \frac{1}{\sqrt{r}} \right) \right] \text{.}
\end{align}
Then, using the relation
\begin{align}
\int_0^{\infty} \frac{\exp(-\tilde{v}_z^2) }{1+r\tilde{v}_z^2} d\tilde{v}_z + r\int_0^{\infty} \frac{\tilde{v}_z^2 \exp(-\tilde{v}_z^2) }{1+r\tilde{v}_z^2} d\tilde{v}_z = \int_0^{\infty} \exp(-\tilde{v}_z^2)   d\tilde{v}_z = \frac{ \sqrt{\pi}}{2}  \text{,}
\end{align}
the integral 
\begin{align} \label{fancy-int-2}
\int_0^{\infty} \frac{\tilde{v}_z^2 \exp(-\tilde{v}_z^2) }{1+r\tilde{v}_z^2} d\tilde{v}_z = \frac{\sqrt{\pi}}{2r} - \frac{\pi}{2r^{3/2}} \exp \left( \frac{1}{r} \right) \left[ 1 - \text{erf} \left( \frac{1}{\sqrt{r} }\right) \right]
\end{align}
is obtained.
Inserting this integral into (\ref{N-r-1}), we obtain the expression for $\mathcal{N}$ in (\ref{N-infty}).

Equation (\ref{kinetic-Chodura-marginal}) is used to obtain the values of the positive constants $u$ and $r$.
For $\tau \leqslant 1$, one inserts the distribution function (\ref{f-inftyz}) into (\ref{kinetic-Chodura-marginal}) and changes variable to $\tilde{w}_z = v_z / v_{\text{t,i}} - u$ to obtain
\begin{align} \label{Chodura-u-1}
 \frac{ v_{\text{t,i}}^2  }{v_{\text{B}}^2} =  \frac{4  \mathcal{N} }{\sqrt{\pi} } \int_{-u}^{\infty}   \exp \left( -  \tilde{w}_z^2  \right) d \tilde{w}_z  = 2  \mathcal{N}  \left[ 1 + \text{erf} \left(  u  \right) \right] \text{.}
\end{align}
Rearranging equation (\ref{Chodura-u-1}) and inserting the value of $\mathcal{N}$ gives equation (\ref{u-def}).
For $\tau > 1$, one changes variable to $\tilde{v}_z = v_z / v_{\text{t,i}} $ in the integral (\ref{kinetic-Chodura-marginal}) to obtain
\begin{align}
 \frac{ v_{\text{t,i}}^2  }{v_{\text{B}}^2} =  \frac{ 4 \mathcal{N}  }{ \sqrt{\pi}  } \int_0^{\infty}  \frac{   \exp \left( - \tilde{v}_z^2   \right) }{   1 +r \tilde{v}_z^2  } d\tilde{v}_z \text{.}
\end{align}
Inserting the value of $\mathcal{N}$ and the integral in equation (\ref{fancy-int-0}) gives equation (\ref{r-def}).

The ion fluid velocity is evaluated using
 \begin{align} \label{uzinfty}
 u_{z\infty} = \frac{1}{n_{\infty}} \int f_{\infty z} ( v_z ) v_{z} dv_z \text{.}
 \end{align}
 For $\tau \leqslant 1$ one has
 \begin{align} \label{uz-u-int}
\frac{  u_{z\infty} }{v_{\text{t,i}}}   
  = \frac{4 \mathcal{N}  n_{\infty} }{\sqrt{\pi}}   \int_{-u}^{\infty}  \left( \tilde{w}_z + u \right)^3   \exp \left( -   \tilde{w}_z^2   \right) d\tilde{w}_z 
 \text{.}
 \end{align}
 The integrals in (\ref{uz-u-int}) evaluate to
 \begin{align}
 \int_{-u}^{\infty}  \left( \tilde{w}_z + u \right)^3   \exp \left( -   \tilde{w}_z^2   \right) d\tilde{w}_z  &  =  \int_{-u}^{\infty}  \left( \tilde{w}_z^3 + 3\tilde{w}_z^2 u + 3 \tilde{w}_z u^2 + u^3 \right)  \exp \left( -   \tilde{w}_z^2   \right) d\tilde{w}_z \nonumber \\
 & =    \frac{  \sqrt{ \pi } u  }{4}  \left( 3 + 2u^2 \right)\left[ 1 + \text{erf} \left( u \right) \right] + \frac{1}{2}  \left( u^2 + 1 \right) \exp(-u^2) \text{,}
 \end{align}
 giving (\ref{flow-u}).
 For $\tau > 1$, one has
  \begin{align} \label{uzinfty-r-int}
\frac{  u_{z\infty} }{v_{\text{t,i}}}  
  = \frac{4 \mathcal{N}  n_{\infty} }{\sqrt{\pi}} \int_0^{\infty}  \frac{  \tilde{v}_z^3 }{ 1 +r \tilde{v}_z^2 } \exp \left( - \tilde{v}_z^2  \right) d\tilde{v}_z \text{.}
 \end{align}
 The integral in equation (\ref{uzinfty-r-int}) is calculated, as before, by expressing $1/(1+r\tilde{v}_z^2)$ as a definite integral,
 \begin{align}
 \int_0^{\infty}  \frac{  \tilde{v}_z^3 }{ 1 +r \tilde{v}_z^2 } \exp \left( - \tilde{v}_z^2  \right) d\tilde{v}_z & =   \int_0^{\infty} d\eta \exp(-\eta)  \int_0^{\infty} \tilde{v}_z^3 \exp \left( - \tilde{v}_z^2 \left( 1 + \eta r\right) \right) d\tilde{v}_z \nonumber \\
 & =   \int_0^{\infty} \frac{ \exp(-\eta) }{ 2 \left( 1 + \eta r \right)^2 } d\eta \text{.}
\end{align}
Then, integrating by parts and changing the integration variable to $\xi = \eta + 1 / r$ gives
\begin{align}
  \int_0^{\infty}  \frac{  \tilde{v}_z^3 }{ 1 +r \tilde{v}_z^2 } \exp \left( - \tilde{v}_z^2  \right) d\tilde{v}_z  & =   \frac{1}{2r} - \frac{1}{2r} \int_0^{\infty} \frac{ \exp(-\eta) }{  1 + \eta r  } d\eta \nonumber \\
 & =   \frac{1}{2r} - \frac{\exp(1/r)}{2r^2} \int_{1/r}^{\infty} \frac{ \exp(-\xi) }{ \xi } d\xi \text{,}
 \end{align}
 Using the definition of the exponential integral in equation (\ref{E1}), we obtain
 \begin{align}
   \int_0^{\infty}  \frac{  \tilde{v}_z^3 }{ \left( 1 +r \tilde{v}_z^2 \right)} \exp \left( - \tilde{v}_z^2  \right) d\tilde{v}_z  = \frac{1}{2r} -  \frac{\exp(1/r)}{2r^2} E_1 \left(\frac{1}{r} \right)  \text{,}
 \end{align}
 leading to equation (\ref{flow-r}).

\bibliographystyle{jpp}

\bibliography{gyrokineticsbibliography}

\end{document}